\documentclass{aa}  

\usepackage{graphicx}
\usepackage{txfonts}
\usepackage[colorlinks = true,
            linkcolor = blue,
            urlcolor  = black,
            citecolor = cyan]{hyperref}
\newcommand{\ppxf}{pPXF}
\newcommand {\ha} {H$\alpha$}
\newcommand {\hb} {H$\beta$}

\newcommand{\nii}{[N\,\textsc{ii}]}
\newcommand{\sii}{[S\,\textsc{ii}]}

\newcommand{\hi}{H\,\textsc{i}}
\newcommand{\nai}{Na\,\textsc{i}}

\newcommand {\kms} {\,{\rm km\,s}^{-1}}

\newcommand {\ergs} {\,{\rm erg\,s}^{-1}}

\newcommand {\kpc} {\,{\rm kpc}}
\newcommand {\Mpc} {\,{\rm Mpc}}
\newcommand {\cmmq}{\,{\rm cm^{-2}}}
\newcommand {\cmmc}{\,{\rm cm^{-3}}}

\newcommand {\de}{^{\circ}}

\newcommand {\um}{\,{$\mu$\rm m}}

\newcommand {\msun}{\,{\rm M}_\odot}

\newcommand{\Myr}{\,{\rm Myr}}

\newcommand{\K}{\,{\rm K}}

\newcommand {\msunyr}{\,{{\rm M}_\odot\,\rm yr}^{-1}}

\def\apjl{ApJ}%
\def\apjs{ApJS}%
\def\ao{Appl.~Opt.}%
\def\aap{A\&A}%
\begin{document}

\title{Shaken, not blown: the gentle baryonic feedback of nearby starburst dwarf galaxies}
\titlerunning{Ionised winds in starburst galaxies}
\authorrunning{A. Marasco et al.}

   \author{A. Marasco\inst{1,2},
          F. Belfiore\inst{1},
          G. Cresci\inst{1},
          F. Lelli\inst{1},
          G. Venturi\inst{3},
          L.\,K. Hunt\inst{1},
          A. Concas \inst{1,4,5},
          A. Marconi\inst{1,5},
          F. Mannucci\inst{1},
          M. Mingozzi\inst{6},
          A.\,F. McLeod \inst{7,8},
          N. Kumari \inst{9},
          S. Carniani \inst{10},
          L. Vanzi \inst{11}
          M. Ginolfi \inst{4,5}
          }

\institute{INAF -- Arcetri Astrophysical Observatory, Largo E. Fermi 5, 50127, Firenze, Italy\\
\email{antonino.marasco@inaf.it}
\and
INAF -- Padova Astronomical Observatory, Vicolo dell?Osservatorio 5, I-35122 Padova, Italy;
\and
Instituto de Astrof\'isica, Facultad de F\'isica, Pontificia Universidad Cat\'olica de Chile, Avda. Vic\~una Mackenna 4860, 8970117 Macul, Santiago, Chile
\and
European Southern Observatory, Karl-Schwarzschild-Strasse 2, D-85748 Garching bei Muenchen, Germany
\and
Dipartimento di Fisica e Astronomia, Università di Firenze, Via G. Sansone 1, 50019, Sesto Fiorentino (Firenze), Italy
\and
Space Telescope Science Institute, 3700 San Martin Drive, Baltimore, MD 21218, USA
\and
Centre for Extragalactic Astronomy, Department of Physics, Durham University, South Road, Durham DH1 3LE, UK
\and 
Institute for Computational Cosmology, Department of Physics, University of Durham, South Road, Durham DH1 3LE, UK
\and
AURA for the European Space Agency (ESA), Space Telescope Science Institute, 3700 San Martin Drive, Baltimore, MD 21218, USA
\and
Scuola Normale Superiore, Piazza dei Cavalieri 7, Pisa I-56126, Italy
\and
Centro de Astro Ingenieria, Pontificia Universidad Catolica de Chile, Santiago, Chile}
\date{Received ; accepted}


\abstract
{
Baryonic feedback is expected to play a key role in regulating the star formation of low-mass galaxies by producing galaxy-scale winds associated with mass-loading factors $\beta\!\sim\!1\!-\!50$. 
We have tested this prediction using a sample of 19 nearby systems with stellar masses $10^7\!<\!M_\star/\msun\!<\!10^{10}$, mostly lying above the main sequence of star-forming galaxies. 
We used MUSE@VLT optical integral field spectroscopy to study the warm ionised gas kinematics of these galaxies via a detailed modelling of their \ha\ emission line. 
The ionised gas is characterised by irregular velocity fields, indicating the presence of non-circular motions of a few tens of $\kms$ within galaxy discs, but with intrinsic velocity dispersion of $40\!-\!60\kms$ that are only marginally larger than those measured in main-sequence galaxies.
Galactic winds, defined as gas at velocities larger than the galaxy escape speed, encompass only a few percent of the observed fluxes.
Mass outflow rates and loading factors are strongly dependent on $M_\star$, star formation rate (SFR), SFR surface density and specific SFR.
For $M_\star$ of $10^8\msun$ we find $\beta\simeq0.02$, which is more than two orders of magnitude smaller than the values predicted by theoretical models of galaxy evolution.
In our galaxy sample, baryonic feedback stimulates a gentle gas cycle rather than causing a large-scale blow out.}

\maketitle
\section{Introduction}\label{s:intro}

Feedback from star formation and/or active galactic nuclei (AGN) is expected to profoundly affect the evolution of low-mass galaxies by launching large-scale winds that can easily escape the shallow gravitation potential of these systems.
While this idea goes back to almost half a century ago \citep{Larson74,Saito79}, in the last decade feedback has been systematically invoked as the main physical mechanism capable of resolving a number of tensions between theoretical models of galaxy evolution in the $\Lambda$ cold dark matter ($\Lambda$CDM) framework and the observed properties of low-mass galaxies \citep[for a review see][and references therein]{BBK17}.
On global scales, feedback offers a natural explanation to the relatively small number density of dwarf galaxies compared to that of low-mass dark matter halos \citep[the so-called `missing-satellite problem'][]{Klypin+99b,Moore+99} by suppressing star formation in the low-mass regime \citep[e.g.][]{Sawala+16}, and shaping the mass-metallicity relation \citep[e.g.][]{MaiolinoMannucci19,Tortora+22}, by efficiently ejecting metal-enriched gas out of low-mass galaxy discs \citep[e.g.][]{Brooks+07}.
On local scales, violent and recurring feedback episodes are expected to produce a flattening of the dark matter profile in the central regions of galaxies \citep[e.g.][]{Governato+12}, leading to slowly rising rotation curves similar to those observed \citep{deBlok10}.
Also, centrally concentrated feedback episodes can selectively remove low-angular momentum material from the galaxy innermost region, leading to the formation of bulge-less, low-mass discs \citep[e.g.][]{Governato+10,Brook+11}.
While these effects were traditionally attributed to feedback from star formation, recent theoretical studies \citep{Silk17,Koudmani+21} supported by observations in the X-ray, optical and near-infrared bands \citep[e.g.][]{Baldassare+17, Baldassare+18, Kaviraj+19, Birchall+20}, have highlighted the importance of feedback from AGN in the evolution of dwarf galaxies, especially in the early Universe. Hereafter, we generally refer to stellar and AGN feedback as `baryonic' feedback.

In spite of its importance in shaping galaxy evolution, a comprehensive and quantitative understanding of baryonic feedback physics is still missing from both the observational and the theoretical side.
Theoretical models of stellar feedback-driven winds must deal with the impracticability of modelling all the affected physical scales at the same time, which range from a few pc, necessary to follow the evolution of single supernova blast-waves, to the several tens of kpc required to track the wind propagation throughout the galaxy halos.
Large-scale cosmological hydrodynamical suites such as EAGLE \citep{Schaye+15} or Illustris TNG \citep{Pillepich+18} do not resolve single supernovae and make use of sub-grid recipes to describe star formation and stellar feedback processes. 
In these models, feedback energy from single `star' elements is deposited onto the surrounding gas in thermal or kinetic forms, driving galactic winds.
These models are optimal to follow the long-term gas cycle in galaxies produced by feedback, but the detailed physical properties of the outflowing gas (e.g. its temperature and density distribution) depend on the feedback implementation, which varies from one simulation to another.
On the other hand, detailed hydrodynamical models of stellar feedback such as that of \citet{KimOstriker18} can accurately track the interaction between single supernova explosions and the multi-phase, magnetised interstellar medium (ISM), leading to realistic predictions for the wind launching conditions, but with the drawback that the long-term evolution of the gas in the outflow remains unknown.
Similar considerations, but with even larger uncertainties, are applicable to feedback from supermassive black holes, for which the impossibility to model sub-pc scale accretion discs in large-scale simulations is combined with severe theoretical uncertainties on the AGN-driven wind propagation mechanism \citep[e.g.][]{King10, KingPounds15, Richings+18, Costa+20}.

While observations are potentially key to constrain feedback and wind propagation models \citep[e.g.][]{CollinsRead22}, they are limited by two factors.
The first is that measurements of wind properties rely on assumptions on the ionisation state, 3D geometry, chemical composition, and kinematics of the gas. 
It is no surprise that reported mass-loading factors (defined as the ratio between the mass outflow rate and the star formation rate) range widely from $0.01$ to $10$ \citep{Veilleux+05}, and can vary up to a factor of $10$ in the same galaxy depending on the assumed conditions \citep{Chisholm+16b, Chisholm+17}.
The second is that feedback-driven outflows have low surface brightness and are multi-phase, thus the study of each phase requires deep observations with a dedicated instrument.
The hot ($T\sim10^6\K$) wind phase, made by gas shock-heated by supernova blast waves, has been observed in the X-ray only in a small number of low-mass systems in the nearby Universe \citep[e.g.][]{Heckman+95,Summers+03,Ott+05}.
Atomic and molecular outflows, originating either from cold ISM entrained in the wind or from the cooling of the hot outflowing gas, can be traced by the \hi\ (atomic) or CO (molecular) emission lines \citep[e.g.][]{Walter+02,KobulnickySkillman08,Bolatto+13, Lelli+14b,DiTeodoro+19, Fluetsch+19,Fluetsch+21} as well as the \nai\ absorption doublet \citep{Schwartz+04,Concas+19}. 
Warm ionised winds can be traced by optical emission lines and are thought to have an origin similar to the colder phase. 
However, until recently \citep[][see below]{McQuinn+19}, only a few observational constraints on ionised winds in dwarf galaxies existed, mostly coming from the characterisation of expanding superbubbles \citep[e.g.][]{Marlowe+95,Martin96,Martin98,vanEymeren+09,Heckman+15}.

Crucially, there is no consensus on which phase should dominate the outflow mass and energy budget.
Models like those of \citet{KimOstriker18} predict that gas at temperatures of $0.5$--$2\times10^4\K$ - thus visible in \hi\ or \ha, depending on the ionisation conditions - dominates the wind mass-loading, whereas most of the wind energy is in the hot phase.
Observations, instead, tend to find winds that are dominated in mass and kinetic power by the molecular phase \citep{Fluetsch+19, Fluetsch+21}. 
This is the case both for galaxies with and without an AGN,
although warm ionised gas may dominate the outflow mass budget in the most luminous quasars \citep{Fiore+17}.

Starburst dwarf galaxies, generally intended as low-mass ($M_\star\!<\!10^{10}\msun$) systems which lie above the main-sequence of star formation \citep{Noeske+07,Popesso+19}, represent an ideal laboratory to study baryonic feedback in the dwarf regime, as they combine large SFRs with shallow gravitational potential wells.
The conditions that trigger the starburst in these systems are highly debated, and both external and internal mechanisms have been proposed such as direct accretion from extra-galactic cold flows \citep{DekelBirnboim06}, tidal perturbations from nearby companions \citep{Noguchi88,Lelli+14a}, wet mergers \citep{Bekki08,Lelli+12a}, torques due to star-forming clumps \citep{Elmegreen+12} or radial flows produced by triaxial dark matter halos \citep{BekkiFreeman02,Marasco+18}. 
Interestingly, \hi\ observations of starburst dwarfs have revealed a balanced mixture of regularly rotating and kinematically disturbed discs, sometimes featuring strong radial motions, but unsettled \hi\ distributions are rare.
These systems have both baryonic and gas fractions similar to those of typical dwarf irregulars, indicating that they did not eject a large amount of gas out of their potential wells \citep{Lelli+14a}.
A similar conclusion was recently reached by \citet{McQuinn+19}, who studied the properties of ionised winds in a sample of $12$ nearby starburst dwarfs using deep \ha\ imaging.
Their results show a very modest spatial extent of all detected ionised material, suggesting that the majority of gas expelled from dwarfs does not escape to the intergalactic medium but remains in the galaxy halo.
However, the study of \citet{McQuinn+19} relies on narrow and broad-band imaging alone, thus misses detailed information on ionised gas kinematics that only optical spectroscopy can provide.

In this study, we make use of the excellent combination of spatial and spectral resolution offered by MUSE to study the kinematics of the ionised gas in a sample of $19$ nearby starburst dwarfs. 
These galaxies are part of a larger sample of $40$ starburst systems with publicly available archival MUSE observations, which make the `DWarf galaxies Archival Local survey for Interstellar medium investigatioN' (\textsc{Dwalin}, Cresci et al. in prep.) sample.
Our goals are to infer ionised mass outflow rates and loading factors, and to study how these are related to a number of galaxy properties such as stellar masses ($M_\star$), star formation rates (SFRs), and mean SFR densities.

This paper is structured as follows.
In Section \ref{s:data} we provide a brief description of the \textsc{Dwalin} sample and present our measurements for $M_\star$ and SFRs.
The MUSE data analysis, which consists in the extraction and modelling of \ha\ velocity profiles aimed at inferring the main properties of the ionised winds, is presented in Sections \ref{s:kinematic} and \ref{s:winds}.
Our results are discussed in the light of other observational and theoretical studies in Section \ref{s:discussion}, and conclusions are drawn in Section \ref{s:conclusions}.


\section{The \textsc{Dwalin} sample}\label{s:data}
\begin{figure*}
\begin{center}
\includegraphics[width=1.0\textwidth]{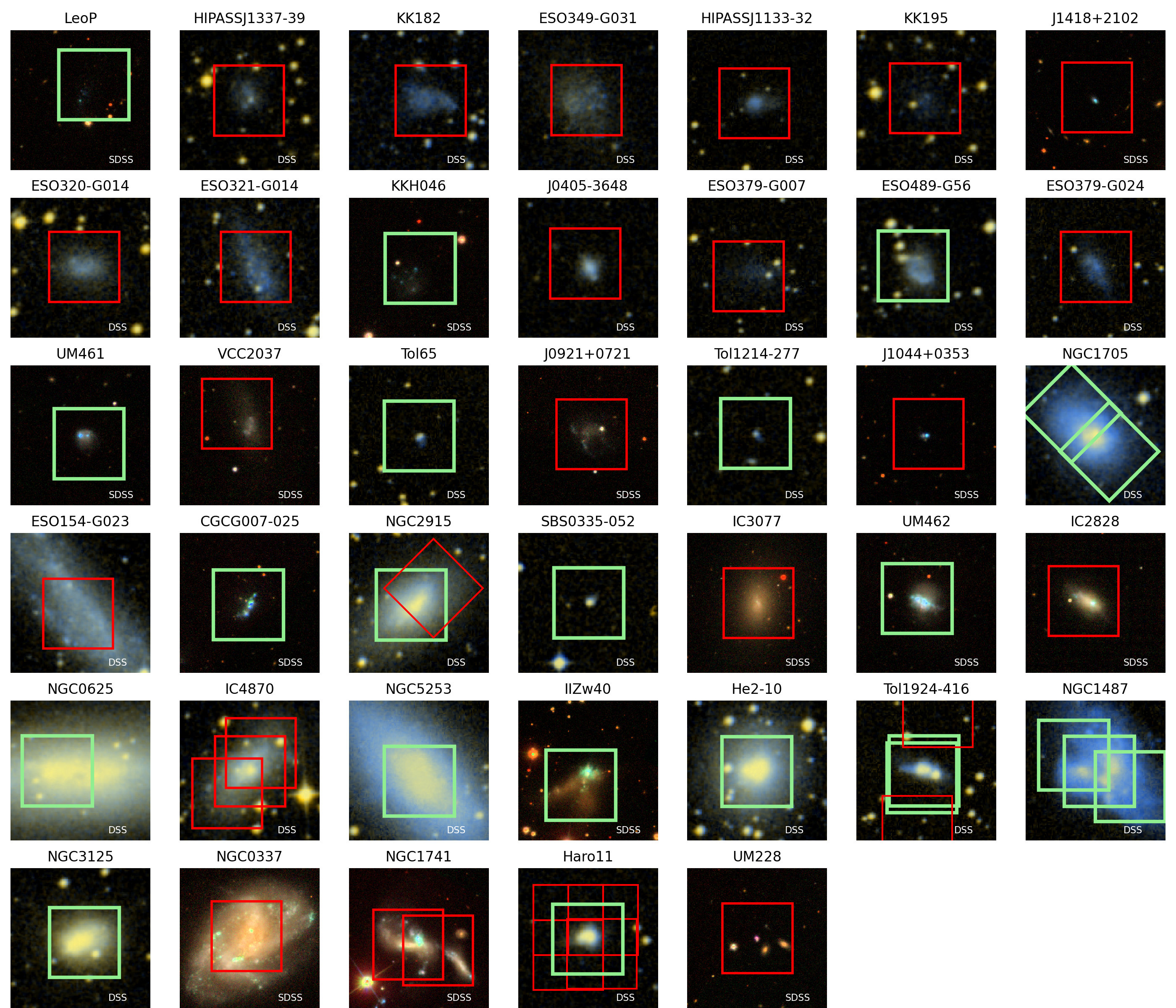}
\caption{Optical DSS or SDSS images ($2'$ on a side) of the \textsc{Dwalin} sample of starburst galaxies, with the MUSE field of view overlaid (squared overlays). Green overlays show the MUSE data that have been re-reduced and analysed and are discussed in the current study. Red overlays are used for the MUSE data that will be analysed in a future study. Galaxies are ordered by their $M_\star$, with lowest-mass galaxies at the top left.}
\label{fig:dwalin_all}
\end{center}
\end{figure*}

\begin{table*}
\caption{Main properties of the \textsc{Dwalin} galaxy sample.}
\label{tab:dwalin} 
\centering
\begin{minipage}{170mm}
\begin{tabular}{lccccccc}
\hline\hline
\noalign{\smallskip}
Galaxy & R.A.\ (J2000) & Dec.\ (J2000)  & $D$ & $\log\frac{M_\star}{\msun}$ & $\log\frac{\rm SFR}{\msunyr}$ & $R_{50}^{SFR}$ & (\small{$M_\star$,SFR})\\
    & h m s & $^\circ$ $'$ $''$ & Mpc & & & kpc & \small{method}\\ 
(1) & (2) & (3) & (4) & (5) & (6) & (7) & (8) \\
\noalign{\smallskip}
\hline
\hline
\noalign{\smallskip}

{\bf CGCG007-025} & 09 44 01.9 & -00 38 32 & $23\pm5^{\rm FM}$ & $8.07\pm0.24$ & $-0.64\pm0.13$ & $0.85\pm0.03$ & 1\\
ESO\,154-G023 & 02 56 50.4 & -54 34 17 & $6.0\pm0.5^{\rm CF3}$ & $8.18\pm0.22$ & $-1.39\pm0.10$ & $3.19\pm0.08$ & 1\\
ESO\,320-G014 & 11 37 53.2 & -39 13 13 & $6.0\pm0.5^{\rm CF3}$ & $6.98\pm0.20$ & $-2.92\pm0.14$ & $0.24\pm0.02$ & 4\\
ESO\,321-G014 & 12 13 49.6 & -38 13 53 & $3.3\pm0.2^{\rm CF3}$ & $6.98\pm0.20$ & $-3.00\pm0.10$ & $0.30\pm0.01$ & 4\\
ESO\,349-G031 & 00 08 13.4 & -34 34 42 & $3.2\pm0.3^{\rm CF3}$ & $6.55\pm0.20$ & $-3.34\pm0.10$ & $0.27\pm0.01$ & 4\\
ESO\,379-G007 & 11 54 43.5 & -33 33 36 & $5.4\pm0.5^{\rm CF3}$ & - & - & - & -\\
ESO\,379-G024 & 12 04 56.7 & -35 44 35 & $5.5\pm0.2^{\rm STD}$ & $6.71\pm0.22$ & $-2.90\pm0.11$ & $0.38\pm0.04$ & 5\\
{\bf ESO\,489-G56} & 06 26 17.6 & -26 15 56 & $6.3\pm0.6^{\rm CF3}$ & $7.47\pm0.15$ & $-3.11\pm0.17$ & $0.63\pm0.04$ & 3\\
HIPASS\,J1133-32 & 11 33 10.9 & -32 57 45 & $5.6\pm0.5^{\rm CF3}$ & $6.69\pm0.21$ & $-3.09\pm0.11$ & $0.27\pm0.03$ & 5\\
HIPASS\,J1337-39 & 13 37 25.3 & -39 53 48 & $5.1\pm0.5^{\rm CF3}$ & $5.99\pm0.27$ & $-2.73\pm0.09$ & $0.25\pm0.02$ & 1\\
{\bf Haro\,11} & 00 36 52.7 & -33 33 17 & $86\pm17^{\rm FM}$ & $10.24\pm0.25$ & $1.78\pm0.16$ & $2.23\pm0.15$ & 1\\
{\bf Henize\,2-10} & 08 36 15.1 & -26 24 34 & $8.2\pm0.8^{\rm CF3}$ & $9.05\pm0.32$ & $0.22\pm0.17$ & $0.69\pm0.01$ & 3\\
IC\,2828 & 11 27 11.2 & +08 43 53 & $13\pm2^{\rm FM}$ & $8.05\pm0.22$ & $-1.49\pm0.11$ & $0.52\pm0.02$ & 1\\
IC\,3077 & 12 15 56.3 & +14 25 59 & $5\pm2^{\rm TFR}$ & $7.78\pm0.20$ & $-3.94\pm0.38$ & $0.51\pm0.01$ & 4\\
IC\,4870 & 19 37 37.6 & -65 48 43 & $8.5\pm0.8^{\rm CF3}$ & $8.53\pm0.20$ & $-1.28\pm0.09$ & $0.64\pm0.03$ & 1\\
{\bf IIZw40} & 05 55 42.6 & +03 23 32 & $14\pm3^{\rm FM}$ & $9.03\pm0.25$ & $0.37\pm0.25$ & $1.31\pm0.11$ & 1\\
J0405-3648 & 04 05 20.3 & -36 49 01 & $9\pm2^{\rm FM}$ & $7.03\pm0.20$ & $-2.31\pm0.10$ & $0.29\pm0.01$ & 4\\
J0921+0721 & 09 21 27.2 & +07 21 53 & $21\pm4^{\rm FM}$ & $8.03\pm0.22$ & $-1.64\pm0.10$ & $1.47\pm0.03$ & 4\\
J1044+0353 & 10 44 58.0 & +03 53 13 & $52\pm10^{\rm FM}$ & $7.83\pm0.24$ & $-0.76\pm0.11$ & $0.80\pm0.08$ & 1\\
J1418+2102 & 14 18 49.9 & +21 02 26 & $29\pm6^{\rm FM}$ & $6.59\pm0.36$ & $-1.32\pm0.17$ & $5.95\pm0.14$ & 3\\
KK182 & 13 05 02.8 & -40 04 59 & $5.9\pm0.5^{\rm CF3}$ & $6.50\pm0.24$ & $-2.47\pm0.10$ & $0.49\pm0.03$ & 1\\
KK195 & 13 21 08.0 & -31 31 48 & $5.6\pm0.3^{\rm STD}$ & - & $-3.33\pm0.70$ & $0.63\pm0.02$ & 4\\
{\bf KKH046} & 09 08 36.5 & +05 17 27 & $12\pm2^{\rm FM}$ & $7.09\pm0.20$ & $-2.37\pm0.10$ & $0.74\pm0.02$ & 4\\
{\bf Leo\,P} & 10 21 45.1 & +18 05 17 & $1.6\pm0.1^a$ & $5.75\pm0.30^a$ & $-4.40\pm0.38$ & $0.07\pm0.01$ & 5\\
NGC\,0337 & 00 59 50.1 & -07 34 41 & $19\pm3^{\rm CF3}$ & $9.75\pm0.21$ & $0.12\pm0.12$ & $3.05\pm0.03$ & 1\\
{\bf NGC\,0625} & 01 35 04.6 & -41 26 10 & $4.0\pm0.4^{\rm CF3}$ & $8.60\pm0.19$ & $-1.20\pm0.14$ & $0.61\pm0.03$ & 1\\
{\bf NGC\,1487} & 03 55 46.1 & -42 22 05 & $9\pm2^{\rm FM}$ & $9.00\pm0.21$ & $-0.68\pm0.10$ & $1.16\pm0.04$ & 1\\
{\bf NGC\,1705} & 04 54 13.5 & -53 21 40 & $5.5\pm0.4^{\rm CF3}$ & $8.13\pm0.24$ & $-1.31\pm0.09$ & $0.21\pm0.01$ & 1\\
NGC\,1741 & 05 01 38.3 & -04 15 25 & $56\pm11^{\rm FM}$ & $9.80\pm0.25$ & $0.85\pm0.14$ & $3.39\pm0.23$ & 1\\
{\bf NGC\,2915} & 09 26 11.5 & -76 37 35 & $4.3\pm0.4^{\rm CF3}$ & $8.21\pm0.20$ & $-1.47\pm0.09$ & $0.44\pm0.01$ & 1\\
{\bf NGC\,3125} & 10 06 33.4 & -29 56 05 & $15\pm3^{\rm FM}$ & $8.91\pm0.24$ & $-0.15\pm0.14$ & $0.68\pm0.04$ & 1\\
{\bf NGC\,5253} & 13 39 56.0 & -31 38 24 & $3.5\pm0.2^{\rm CF3}$ & $8.64\pm0.24$ & $-0.26\pm0.15$ & $0.32\pm0.01$ & 1\\
{\bf SBS\,0335-052} & 03 37 44.1 & -05 02 40 & $59\pm12^{\rm FM}$ & $8.32\pm0.24$ & $0.07\pm0.13$ & $0.86\pm0.01$ & 1\\
{\bf Tol\,1214-277} & 12 17 17.1 & -28 02 33 & $100\pm20^{\rm FM}$ & $7.82\pm0.30$ & $-0.45\pm0.10$ & $1.46\pm0.03$ & 1\\
{\bf Tol\,1924-416} & 19 27 58.2 & -41 34 32 & $33\pm6^{\rm FM}$ & $8.87\pm0.24$ & $0.31\pm0.11$ & $0.96\pm0.05$ & 1\\
{\bf Tol\,65} & 12 25 46.5 & -36 14 01 & $33\pm6^{\rm FM}$ & $7.41\pm0.24$ & $-1.04\pm0.11$ & $0.54\pm0.08$ & 1\\
UGC\,07983 & 12 49 47.0 & +03 50 32 & $6\pm1^{\rm FM}$ & $7.08\pm0.20$ & $-3.02\pm0.11$ & $0.42\pm0.01$ & 4\\
UM\,228 & 00 21 01.0 & +00 52 48 & $423\pm84^{\rm FM}$ & $10.60\pm0.29$ & $1.37\pm0.12$ & $6.58\pm0.97$ & 1\\
{\bf UM\,461} & 11 51 33.3 & -02 22 22 & $19\pm4^{\rm FM}$ & $7.56\pm0.24$ & $-1.21\pm0.12$ & $0.53\pm0.03$ & 1\\
{\bf UM\,462} & 11 52 37.2 & -02 28 10 & $19\pm4^{\rm FM}$ & $8.36\pm0.25$ & $-0.60\pm0.10$ & $0.64\pm0.01$& 1\\
VCC\,2037 & 12 46 15.3 & +10 12 20 & $9.6\pm0.9^{\rm STD}$ & $7.66\pm0.16$ & $-2.29\pm0.09$ & $0.57\pm0.05$ & 1\\
\noalign{\smallskip}
\hline\hline
\noalign{\smallskip}
\end{tabular}
Notes. (1) Galaxy name. Systems studied in this work are highlighted in boldface. (2)-(3) Celestial coordinates in J2000 from NED; (4) galaxy distance from the Extragalactic Distance Database, based on: CF3 - Cosmicflow-3 catalogue \citep{Tully+16}; FM - flow model \citep{Kourkchi+20}; STD - stellar distances from \citet{Jacobs+09} and \citet{Anand+21}; TFR - Tully-Fisher relation distance from \citet{Kourkchi+22}; (5)-(6) $M_\star$ and SFR as determined in this work; (7) half-light radius in the \emph{GALEX} NUV band or, when this is not available, in the WISE W4 band; (8) method used to determine $M_\star$ and SFR, following the notation of Section \ref{ss:Ms_SFR}. $^a$ distance and $M_\star$ for Leo\,P are from \citet{McQuinn+15}.
\end{minipage}
\end{table*}


\textsc{Dwalin} is a sample of $40$ nearby galaxies that is specifically designed to study the gas properties in low-mass, highly star-forming systems. 
All galaxies in \textsc{Dwalin} have archival MUSE data, and have been selected either 1) from the Herschel Dwarf Galaxy Survey \citep[DGS;][]{Madden+13,Cormier+15}, a survey of nearby ($D\!<\!200\Mpc$) low-metallicity galaxies using far-infrared (FIR) and sub-millimetre data from the Herschel Space Observatory, or 2) from the \cite{Karachentsev+13} catalogue of galaxies in the local Volume (distance $<$ 11 Mpc) and $\log(M/M_\star)< 9.0$.
A detailed description of the \textsc{Dwalin} sample will be provided by Cresci et al.\,(in prep). 

The main properties of our sample are listed in Table \ref{tab:dwalin}, along with our new measurements of $M_\star$ and SFRs determined as described in Section \ref{ss:Ms_SFR}.
Most \textsc{Dwalin} galaxies have $M_\star\!<\!2\times10^9\msun$, but the sample spans more than five dex in $M_\star$ and SFR.
Distances in \textsc{Dwalin} are taken from the Extragalactic Distance Database\footnote{\url{http://edd.ifa.hawaii.edu/}} \citep[EDD;][]{Tully+09}, which collects and homogenises distance measurements from a variety of sources.
Specifically, for 16 galaxies we have used the Cosmicflows-3 distance catalogue \citep[CF3,][]{Tully+16}, which provides weighted distances for about 18,000 nearby galaxies using multiple velocity-independent methods such as Cepheids, tip of red giant branch (TRGB), type Ia supernovae, Tully-Fisher relation \citep[TFR,][]{TullyFisher77}, and others.
Some \textsc{Dwalin} galaxies have velocity-independent distances but do not appear in the CF3: for these objects we have used stellar distances from \citet{Jacobs+09} and \citet{Anand+21}, or TFR distances from \citet{Kourkchi+22}.
For Leo\,P we have adopted the estimate of \citet{McQuinn+15} based on the luminosity of horizontal branch stars and 10 RR Lyrae candidates.
Finally, for the 20 galaxies that have no velocity-independent distances we have adopted estimates based on local 3D flow models \citep{Kourkchi+20}. 
Distance uncertainties are taken from the EDD, with the exception of those determined from the flow models, for which we have assumed an error of $20\%$ (about twice the typical uncertainty of the CF3 measurements).

An atlas of the \textsc{Dwalin} sample from the Sloan Digital Sky Survey \citep[SDSS;][]{York+00} and the Digital Sky Survey (DSS) is shown in Fig.\,\ref{fig:dwalin_all}, with the field of view of the archival MUSE pointings overlaid.
As a preliminary step, we planned to re-reduce the archival MUSE data for all $40$ \textsc{Dwalin} galaxies, with the goal of providing a uniform analysis of the ionised gas kinematics and wind properties across the sample.
This step required considerable computational and human resources and, at the current time, is still in progress.
So far we have re-reduced the MUSE data for $19$ galaxies (green frames in Fig.\,\ref{fig:dwalin_all}), randomly selected from \textsc{Dwalin}: these systems comprise the \textsc{Dwalin-19} sub-sample.
As the midway point of the analysis of the complete sample, in this study we focus on the ionised gas properties of this sub-sample. 
The analysis of the $21$ remaining \textsc{Dwalin} galaxies will be presented in a later paper.

The MUSE data reduction has been carried out with the MUSE pipeline \citep{Weilbacher+20} v2.8.1, using the ESO Recipe flexible execution workbench \citep[Reflex,][]{Freudling+13}, which gives a graphical and automated way to execute with EsoRex the Common Pipeline Library (CPL) reduction recipes, within the Kepler workflow engine \citep{Altintas+06}.
Details on the MUSE data reduction of individual galaxies will be provided by Cresci et al.\,(in prep.), while the data analysis is presented in Section \ref{s:kinematic}.

\subsection{Stellar masses and star formation rates} \label{ss:Ms_SFR}
\begin{figure*}
\begin{center}
\includegraphics[width=0.65\textwidth]{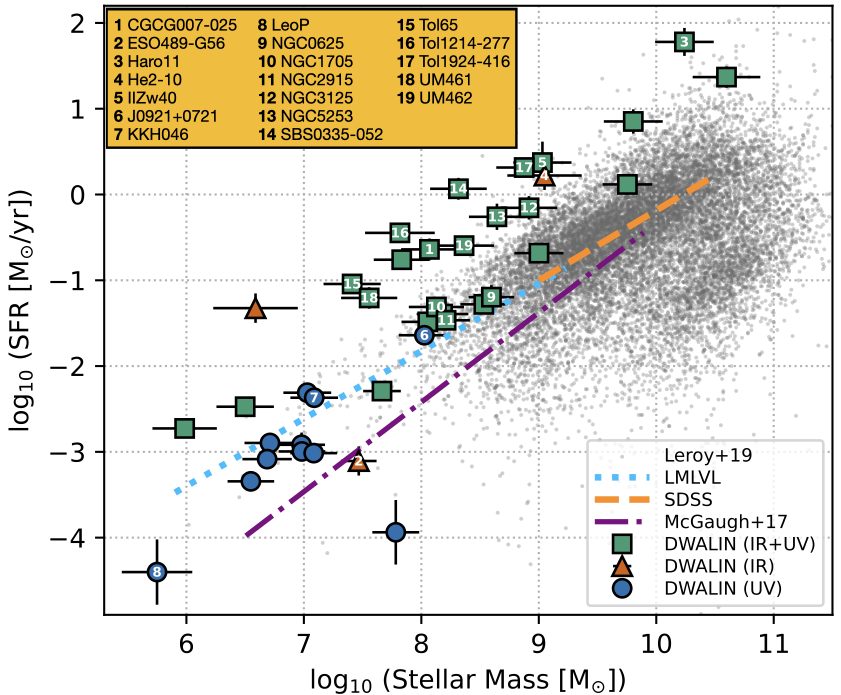}
\caption{SFR vs. $M_\star$ plot for the \textsc{Dwalin} galaxy sample. Green squares, brown triangles and blue circles indicate galaxies whose SFR measurements come from the combination of UV and MIR data, MIR data alone or UV data alone, respectively (see text for the details). ID numbers are shown for the \textsc{Dwalin-19} sample studied with MUSE in this work. Gray dots show the sample of $\approx15,750$ nearby galaxies studied by \citet{Leroy+19}. We also show the SFR-$M_\star$ trends at $z\!\simeq\!0$ from the SDSS \citep[][dashed line]{Chang+15}, from the LMLVL sample \citep[][dotted line]{Berg+12}, and the relation derived by \citet{McGaugh+17} for a sample of low-surface brightness galaxies (dot-dashed line).}
\label{fig:SFR_Ms}
\end{center}
\end{figure*}

In order to determine homogeneous and reliable estimates of $M_\star$ and SFRs for all galaxies in the \textsc{Dwalin} sample, we employ the approach outlined by \citet{Leroy+19} (see their Appendix) based on the combined use of infrared and ultraviolet data. 
We make use of archival, publicly available near- and mid- infrared (NIR and MIR) images from the Wide-field Infrared Survey Explorer \citep[WISE;][]{Wright+10} in bands W1 ($3.4$\um) and W4 ($22$\um), from the IRAC $3.6$\um\ and MIPS $24$\um\ camera onboard of the \emph{Spitzer Space Telescope} \citep{Werner+04}, and of near- and far- ultraviolet (NUV and FUV) images from the Galaxy Evolution Explorer \citep[\emph{GALEX;}][]{GildePaz+07}.
The photometric analysis of these images is performed using our own routines, which are described in detail in Appendix \ref{app:photometry}.

\citet{Leroy+19} outlined a series of methods to determine the integrated $M_\star$ and SFR in nearby galaxies using only WISE and/or \emph{GALEX} data.
All these methods are calibrated on measurements from the \emph{GALEX}-SDSS-WISE Legacy Catalogue \citep[GSWLC;][]{Salim+16,Salim+18}, which combined \emph{GALEX} and WISE photometry with SDSS observations to infer $M_\star$ and SFR for about 600,000 galaxies via population synthesis modelling with the CIGALE code \citep{Boquien+19}.
The impressive size of this sample and the refined spectral modelling, which accounts for energy balance and contamination by emission lines to the broad-band photometry, make GSWLC an excellent benchmark to assess how global galaxy parameters are recovered when only a subset of the full photometric data is available.

Following \citet{Leroy+19}, we have used five methods to compute $M_\star$ and SFR depending on the data available. These are the following, in descending order of preference:
\begin{enumerate}
\item SFR from a combination of FUV and W4 luminosities, thus accounting for both the obscured and the unobscured SF components. Then $M_\star$ is derived from W1 luminosities using a mass-to-light ratio ($\Psi_\star$) that depends on the specific SFR \citep[eq.\,24 in][]{Leroy+19}.
\item Same as method (1) but using NUV if FUV data are not available. 
\item Same as method (1) but using only WISE W4 luminosities if NUV and FUV data are not available, thus accounting only for the obscured component of the SF.
\item SFR from FUV luminosities only, if W4 data are not available, thus accounting only for the unobscured component of the SF. Then $M_\star$ is derived from W1 data using a constant $\Psi_\star\!=\!0.35$.
\item Same as method (4) but using NUV luminosities if FUV data are not available.
\end{enumerate}
We stress that each of the methods listed above has been calibrated separately by \citet{Leroy+19}, thus the coefficients adopted in the calculation of $M_\star$ and SFR vary from one method to another.
In addition to the procedures listed above, we have used IRAC $3.6$\um\ (MIPS $24$\um) luminosities as a replacement for W1 (W4) luminosities when these were not available, after having assessed the excellent agreement between our \emph{Spitzer} and WISE photometry in similar bands.
In fact, we have verified that \emph{Spitzer} and WISE data are practically interchangeable, both providing $M_\star$ and SFR measurements that are compatible within their uncertainties (computed as described in Appendix \ref{app:photometry}).
However, \emph{Spitzer} data are less-than-optimal to use for our $M_\star$ and SFR estimates, given that the procedures of \citet{Leroy+19} are specifically calibrated on WISE data.
Our NUV and FUV measurements are corrected for Galactic extinction using the reddening map of \citet{Schlegel98}.
We make a single exception to our procedure: we take the $M_\star$ of Leo\,P from \citet{McQuinn+15}, given that this very faint system is only barely detected in our NIR images.

The resulting $M_\star$ and SFRs for the \textsc{Dwalin} sample are listed in Table \ref{tab:dwalin}. 
In Fig.\,\ref{fig:SFR_Ms} we compare our measurements with those of \citet{Leroy+19} for a sample $\approx15,750$ nearby galaxies, and with the main-sequence trends at $z\!\sim\!0$ from the Sloan Digital Sky Survey \citep[SDSS;][]{Chang+15}, from the Low-Mass Local Volume Legacy (LMLVL) sample \citep{Berg+12}, and from \citet{McGaugh+17} for a sample of late-type, low surface brightness galaxies.
The majority of the \textsc{Dwalin} galaxies are located above the main-sequence of star formation. 
A possible exception is given by the faintest systems ($M_{\star}<10^8\msun$) for which, however, information on the obscured component of the SF are missing because W4 or MIPS data are unavailable (purple circles in Fig.\,\ref{fig:SFR_Ms}).
We stress that most of these faint galaxies may still lie in the starburst region for their low $M_\star$, assuming a steepening in the main-sequence relation at $M_\star \lesssim 10^9 M_\odot$ as suggested by \citet{McGaugh+17}.
We also notice that the \textsc{Dwalin-19} sub-sample (numbered systems in Fig.\,\ref{fig:SFR_Ms}) spans the same dynamical range of its parent sample.

Different studies have suggested that free-free radiation and hot dust emission can contribute to the $3$--$5$\um\ luminosity in dwarf galaxies \citep[e.g.][]{SmithHancock09}.
In this study we have assumed that such contribution is negligible.
Correcting for these effects would lead to slightly lower estimates for $M_\star$, shifting the \textsc{Dwalin} sample even further away from the main sequence of star formation. 

We have adopted the method of \citet{Leroy+19} so that the \textsc{Dwalin} sample and GSWLC sample can be compared directly.
We point out, however, that M$_\star$ calibrations from different population synthesis models \citep{Meidt+14,McGaughSchombert14,Herrmann+16,Norris+16,Hunt+19,Schombert+19,Schombert+21}, color-magnitude diagrams of resolved stellar populations \citep{Eskew+12,Zhang+17} and dynamical arguments \citep{McGaughSchombert15,Lelli+16a,Lelli+16b} give systematically higher $M_\star$ by a factor of $\sim1.5\!-\!2.0$.

\section{Ionised gas kinematics in \textsc{Dwalin}}\label{s:kinematic}
We now describe in detail the steps taken to infer the global kinematic properties of the ionised gas in the \textsc{Dwalin-19} sample.
We illustrate our procedure on the MUSE data of He\,2-10 \citep[previously studied by][]{Cresci+17}, but we proceed in the same way for all the \textsc{Dwalin-19} galaxies.

\subsection{Continuum subtraction and velocity cube creation}
The first step of our procedure consists in the subtraction of the continuum from the MUSE spectra.
This operation must be highly accurate, since even an error of few percents can artificially enhance or suppress the faint wings in the line profiles that can be associated with the outflow component.

Stellar absorption features are quite weak in the \textsc{Dwalin-19} sample but nonetheless visible, especially around the \hb\ line in the most massive systems. 
The stellar continuum is subtracted following a procedure similar to that outlined by \citet{Cresci+17}, which consists in first enhancing the stellar signal via a Voronoi tessellation \citep{CappellariCopin03} of the MUSE data (we impose S/N $\!>\!20$ on the continuum at $5100\!<\!\lambda/\AA\!<\!5500$, on average per $1.25\AA$ spectral channel), and then fitting the resulting binned cube via a multi-component model using the \ppxf\ software \citep{Cappellari17}.
The model adopted is made by a combination of E-MILES stellar population model templates \citep{MILES,Rock+16}, which cover the entire MUSE wavelength range, Gaussian features to model the main optical emission lines, and an additive third-order polynomial to account for any additional `smooth' stellar feature extended over the whole $\lambda$ range.
Multi-Gaussian fitting could be employed in order to achieve a finer  modelling of the emission lines but, at this stage, we are primarily interested in removing the stellar continuum thus we use a single Gaussian component per line. 
The outcome of this process is a model cube for the stellar continuum matched to the binned data.
This cube is then subtracted from the original (unbinned) data by re-scaling the model spectrum of each Voronoi cell to match the intensity level of individual spaxels within that cell. 

This procedure is generally robust, but is subject to small imperfections since the additional polynomial term included in our model may not be adequate to describe with sufficient accuracy the residual continuum flux.
This may produce spurious, extended wings in the line profile, mimicking the presence of broad components that could be interpreted as outflows.
We deal with this problem by means of a `local' refinement of our continuum subtraction, using an approach that is tailored around the emission lines of interest.
Specifically, we fit third order polynomials to the continuum-subtracted spectra only around small ($120\AA$-wide) spectral windows, centred around each emission line of interest, after careful masking of all the principal lines.
Visual inspection of the spectra confirms that this local approach improved considerably the continuum subtraction where we need it the most, and we employ it on the four emission lines that will be used in the rest of the analysis: \ha, \hb, \sii$\lambda6716$ and \sii$\lambda6731$.

After this additional correction, velocity cubes for the individual lines are extracted and studied separately using the multi-Gaussian decomposition method discussed below.
\ha\ and \hb\ cubes span a velocity range of $\pm600\kms$, while we build a unique cube for the \sii\ doublet, centred around the doublet centre and encompassing $\pm900\kms$.
These velocity ranges are adequate to capture virtually all the emission coming from these bright lines, while minimising the contamination from the nearby fainter lines.

\subsection{Emission-line modelling}\label{ss:multigau}
We model the velocity profiles in our cubes using a combination of Gaussian components, which we then analyse a-posteriori to assess the presence of outflows in our data.
This approach is preferred for the \textsc{Dwalin-19} galaxies which, as we show below, do not possess the highly-regular velocity fields that are typical of more massive spirals and would permit a geometric modelling of the data, e.g. via tilted ring methods \citep[e.g.][]{Rogstad+74,Barolo,Bouche+15,Concas+22}.
We proceed by distinguishing between a reference line (\ha), which is modelled first, using a complete multi-Gaussian decomposition, and `secondary' lines (all the others) for which only the amplitudes of the Gaussian components are fitted, while the mean velocities and widths of the various components are fixed to those determined for the reference line.
Adopting this separation has two advantages. 
The first is that it improves the modelling of fainter lines like the \sii\ doublet, for which an unconstrained fit may give unpredictable results, especially in the lowest S/N regions.
The second, and more relevant advantage, is that it allows us to use the same components for very different lines, which is useful in the calculation of line ratios.

In each spaxel, the \ha\ line is modelled with a variable number of Gaussian components ranging from one to four.
For each component we fit the amplitude, the mean velocity, and the `intrinsic' velocity dispersion $\sigma_{\rm int}$, defined as
\begin{equation}\label{eq:sigma}
    \sigma_{\rm obs}^2 \equiv \sigma_{\rm int}^2 + \sigma^2_{\rm MUSE}(\lambda)
\end{equation}
where $\sigma_{\rm obs}$ is the observed velocity dispersion of the \ha\ line in our velocity cube and $\sigma_{\rm MUSE}$ is the $\lambda$-dependent instrumental broadening, which is equal to $\sim50\kms$ (FWHM of $116\kms$) around the \ha\ line \citep[see eq.\,(8) of ][]{Bacon+17}. 
Following \citet{Marasco+20}, the optimal number of components is decided spaxel-by-spaxel on the basis of a Kolgomorov-Smirnov (KS) test on the residuals of the fits determined for $n$ and $n+1$ (with $1\!<\!n\!<\!3$) components: if the two residual distributions are statistically different\footnote{This is decided using a $p$-value threshold, $0\!\le\!p\!<\!1$ where $0$ ($1$) minimises (maximises) the number of components required. Here we request an accurate modelling of the fainter line details and set $p\!=\!0.95$.}, then the $n+1$ component model is preferred.
Thus this method is sensitive to the relative improvements in the data fitting due to the use of increasingly complex models, but is independent of the `goodness' of the fit in absolute terms.

\begin{figure}
\begin{center}
\includegraphics[height=0.86\textheight]{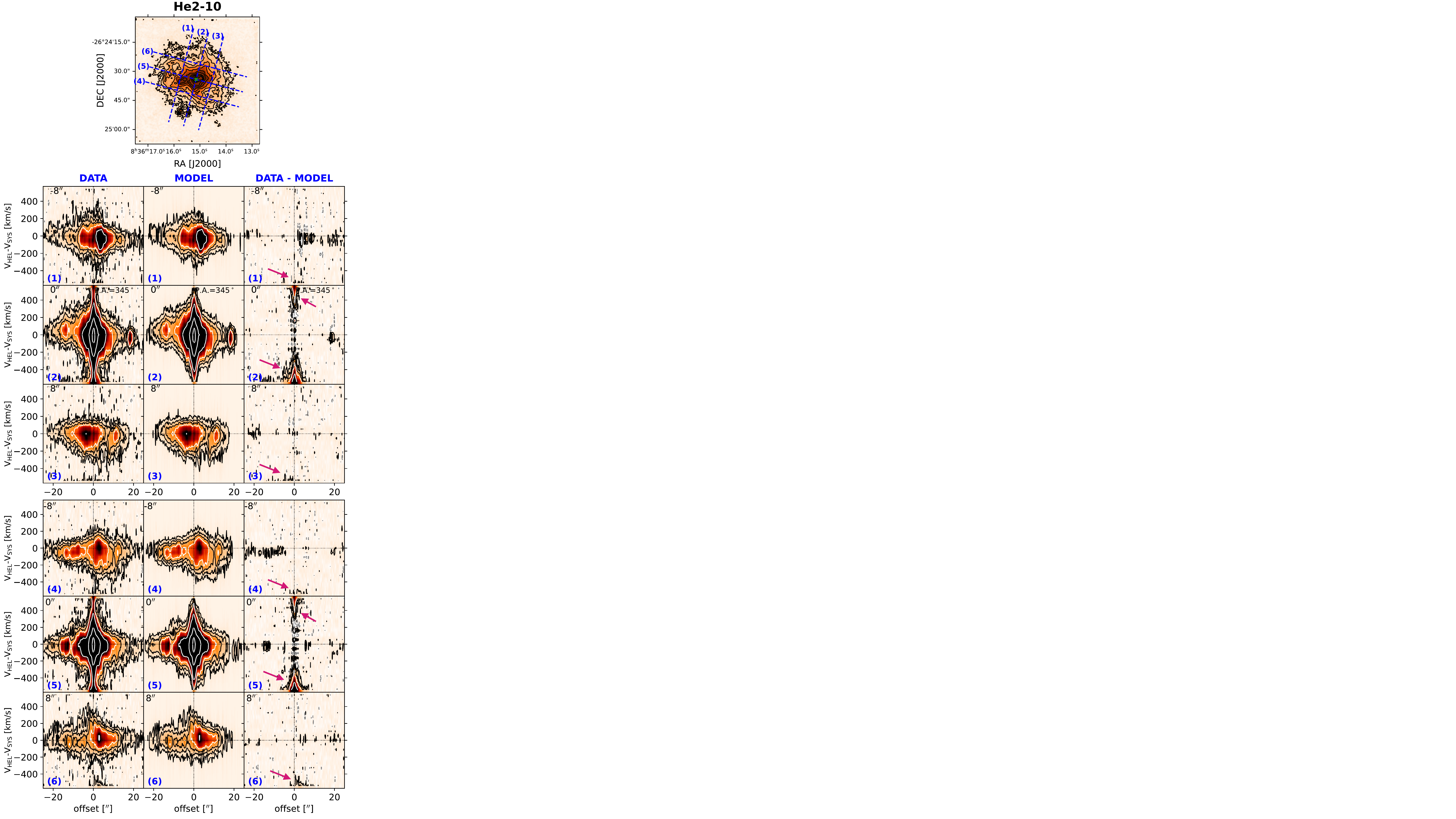}
\caption{Multi-Gaussian modelling of the \ha\ line in He\,2-10. The top panel shows the total \ha\ intensity map and the six representative slices (blue-dashed segments) used to extract the position-velocity plots. These are shown in the panels below (left column -- data, middle column -- model, right column -- residual), using a width of three spaxels. Iso-intensity contours are at $2$, $5$, $10$, $20$, $10^2$, $10^3$, $10^4$ times the rms-noise $\sigma$. An additional contour at $-2\sigma$ is shown in grey. Arrows mark the regions where emission from the \nii\ doublet leaks into the \ha\ cube.}
\label{fig:pv}
\end{center}
\end{figure}

As the emission from the \nii\ doublet can potentially contaminate the \ha\ line (arrows in Fig.\,\ref{fig:pv}), we do not model the \nii\ a posteriori as we do for all the other secondary lines.
Instead, we have included a single extra parameter in the \ha\ fit to each spaxel in order to account for the amplitude of the \nii$\lambda6583$ line, under the assumption that it can be modelled as a re-scaled version of the \ha.
Assuming a \nii$\lambda6583$/\nii$\lambda6548$ ratio of $3$ fully constrains also the fainter line of the doublet.

\begin{figure*}
\begin{center}
\includegraphics[width=0.9\textwidth]{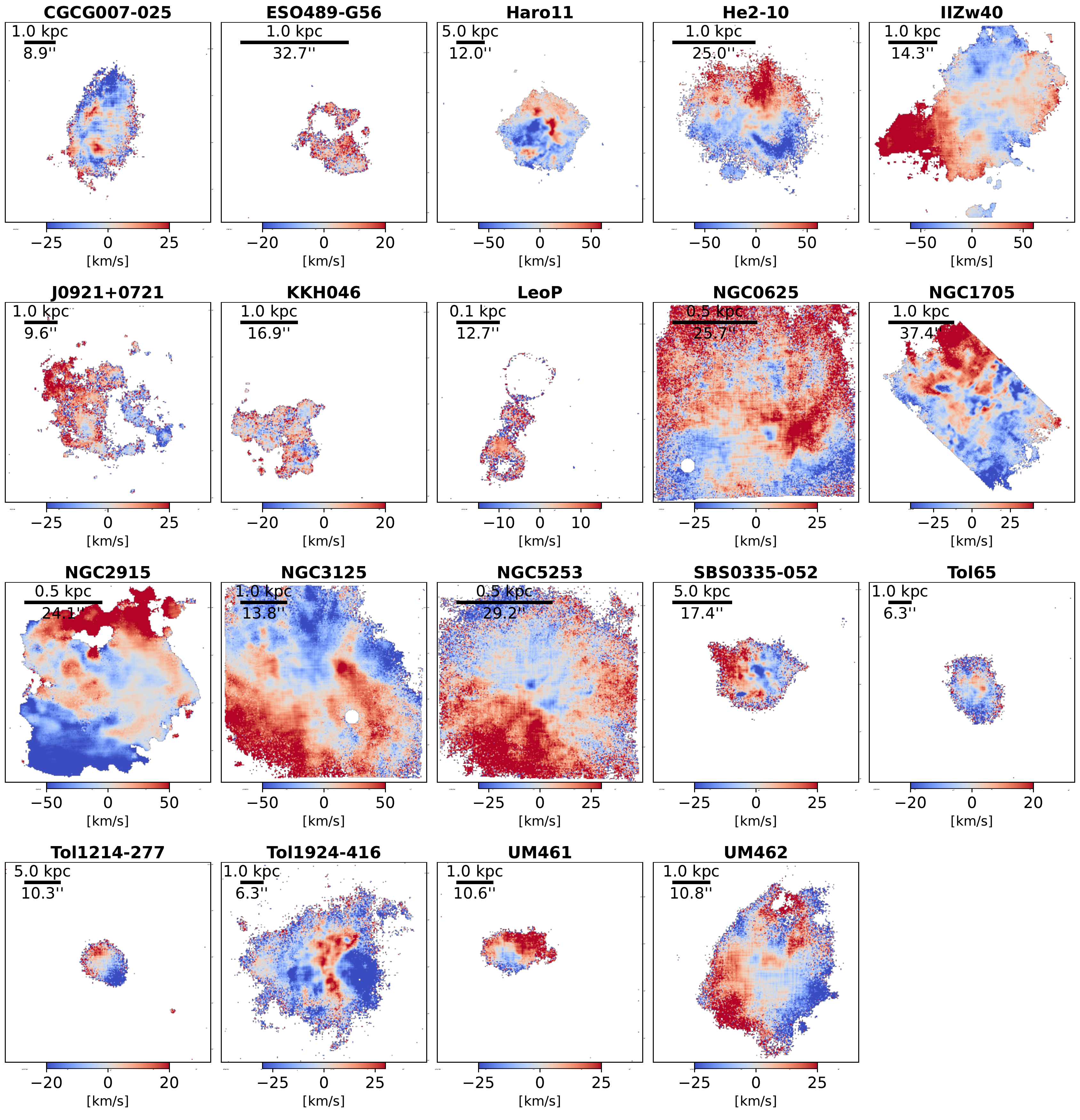}
\caption{\ha\ velocity fields of the \textsc{Dwalin-19} galaxies as derived from our multi-Gaussian modelling of the MUSE data.}
\label{fig:moment1}
\end{center}
\end{figure*}

\begin{figure*}
\begin{center}
\includegraphics[width=0.9\textwidth]{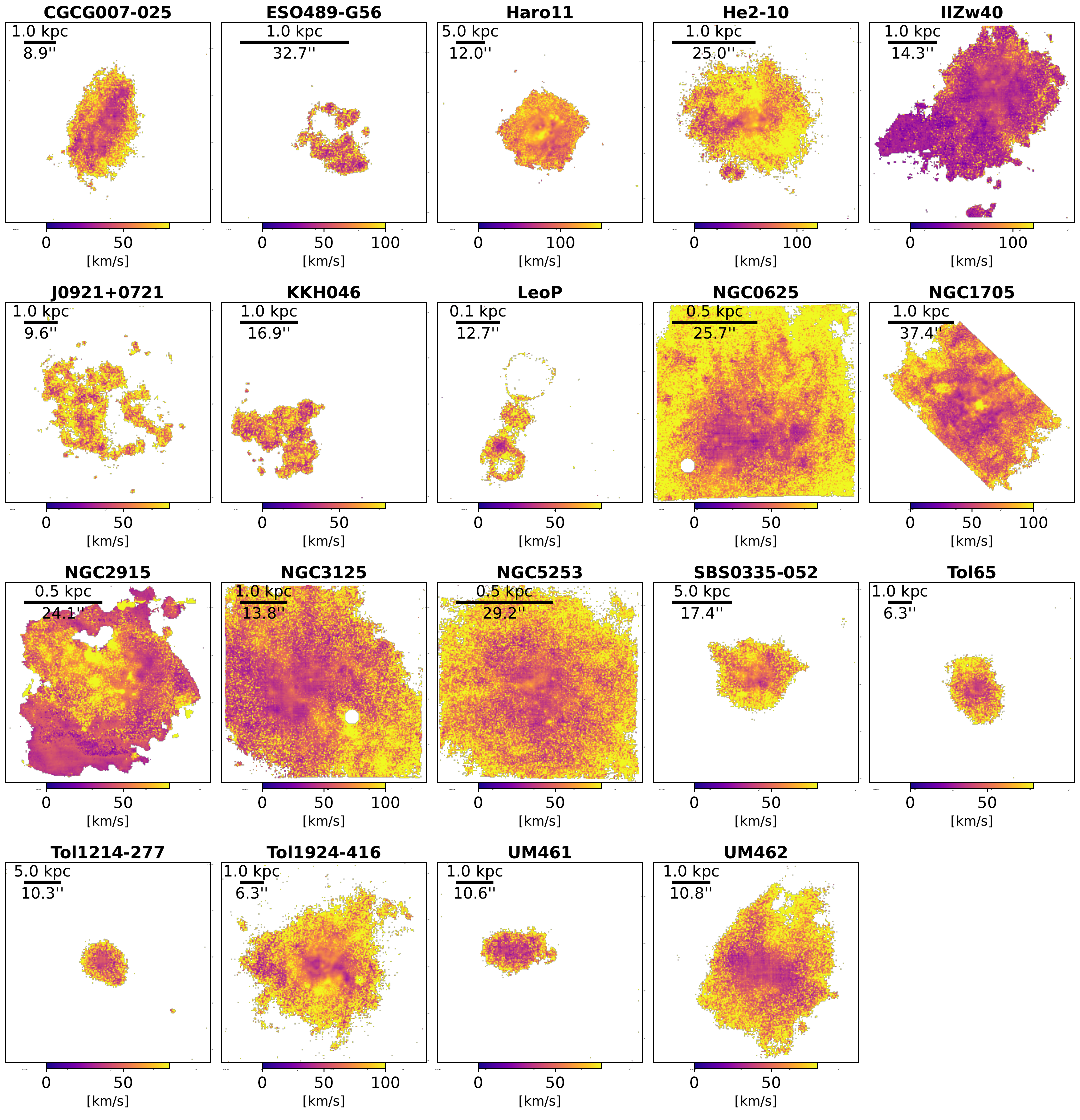}
\caption{\ha\ velocity dispersion fields of the \textsc{Dwalin-19} galaxies as derived from our multi-Gaussian modelling of the MUSE data. Only the intrinsic line broadening is shown (see Section \ref{ss:multigau}).}
\label{fig:moment2}
\end{center}
\end{figure*}

To illustrate the outcome of this procedure, Fig.\,\ref{fig:pv} shows a series of position-velocity plots taken along different slices for the \ha\ data (left column), model (middle column), and residual (data-model, right column) velocity cubes of He\,2-10.
The model provides an excellent description of the data over the whole \ha\ intensity range which, in this particular case, spans about four orders of magnitude.

The treatment of the instrumental broadening $\sigma_{\rm MUSE}$ in eq.\,(\ref{eq:sigma}) requires a separate discussion.
A caveat associated with the implementation of $\sigma_{\rm MUSE}$ is that the MUSE line spread function (LSF) is not actually Gaussian, but instead more squared in shape. 
However, as we need to model a few millions velocity profiles with multiple components, fitting complex numerical LSF profiles is highly unpractical and the use of simpler Gaussian shapes becomes mandatory.
Although \citet{Bacon+17} state that a Gaussian approximation of the MUSE LSF is perfectly valid for most applications, in practice we found that some of the emission lines at the longest wavelength in our spectra were narrower than what the instrumental broadening alone would allow (that is, $\sigma_{\rm obs}<\sigma_{\rm MUSE}$ in eq.\,\ref{eq:sigma}).
We bypassed this issue by multiplying the $\sigma_{\rm MUSE}$ predicted by \citet{Bacon+17} by a factor $\mathcal{K}\!=\!0.8$, which we found to be the approximate value for which spectral lines at different $\lambda$ have similar $\sigma_{\rm int}$ when modelled separately with a single Gaussian component defined with eq.\,(\ref{eq:sigma}). 
Different values of $\mathcal{K}$ would imply that $\sigma_{\rm int}$ smoothly varies across the wavelength range, which is hardly justifiable, given that the various lines probe the same phase of the ISM (warm ionised gas) with similar temperature and excitation conditions.

\subsection{Ionised gas kinematics}\label{ss:momentmaps}

The multi-Gaussian decomposition allows us to describe the position-velocity distribution of the ionised gas with a virtually infinite velocity resolution. 
We use these convenient analytical proxies for the actual velocity profiles to build \ha\ intensity maps, velocity fields (moment-1 maps, Fig.\,\ref{fig:moment1}) and velocity dispersion fields (moment-2 maps, Fig.\,\ref{fig:moment2}) for all \textsc{Dwalin-19} galaxies.

As shown in Fig.\,\ref{fig:moment1}, in the \textsc{Dwalin-19} sample, the ionised gas is characterised by complex velocity fields that cannot be ascribed to pure regularly rotating discs (Fig.\,\ref{fig:moment1}).
Velocity gradients are visible in many galaxies but are often very irregular, indicating the presence of strong, localised non-circular motions that are likely induced by baryonic feedback and/or past interaction events.
This kinematic complexity makes a modelling approach based on simple geometry and large-scale motions (such as in a tilted-ring model) unsuitable, which is why we opted for a different methodology.
However, in spite of this complexity, the typical line-of-sight velocities of the ionised gas are of the order of some tens $\kms$ in the \textsc{Dwalin-19} sample, suggesting the absence of large-scale bulk flows that are sufficiently powerful to expel a vast amount of material out of these galaxies, at least on the spatial scales probed by MUSE.
Fig.\,\ref{fig:moment2} shows that the typical velocity dispersion for the ionised gas is in the range $40$--$60\kms$ (Fig.\,\ref{fig:moment2}), with the exceptions of He\,2-10 and Haro\,11, both reaching values slightly above $\sim100\kms$.  
Thus the ionised gas in these systems is slightly more turbulent than that of typical star-forming galaxies where $\sigma\approx30\pm10\kms$ \citep[e.g.][]{Epinat+10,Green+14}, not surprising given their higher-than-average SFR \citep[e.g.][]{Bacchini+20}.
Clearly, gas turbulence can also be increased by recent mergers/interactions events, which may trigger the starburst episode in the first place. 
In fact, Haro\,11 is thought to be a merger system \citep{Ostlin+15}, and He\,2-10 shows signatures of a recent accretion event \citep{Vanzi+09}.
However, the \textsc{Dwalin-19} sample also features systems that are highly isolated and whose \ha\ velocity dispersion is above average, such as KKH046 and NGC\,2915.

The kinematic complexity shown by the \textsc{Dwalin-19} galaxies is likely to stem from a combination of internal and environmental mechanisms.
However, even assuming a scenario where baryonic feedback is the only driver of the observed kinematics, we would conclude that its impact on the ISM is limited to promoting non-circular motions of a few tens of $\kms$ over scales of some hundred parsecs, and to producing only a marginal enhancement of the gas velocity dispersion.
The impression is that most of the gas is `shaken' within the ISM, rather than being violently expelled from the galaxy as a result of feedback processes.
Similar results are found for the \hi\ kinematics on larger spatial scales, which show complex velocity fields in $\sim50\%$ of starburst dwarfs but with typical line-of-sight velocities of just tens of $\kms$ \citep{Lelli+14b}.
Quantifying the amount of gas that is actually outflowing requires a more careful investigation of the line profiles, which we present below.

\section{Ionised winds in \textsc{Dwalin}}\label{s:winds}
The above results indicate that the \textsc{Dwalin-19} galaxies are not simple rotating discs; they also lack unambiguous, spatially resolved evidence for a large-scale bulk motion in their ionised gas. 
In these conditions, the signature of a galactic wind can be identified through the presence of wings and/or secondary (broad) components in the line profiles.

However, even after the identification of such features, two key decisions must be made: what is the wind speed, and what is the fraction of flux associated with the wind.
These choices largely affect the wind properties but are often arbitrarily made.
With the purpose of providing a more physically motivated definition of a galaxy wind, we adopt basic prescriptions to define the flux and velocity associated with the wind component based on simple dynamical and geometrical considerations. 

\subsection{Towards a more physically motivated selection of the wind}\label{ss:wind_definition}
We define `wind' as the material whose velocity, measured with respect to the systemic velocity of the galaxy, exceeds the local escape velocity $v_{\rm esc}$, defined as the minimum speed required to bring a test particle from its original location out to the halo's virial radius\footnote{This gives a slightly lower velocity than using the traditional definition of $v_{\rm esc}$ that requires the particle to reach infinite distance.}.
This choice is motivated by the idea that material ejected with a speed higher than $v_{\rm esc}$ leaves the galaxy's virialised region and is not longer bound to that system.
Clearly, this is a strong simplification of the process of gas cycling induced by feedback, as it ignores hydrodynamical effects that can alter the purely `ballistic' dynamics of the cycle.
On the one hand, since drag from the pre-existing CGM slows down the cloud speed, an initial velocity higher than $v_{\rm esc}$ is needed to expel clouds from the virialised region.
On the other hand, even for low outflow speeds, the development of hydrodynamical instabilities due to the cloud-CGM interaction can fragment the outflowing cloud into small cloudlets that, via thermal conduction, can evaporate into the CGM \citep[e.g.][]{Armillotta+17}.
As the outcome of these processes depends on the physical condition of the outflowing gas and on the detailed properties of the CGM, we prefer to neglect them in favour of a simpler and easy-to-model ballistic interpretation of the gas flow.

The $v_{\rm esc}$ radial profile of each galaxy is determined using a mass model consisting of a dark matter (DM) halo, a stellar disc and a gaseous disc. 
For the DM halo, we assume a Navarro-Frenk-White \citep[NFW,][]{NFW} profile, with a virial mass $M_{200}$ determined from $M_\star$ via the stellar-to-halo mass relation (SHMR) of \citet{Moster+13}, and a concentration $c$ set by the $M_{200}$-$c$ relation of \citet{DuttonMaccio14}, thus consistent with a $\Lambda$CDM Universe.
The stellar disc is modelled with a double-exponential profile, with scale-length $R_{\rm d}$ determined either from the IRAC $3.6$\um\ data or, when these are not available, from the W1 data\footnote{We take $R_{\rm d}$ equal to the half-light radius divided by $1.68$, which is correct for a purely exponential disc.}, and scale-height given by $R_{\rm d}/5$ \citep[e.g.][]{vanderKruitFreeman2011}.
Given the lack of a complete database of atomic and molecular gas observations for \textsc{Dwalin}, we used a $M_\star$-based proxy for the cold gas mass \citep[eq.\,(7) in][]{Chae+21}, and assumed a size of the gaseous disc equal to twice that of the stars.
This formulation is overall consistent (to within $0.2$--$0.5$ dex) with the gas content in low-mass galaxies that would be inferred from the $M_\star$+SFR empirical approach by \citet{Hunt+20}. Both of these approaches are only rough approximations, and in the calculation of $v_{\rm esc}$, the baryon distribution plays a very marginal role as the escape speed is primarily affected by the overall depth of the gravitational potential rather than by the potential gradient (unlike the case of the circular velocity, for instance).
In practice, the key ingredient is the assumed SHMR, for which we have used the model of \citet{Moster+13} that is compatible with dynamical estimates for $M_{\star}$ and $M_{200}$ in the mass range spanned by our sample \citep[e.g.][]{Katz+17,Posti+19a}.
While gas-rich dwarfs may feature cores in their central mass distribution \citep[e.g.][]{deBlok10}, we have verified that the use of a cored DM profile like that of \citet{Burkert95} makes little difference in the resulting $v_{\rm esc}$ profile.

\begin{figure*}
\begin{center}
\includegraphics[width=\textwidth]{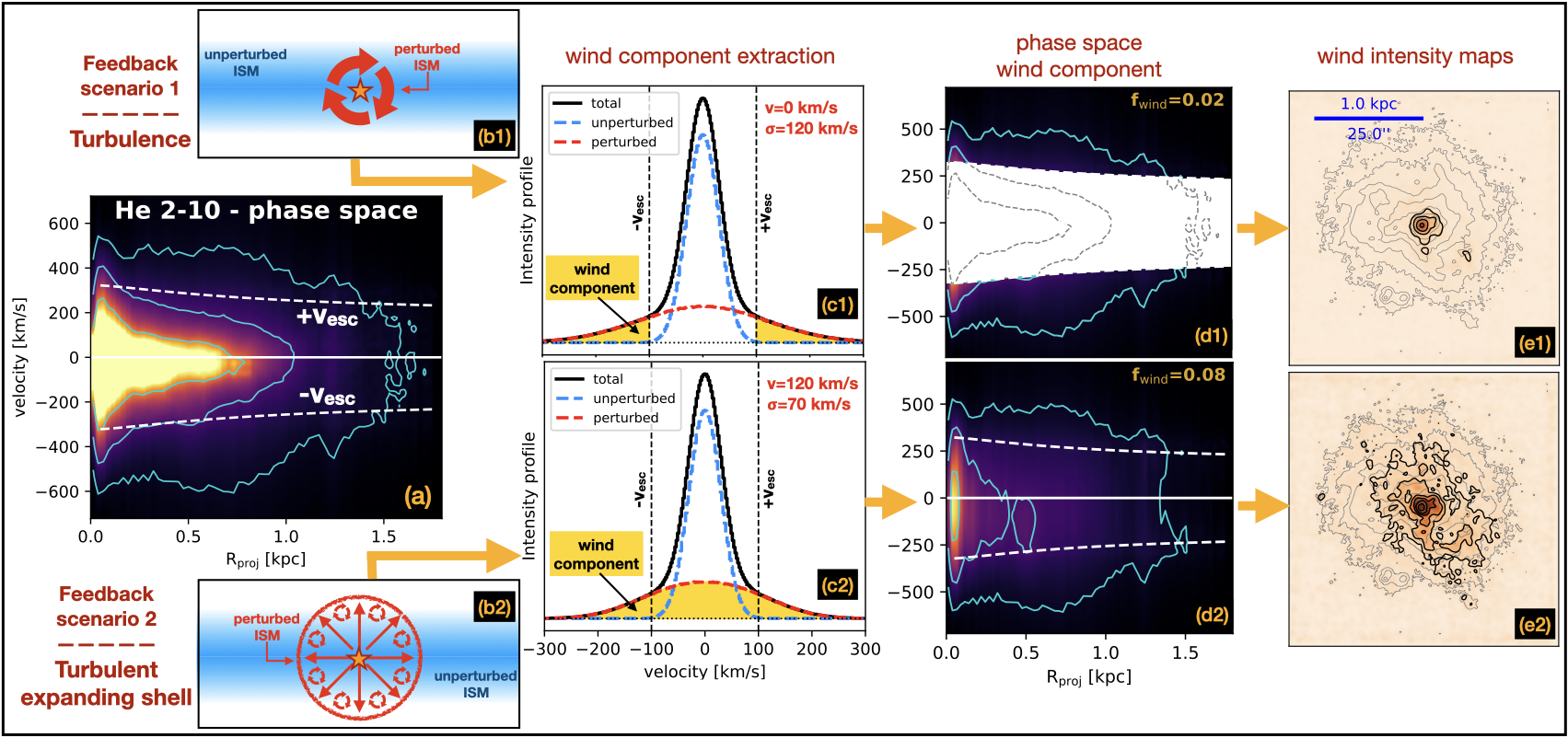}
\caption{Scheme illustrating the extraction of the wind component from the \ha\ velocity cube of He\,2-10. 
\emph{(a)}: phase-space distribution for the whole \ha\ emission. Instrumental broadening is removed via the multi-Gaussian decomposition. Cyan contours encompass $50\%$, $70\%$ and $90\%$ of the total \ha\ flux. White-dashed lines show the escape speed radial profile (see Section \ref{ss:wind_definition}).
\emph{(b)}: the two feedback scenarios of Section \ref{ss:feedback_scenarios}. In (b1) feedback injects turbulence in the ISM, but produces no bulk flow. In (b2) feedback produces the spherical expansion of a turbulent shell of gas (TES).
\emph{(c)}: illustration of the effects of feedback on a single velocity profile. For both scenarios we assume $v_{\rm esc}$ of $100\kms$, a systemic velocity of $0\kms$, and velocity dispersion of $30\kms$ for the unperturbed gas (blue-dashed curve). The component perturbed by feedback (red-dashed curve) is derived by assuming a shell expansion speed $v$ and a velocity dispersion $\sigma$ indicated in the top-right corner. Both scenarios feature very similar broad components. In the turbulent case (c1), only the flux at $|v|\!>\!v_{\rm esc}$ (highlighted in yellow) is eligible as `wind'. In the TES case, given that $v\!>\!v_{\rm esc}$, the whole component will be associated with a wind.
\emph{(d)}: phase-space plots for the wind component alone, extracted by processing each \ha\ velocity profile as discussed in Section \ref{ss:feedback_scenarios}. Contours are defined as in panel (a). 
\emph{(e)}: integrated \ha\ intensity maps of the wind component. Iso-intensity contours for the total \ha\ emission are shown in grey, as a reference.}
\label{fig:phase_space_method}
\end{center}
\end{figure*}

The $v_{\rm esc}$ curve can be compared with the position-velocity distribution of the ionised gas using a phase-space plot. 
Panel (a) in Fig.\,\ref{fig:phase_space_method} shows an example of such plot for the \ha\ line of He\,2-10, where we have populated the position-velocity space using the various components of our multi-Gaussian modelling considering only the intrinsic broadening in the line profiles.
From this diagram, it is clear that some fraction of the ionised gas is located in regions beyond the escape speed curves, identified by white-dashed lines in the panel.
However, as we discuss below, the selection of a wind component from this diagram depends on the 3D geometry and kinematics of the wind itself, which must be assumed.

\subsection{Feedback-driven turbulence and expansion}\label{ss:feedback_scenarios}
We propose two different approaches to select the wind component from the analysis of the phase-space plots. The first is more conservative in terms of wind mass, while the second is more generous, corresponding to diverse modes by which the energy and momentum injected by feedback is imparted to the ISM.
The use of both approaches allows us to better assess the uncertainties in the outflow rate estimates.
These two idealised feedback scenarios, along with their implications in terms of wind component selection, are illustrated in panels (b) and (c) of Fig.\,\ref{fig:phase_space_method}.

The first approach is based on a scenario where baryonic feedback augments the turbulence of the surrounding gas without affecting its bulk motion.
In this `turbulence' scenario, sketched in panel (b1) of Fig.\,\ref{fig:phase_space_method}, the wind will be made solely by those cloudlets that are randomly scattered at sufficiently large ($>\!|v_{\rm esc}|$) velocities (panel c1 in Fig.\,\ref{fig:phase_space_method}).
This scenario can be applied to our data by selecting only the portion of the emission at $|v|\!>\!|v_{\rm esc}|$ in the phase space (panel d1 in Fig.\,\ref{fig:phase_space_method}). The resulting wind flux is multiplied by a factor $\sqrt{3}$ to take into account that we only see projected velocities.

A completely converse model is the one where feedback does not affect the gas turbulence but only its bulk motion.
Let us consider a spherical expanding shell of gas with a constant velocity $v_{\rm wind}$ slightly larger than $v_{\rm esc}$.
One can demonstrate that, in the absence of extinction, the line-of-sight velocity profile produced by a (spatially unresolved) expanding, homogeneous shell of non-turbulent gas is a perfectly flat distribution confined within $\pm v_{\rm wind}$.
This occurs because we observe the intrinsic $\pm v_{\rm wind}$ only along two opposite shell elements aligned with the line of sight, while we see a fraction of $\pm v_{\rm wind}$ for all the other shell elements due to projection effects.
Trivially, the fact that flat velocity profiles are never observed suggests that this model is not realistic and that some amount of turbulence is always injected within the expanding gas, producing a smoother profile.
In this `turbulent expanding shell' (hereafter TES) scenario, selecting only the flux at $|v|\!>\!|v_{\rm esc}|$ would largely underestimate the wind mass, since we know that - by construction - the whole shell moves faster than $v_{\rm esc}$ (panels b2 and c2 in Fig.\,\ref{fig:phase_space_method}).
We apply these considerations to our data by inspecting one by one the components of the multi-Gaussian fit, and assign them to the wind component if they have at least a fraction $f_{\rm esc}$ of their flux beyond the $v_{\rm esc}$. 
We use $f_{\rm esc}\!=\!0.05$ as our fiducial value, but in Section \ref{ss:final} we explore values uniformly distributed between $0.01$ and $0.1$ to assess the uncertainty of this method.

Fig.\,\ref{fig:phase_space_method} illustrates the application of these two feedback scenarios to the \ha\ data of He\,2-10, showing the phase-space plots (panels d1 and d2) and the integrated \ha\ intensity maps (panels e1 and e2) associated with the wind component alone.
As expected, the two scenarios predict a different flux for the wind, equal to $2\%$ of the total for the turbulence case and $8\%$ of the total for the TES case.
With respect to the turbulence case, the TES scenario outputs a more defined wind structure, which features a spiral-like wind morphology. 
Even though the \ha\ flux of the wind component is small, its half-light radius is about twice that determined for the total \ha\ distribution.
Typically, we find that \ha\ winds in the \textsc{Dwalin-19} sample have half-light radii that are $50$--$80\%$ larger than those of the total \ha\ distribution.
This suggests a radial expansion of the material ejected from the galaxy by baryonic feedback, in agreement with expectations from ballistic models of the galactic fountain \citep[e.g.][]{fraternali17}.
Plots similar to those presented in Fig.\,\ref{fig:phase_space_method} are shown for all \textsc{Dwalin-19} galaxies in Appendix \ref{app:phasespace_wind_plots}.

Our wind selection is performed on the reference line (\ha), but we apply it to the secondary lines too, using the following criteria: for the turbulent scenario we simply select the emission at $v\!>\!v_{\rm esc}$, as we have already done for the \ha\ line, while for the TES case we rely on our multi-Gaussian modelling (Section \ref{ss:multigau}) and select all Gaussian components in the secondary lines that are associated with the wind in the reference line.
This approach allows us to have a self-consistent wind definition across different lines, which is important for the computation of the wind properties, as we discuss below.

\subsection{Outflow rates and mass-loading factors}\label{ss:final}
\begin{sidewaystable*}
\caption{Properties of the warm ionised wind in the \textsc{Dwalin-19} sample. Results for the two feedback scenarios discussed in Section \ref{ss:feedback_scenarios} are shown separately.}
\label{tab:final} 
\centering
\def\arraystretch{1.1}
\resizebox{0.95\textwidth}{!}{
\begin{tabular}{l|cccccc|cccccc}
\hline\hline\noalign{\vspace{5pt}}
\multicolumn{1}{l}{} & \multicolumn{6}{c}{Scenario 1 - turbulence} & \multicolumn{6}{c}{Scenario 2 - turbulent expanding shell (TES)}\\
\noalign{\smallskip}
\hline
\noalign{\smallskip}
Galaxy & A(\ha) & $\log_{10}\left(\frac{n_{\rm e}}{\cmmc}\right)$ & $r_{\rm wind}$ & $v_{\rm wind}$ & $\log_{10}\left(\frac{\dot{M}_{\rm wind}}{\msunyr}\right)$ & $\log_{10}\beta$ & A(\ha) & $\log_{10}\left(\frac{n_{\rm e}}{\cmmc}\right)$ & $r_{\rm wind}$ & $v_{\rm wind}$ & $\log_{10}\left(\frac{\dot{M}_{\rm wind}}{\msunyr}\right)$ & $\log_{10}\beta$ \\
 & & & [$\kpc$] & [$\kms$] & & & & & [$\kpc$] & [$\kms$] & & \\
 & (1) & (2) & (3) & (4) & (5) & (6) & (7) & (8) & (9) & (10) & (11) & (12)\\ 
\noalign{\smallskip}
\hline\noalign{\vspace{5pt}}
CGCG\,007-025 & $0.52\pm0.01$ & $2.59\pm0.01$ & $1.04$ & $163$ & $-2.98\pm0.01$ & $-2.34\pm0.13$ & $0.00\pm0.08$ & $2.53\pm0.05$ & $0.52\pm0.13$ & $168\pm1$ & $-2.59\pm0.22$ & $-1.95\pm0.26$\\
ESO\,489-G56 & $4.02\pm0.03$ & $2.46\pm0.05$ & $0.61$ & $138$ & $-1.89\pm0.05$ & $1.21\pm0.18$   & $4.16\pm0.05$ & $2.45\pm0.05$ & $0.51\pm0.04$ & $140\pm1$ & $-1.76\pm0.07$ & $1.35\pm0.18$\\
Haro\,11 & $0.92\pm0.01$ & $2.57\pm0.01$ & $2.01$ & $453$ & $-1.22\pm0.01$ & $-3.00\pm0.16$      & $0.92\pm0.08$ & $2.51\pm0.55$ & $2.09\pm1.13$ & $450\pm33$ & $-1.01\pm1.16$ & $-2.79\pm1.17$\\
Henize\,2-10 & $2.39\pm0.01$ & $2.55\pm0.01$ & $0.46$ & $288$ & $-1.08\pm0.01$ & $-1.30\pm0.17$  & $2.50\pm0.05$ & $2.49\pm0.02$ & $0.22\pm0.20$ & $309\pm16$ & $-0.28\pm0.65$ & $-0.50\pm0.67$\\
IIZw40 & $2.53\pm0.01$ & $2.64\pm0.01$ & $0.32$ & $233$ & $-1.71\pm0.01$ & $-2.08\pm0.24$        & $2.71\pm0.28$ & $2.64\pm0.07$ & $0.21\pm0.02$ & $234\pm1$ & $-1.14\pm0.04$ & $-1.51\pm0.25$\\
J0921+0721 & $2.91\pm0.02$ & $2.71\pm0.02$ & $2.04$ & $150$ & $-2.46\pm0.01$ & $-0.82\pm0.10$    & $2.98\pm0.03$ & $2.64\pm0.05$ & $1.94\pm0.03$ & $150\pm1$ & $-2.32\pm0.09$ & $-0.68\pm0.13$\\
KKH\,046 & $2.33\pm0.02$ & $2.61\pm0.02$ & $1.15$ & $115$ & $-3.11\pm0.02$ & $-0.74\pm0.10$      & $2.28\pm0.40$ & $2.47\pm0.22$ & $1.06\pm0.08$ & $116\pm1$ & $-2.94\pm0.22$ & $-0.57\pm0.24$\\
Leo\,P & $2.96\pm0.02$ & $2.77\pm0.02$ & $0.17$ & $87$ & $-4.34\pm0.02$ & $0.06\pm0.38$          & $1.76\pm0.88$ & $2.68\pm0.06$ & $0.06\pm0.05$ & $95\pm3$ & $-4.17\pm0.05$ & $0.24\pm0.38$\\
NGC\,0625 & $0.18\pm0.01$ & $2.32\pm0.01$ & $0.56$ & $202$ & $-3.20\pm0.01$ & $-2.00\pm0.14$     & $0.00\pm0.01$ & $1.79\pm0.17$ & $0.57\pm0.01$ & $202\pm1$ & $-2.56\pm0.23$ & $-1.36\pm0.27$\\
NGC\,1705 & $0.56\pm0.01$ & $2.56\pm0.01$ & $0.64$ & $176$ & $-3.06\pm0.01$ & $-1.75\pm0.09$     & $0.49\pm0.10$ & $2.45\pm0.10$ & $0.56\pm0.04$ & $177\pm1$ & $-2.79\pm0.12$ & $-1.48\pm0.15$\\
NGC\,2915 & $0.00\pm0.01$ & $2.29\pm0.01$ & $0.23$ & $188$ & $-3.43\pm0.01$ & $-1.97\pm0.09$     & $0.29\pm0.04$ & $2.13\pm0.05$ & $0.16\pm0.01$ & $189\pm1$ & $-2.83\pm0.12$ & $-1.37\pm0.15$\\
NGC\,3125 & $1.11\pm0.01$ & $2.63\pm0.01$ & $0.93$ & $233$ & $-2.41\pm0.01$ & $-2.25\pm0.14$     & $1.30\pm0.18$ & $2.57\pm0.07$ & $0.79\pm0.02$ & $236\pm1$ & $-1.90\pm0.11$ & $-1.75\pm0.17$\\
NGC\,5253 & $0.78\pm0.01$ & $2.76\pm0.01$ & $0.34$ & $225$ & $-2.52\pm0.01$ & $-2.26\pm0.15$     & $0.76\pm0.12$ & $2.77\pm0.03$ & $0.18\pm0.05$ & $229\pm1$ & $-2.00\pm0.19$ & $-1.74\pm0.25$\\
SBS\,0335-052 & $2.27\pm0.01$ & $3.94\pm0.03$ & $1.31$ & $178$ & $-2.97\pm0.03$ & $-3.04\pm0.13$ & $2.17\pm0.16$ & $4.42\pm0.11$ & $0.69\pm0.01$ & $187\pm1$ & $-3.07\pm0.08$ & $-3.14\pm0.15$\\
Tol\,65 & $1.86\pm0.01$ & $2.55\pm0.02$ & $1.50$ & $126$ & $-2.43\pm0.02$ & $-1.39\pm0.11$       & $0.85\pm0.15$ & $2.57\pm0.03$ & $0.51\pm0.01$ & $136\pm1$ & $-2.27\pm0.04$ & $-1.23\pm0.11$\\
Tol\,1214-277 & $1.23\pm0.01$ & $3.09\pm0.01$ & $2.12$ & $139$ & $-2.28\pm0.01$ & $-1.83\pm0.10$ & $0.61\pm0.25$ & $2.98\pm0.10$ & $1.58\pm0.08$ & $143\pm1$ & $-1.95\pm0.20$ & $-1.50\pm0.22$\\
Tol\,1924-416 & $1.82\pm0.01$ & $2.61\pm0.01$ & $1.79$ & $213$ & $-1.60\pm0.01$ & $-1.92\pm0.11$ & $1.65\pm0.45$ & $2.53\pm0.09$ & $0.71\pm0.14$ & $230\pm3$ & $-0.81\pm0.16$ & $-1.12\pm0.20$\\
UM\,461 & $0.26\pm0.01$ & $2.78\pm0.01$ & $1.09$ & $135$ & $-3.50\pm0.01$ & $-2.30\pm0.12$       & $0.00\pm0.01$ & $2.77\pm0.05$ & $0.58\pm0.04$ & $140\pm1$ & $-3.27\pm0.13$ & $-2.06\pm0.18$\\
UM\,462 & $0.94\pm0.01$ & $2.63\pm0.01$ & $1.24$ & $180$ & $-2.79\pm0.01$ & $-2.20\pm0.11$       & $0.03\pm0.20$ & $2.47\pm0.09$ & $0.79\pm0.08$ & $186\pm1$ & $-2.55\pm0.16$ & $-1.96\pm0.19$\\
\noalign{\vspace{2pt}}\hline
\noalign{\vspace{5pt}}
\multicolumn{13}{p{1\textwidth}}{\textbf{Notes.} (1) Internal extinction, from Balmer decrement; (2) electron density, from \sii\ line ratio; (3) wind half-light radius; (4) wind speed, assumed to be equal to the escape speed at $r\!=\!r_{\rm wind}$; (5) wind outflow rate, computed via eq.\,\ref{eq:mout_dot}; (6) mass-loading factor; (7)-(12) as in columns 1-to-6, but for the TES feedback scenario.}\\
\end{tabular}}
\end{sidewaystable*}

Our calculations for the (ionised) gas outflow rates follow those outlined in a number of previous studies \citep[e.g.][]{Liu+13, Cresci+15, Cresci+17,Marasco+20,Tozzi+21}.
The main difference is that, in this work, we present separate computations for the two feedback scenarios described above, which lead to different estimates of the wind rates: a more conservative one for the turbulence scenario, a less conservative one for the TES case.
In the calculations that follow, all wind properties (flow rate, electron density, extinction) are determined using integrated fluxes of the wind component. 

Assuming that the ionised wind can be described as a collection of ionised gas clouds all having the same electron density $n_{\rm e}$, and that the ionisation conditions do not vary across the field of view, its mass can be computed from the luminosity of the wind component of the \ha\ line, $L_{\rm wind}^{{\rm H}\alpha}$, as
\begin{equation}\label{eq:mout_ha}
M_{\rm wind} = 3.2\times10^5 \left(\frac{L_{\rm wind}^{{\rm H}\alpha}}{10^{40}\ergs} \right) \left(\frac{100\cmmc}{n_{\rm e}}\right)\msun\,.
\end{equation}

To determine $L_{\rm wind}^{{\rm H}\alpha}$, the \ha\ flux of the wind component must be corrected for internal dust extinction, $A({\rm H}\alpha)$.
We determine $A({\rm H}\alpha)$ from the Balmer decrement, assuming an intrinsic \ha/\hb\ of $2.86$ for a temperature $T\!=\!10^4\K$ \citep{OF06}\footnote{We have verified that our results do not change if we assume a wind temperature of $2\times10^4\K$, which gives intrinsic \ha/\hb\ of $\sim2.75$}, a \citet{Calzetti+00} extinction law and, importantly, computing the observed \ha/\hb\ total flux ratio for the wind component alone.
This is relevant because, as we discuss below, the properties of wind material can differ from the average properties of the galactic ISM.
Similarly, we determine $n_{\rm e}$ in eq.\,\ref{eq:mout_ha} from the \sii$\lambda6716$/\sii$\lambda6731$ flux ratio computed for the wind component using the prescription of \citet{Sanders+16}.

\begin{figure}
\begin{center}
\includegraphics[width=0.37\textwidth]{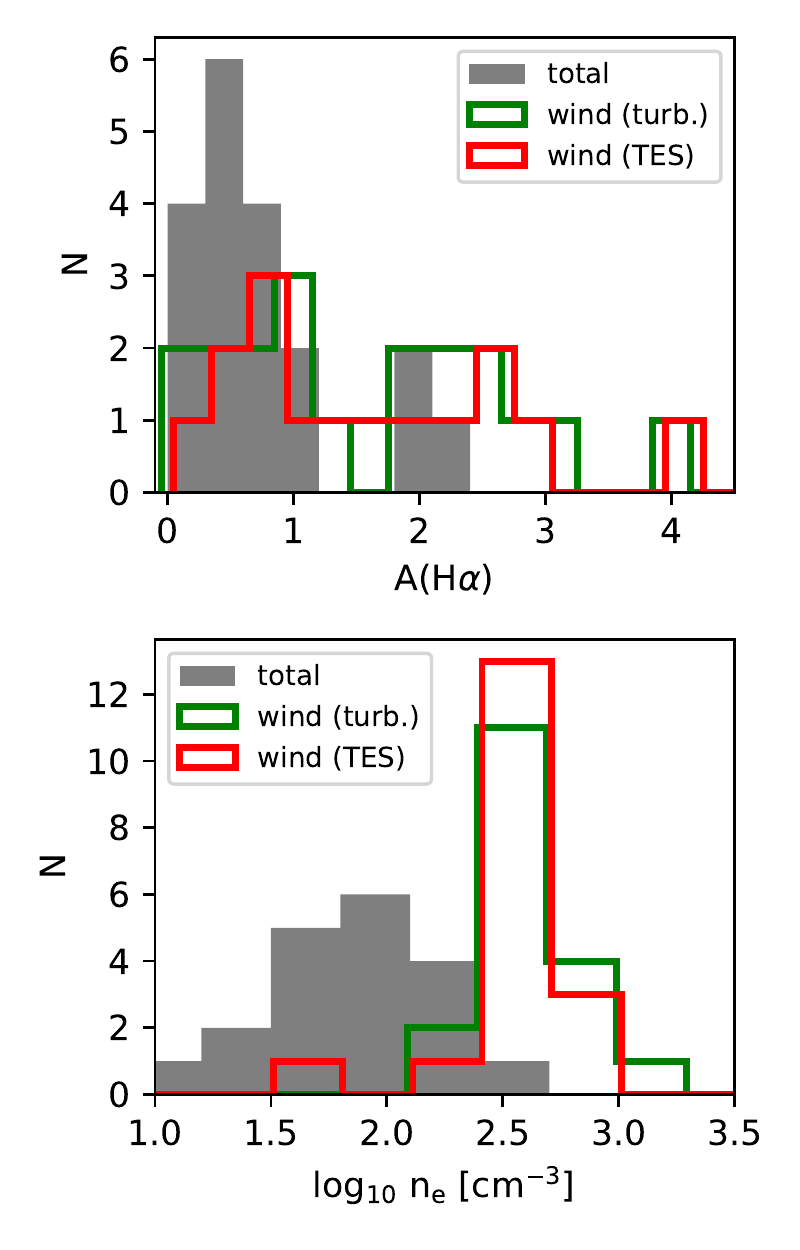}
\caption{Extinction $A({\rm H}\alpha)$ (\emph{top panel}) and electron density $n_{\rm e}$ (\emph{bottom panel}) distribution in our sample. The grey-shaded histograms show mean values representative for the whole galaxy. The solid green and red lines refer to the wind component computed in the turbulence and TES scenarios, respectively. Winds are characterised by higher $A({\rm H}\alpha)$ and $n_{\rm e}$ than the rest of the galaxy.}
\label{fig:wind_AHa_ne}
\end{center}
\end{figure}

The distribution of the resulting $A({\rm H}\alpha)$ and $n_{\rm e}$ are shown in Fig.\,\ref{fig:wind_AHa_ne}, where we compare the values determined for the wind in each galaxy (red and green histograms) with those inferred for the entire galaxy (grey-shaded histograms).
Clearly, the ionised wind has on average a higher electron density and extinction compared to the rest of the galaxy.
A higher $n_{\rm e}$ for the (stellar- or AGN-driven) ionised winds has been found in several other studies \citep{Arribas+14,Perna+17,Rose+18,Mingozzi+19,Davies+20,Fluetsch+21}, and can be driven by the compression of the gas caused by the expanding superbubbles in the star forming disc \citep[e.g.][]{Keller+14}.
In particular, our findings agree well with measurements of local ultra-luminous infrared galaxies from \citet{Fluetsch+21}, where the typical $n_{\rm e}$ is $\sim150\cmmc$ in the disc and $\sim500\cmmc$ in the outflow.

Measurements for dust extinction in the wind are more debated in the literature, with different groups finding both higher \citep{Holt+11,VillarMartin+14} and lower \citep{Rose+18,Mingozzi+19,Fluetsch+21} values in the outflow than in the rest of the ISM.
Visual inspection of the spaxel-by-spaxel distribution of $A({\rm H}\alpha)$ for the wind in single \textsc{Dwalin-19} galaxies often shows a peak at values closer to (or below) that of the ISM ($0.5$--$1$ mag), followed by a tail that extends towards very large values and increases the mean $A({\rm H}\alpha)$ for this component.
Our findings of a higher mean extinction agree qualitatively with predictions from radiation pressure-driven models of stellar feedback \citep[e.g.][]{IshibashiFabian16}. 
High density and extinction are also requirements for the formation of molecules in the wind \citep[][]{Richings+18}. 
We show below that ionised winds in \textsc{Dwalin-19} have much lower mass-loading factors than what cosmological models would predict, thus the presence of a significant molecular component in the outflow can potentially mitigate this tension.
Our measurements for the wind extinction and electron density in each galaxy are listed in Table \ref{tab:final}.

The mass outflow rate, $\dot{M}_{\rm wind}$, at a given radius $r_{\rm wind}$ is derived using the simplified assumptions of spherical (or multi-conical) geometry and a constant outflow speed $v_{\rm wind}$.
Following \citet{Lutz+20}, we have
\begin{equation}\label{eq:mout_dot}
\dot{M}_{\rm wind} = 1.03\times10^{-9} \left(\frac{v_{\rm wind}}{\kms}\right) \left(\frac{M_{\rm wind}}{\msun}\right) \left(\frac{\kpc}{r_{\rm wind}}\right)\mathcal{H}\msunyr
,\end{equation}
where $\mathcal{H}$ is a multiplicative factor that depends on the adopted outflow history. 
We take $\mathcal{H}\!=\!1$, adequate for a temporally constant $\dot{M}_{\rm wind}$ during the flow time $r_{\rm wind}/v_{\rm wind}$, which is typically $\sim6\Myr$ in our sample.
For simplicity, we consider $r_{\rm wind}$ to be equal to the half-light radius of the wind component, which can be easily determined from our phase-space diagrams, and $v_{\rm wind}$ to be equal to the escape speed computed at $r\!=\!r_{\rm wind}$.
Thus in our approach the wind speed is not directly measured from the data but is assumed from our mass model, for consistency with the wind selection method (Section \ref{ss:wind_definition}).
Wind speeds, radii and outflow rates are listed in Table \ref{tab:final} for each galaxy in the \textsc{Dwalin-19} sample.

The uncertainties quoted in Table \ref{tab:final} come from multiple sources.
For the turbulence scenario, uncertainties on $A({\rm H}\alpha)$ and $n_{\rm e}$ originate from statistical errors on the multi-Gaussian models of the Balmer and \sii\ lines used to infer such quantities.
These are propagated to $\dot{M}_{\rm wind}$ and $\beta$ estimates, although the errors on the latter are dominated by uncertainties in the SFRs.
For the TES scenario, instead, we also account for $f_{\rm esc}$, which we take free to vary between $0.01$ and $0.1$.
$f_{\rm esc}$ is the dominant source of uncertainty for all quantities in the TES case.

\begin{figure*}
\begin{center}
\includegraphics[width=0.70\textwidth]{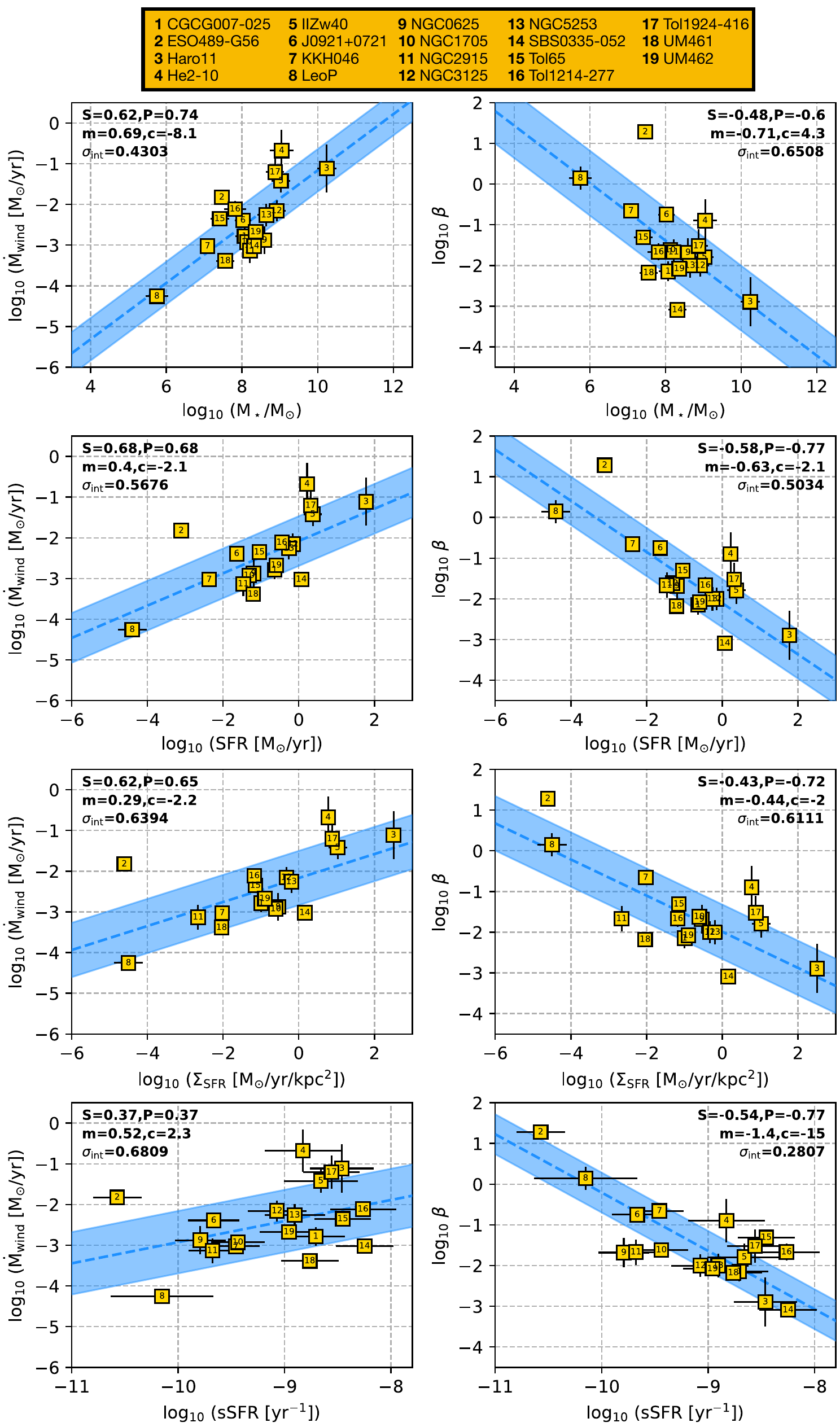}
\caption{Summary of the wind properties in the \textsc{Dwalin-19} sample. Ionised gas outflow rate $\dot{M}_{\rm wind}$ (\emph{left-hand panels}) and mass-loading factors $\beta$ (\emph{right-hand panels}) as a function of galaxy $M_\star$ (\emph{first row}), SFR (\emph{second row}), mean SFR surface density (\emph{third row}) and specific SFR (\emph{fourth row}). Each galaxy is labelled with an ID number (see legend on top).
Data points are based on the average of values from the turbulence and TES scenarios; error-bars account both for the difference between the two scenarios and for the uncertainties associated with each of them (see text).
Spearman (S) and Pearson (P) correlation coefficients are reported on top of each panel. Blue-dashed lines show the best-fit linear relations determined with the \textsc{LtsFit} Python package \citep{Cappellari+13}, with the blue-shaded region showing the resulting intrinsic scatter $\sigma_{\rm int}$. Values for the line slope $m$, intercept $c$ and $\sigma_{\rm int}$ are reported in each panel.}
\label{fig:outflow_properties}
\end{center}
\end{figure*}

Fig.\,\ref{fig:outflow_properties} summarises the properties of the ionised winds in the \textsc{Dwalin-19} sample, showing mass outflow rates $\dot{M}_{\rm wind}$ (left-hand column) and mass-loading factors $\beta\!\equiv\!\dot{M}_{\rm wind}/{\rm SFR}$ (right-hand column), derived as mean values of the two methods discussed above, as a function of galaxy $M_\star$ (top row), SFR (second row), mean SFR density ($\Sigma_{\rm SFR}$\footnote{Defined as SFR$/\pi R^2_{\rm SFR}$, where $R_{\rm SFR}$ is taken from Table \ref{tab:dwalin}.}, third row), and SFR-to-$M_\star$ ratio (or specifc SFR, sSFR, bottom row).
Error-bars on $\dot{M}_{\rm wind}$ measurements come from the quadratic sum of two uncertainties: the first is given by half the difference between the values determined in the turbulence and TES scenarios, the second is the largest error-bar of the two scenarios, provided by Table \ref{tab:final}.
The former uncertainty is typically dominant since, as expected, the TES scenario outputs outflow rates that are larger on average by a factor of $2$, and up to a factor of $6$, than the turbulence case (see Table \ref{tab:final}).

We find that galaxies in the \textsc{Dwalin-19} sample have ionised gas outflow rates ranging from $10^{-4}$ to $10^{-1}\msunyr$, corresponding to loading factors of $10^{-3}$-$10^{1}$, with a median $\beta$ of $0.02$.
These values are remarkably small compared to those predicted by galaxy evolutionary models (see Section \ref{ss:comparison_theory} for further discussion).
Also, Fig.\,\ref{fig:outflow_properties} clearly shows that the properties of the ionised wind are tightly related to those of the host galaxy. 
$\dot{M}_{\rm wind}$ ($\beta$) correlate (anti-correlate) with $M_\star$, SFR and $\Sigma_{\rm SFR}$, with Spearman (S) and Pearson (P) correlation coefficients typically in the range $0.4\!-\!0.7$ (in modulus).
Specific SFRs show instead a somewhat weaker correlation with $\dot{M}_{\rm wind}$, but remain highly anti-correlated with $\beta$.
We point out that these correlations are at least partially driven by the fact that the quantities compared are not fully independent, as $\dot{M}_{\rm wind}\! \propto\! v_{\rm wind} \!=\! f(M_\star)$, and $\beta\propto$ (SFR)$^{-1}$ by construction.
On the other hand, similar trends have been reported in other studies that make use of absorption line techniques to measure outflow rates \citep[][]{Chisholm+17,Xu+22}.
We used the \textsc{LtsFit} Python package from \citet{Cappellari+13} to make a linear fit to these relations in the logarithmic space, finding an intrinsic perpendicular scatter ranging from $0.28$ dex (for the $\beta$-sSFR relation) to $0.68$ dex (for the $\dot{M}_{\rm wind}$-sSFR relation).
We stress that most of the observed scatter is driven by two galaxies, ESO\,489-G56 and He\,2-10 (ID number 2 and 4, respectively), for which SFRs have been determined without UV information (Table \ref{tab:dwalin}).
More accurate measurements for their SFRs may bring these two systems in better agreement with the general trends, with possible further reduction of the scatter.


\section{Discussion}\label{s:discussion}
\subsection{Comparison with previous studies}\label{ss:comparison_observations}
Previous studies based on \hi, \ha\ or \nai\ D data have found that gas in dwarf galaxies exhibits outflow velocities that are too small to escape the gravitational potential well of their host \citep{Martin96,Schwartz+04,vanEymeren+09,vanEymeren+09b,vanEymeren+10,Lelli+14a}.
Our work confirms this result: we find that the fraction of gas that can potentially escape the virial radius is $\lesssim$ a few percent.

However, more recent estimates for mass outflow rates and loading factors of warm ionised gas in star forming galaxies are quite controversial, and this can be appreciated by comparing the various observational studies shown in Fig.\,\ref{fig:outflow_comparison}.
Works that make use of UV absorption lines from the Cosmic Origins Spectrograph on board of the \emph{Hubble Space Telescope} tend to find values for $\beta$ in the range $1\!-\!10$ for $M_\star\!\sim\!10^8\msun$ systems \citep[e.g.][brown circles in Fig.\,\ref{fig:outflow_comparison}]{Chisholm+17}, which are factors $100\!-\!1000$ larger than those reported in this study.
An extreme case is that of the galaxy Haro\,11, for which \citet{Chisholm+17} report $\log{\beta}\!=\!0.08\pm0.15$, whereas our estimate is $\log{\beta}\!\sim\!-3.4$, about $2\times10^3$ times smaller.
Comparatively large outflow rates are also found by \citet{Xu+22}, who infer $\sim3.3\msunyr$ for CGCG007-025 and $\sim1.8\msunyr$ for Haro\,11, which are factors $\sim70$ and $\sim2600$ larger than our measurements, respectively.
On the other hand, kinematic modelling of stacked optical emission lines in gravitationally lensed star-forming galaxies at $1.2\!<\!z\!<\!2.6$ have shown that objects with $8.0\!<\!\log(M_\star/\msun)\!<\!9.6$ have velocity profiles consistent with those expected from regularly rotating discs, suggesting a typical $\log(\beta)\!<\!-1.6$ for these systems \citep[][purple stars in Fig.\,\ref{fig:outflow_comparison}]{Concas+22}, in excellent agreement with our local analysis.
Also, spatially resolved gas kinematics in the small ($M_\star\sim10^5\msun$) starburst dwarf Pox\,186 indicate a mass-loading factors of $0.5$ \citep{Eggen+21}, compatible with the trends shown in Fig.\,\ref{fig:outflow_properties} for galaxies at such low $M_\star$.

Pinpointing the dominant source of these discrepancies is not trivial, since the quoted studies largely differ in terms of methodology, sample selection and atomic species considered.
By construction, absorption line studies infer flow rates from a small number of pencil-beam observations along sparse sight-lines, thus lack any spatial information and rely on strong assumptions on the geometry, kinematics and filling factor of the wind.
Even when hundreds of sight-lines are available, as in the case of the Milky Way, the interpretation of the gas flow outside the disc varies depending on such assumptions \citep[e.g.][]{Clark+22,Marasco+22}.
Overall, the impression is that low values of $\beta$ are found when gas kinematics are modelled in some detail, like in the current study or in that of \citet{Concas+22}.

The study of \citet{McQuinn+19} provided one of the first systematic investigations of ionised galactic winds in nearby, low-mass ($M_\star\!\sim\!10^7\!-\!10^{9.3}\msun$) starburst galaxies from \ha\ narrow-band imaging.
The main results of that work are that winds are spatially confined within the innermost $10\%$ of galaxy virial radii, indicating that most of material expelled from dwarf galaxies remains in the halo and can be eventually re-accreted onto their discs.
These findings align with ours, and support a scenario where baryonic feedback in dwarfs stimulates a gentle gas cycle rather than producing a massive blowout.
However, in \citet{McQuinn+19}, typical values quoted for mass-loading factors $\beta$ are in the range $0.5\!-\!3$ (grey diamonds in Fig.\,\ref{fig:outflow_comparison}), approximately a factor $100$ larger than those inferred in our study. 

A key difference between our work and that of \citet{McQuinn+19} is the selection of the wind component.
Both studies rely on the \ha\ line to characterise the wind, but while our approach focuses on the \ha\ kinematics in relation to the galaxy escape speed (Section \ref{ss:wind_definition}), the criterion used by \citet{McQuinn+19} is based on the \ha\ morphology, and specifically on the \ha\ radial extent compared to that of the \hi\ component: ionised gas located beyond an \hi\ surface density contour-level of $5\times10^{20}\cmmq$ is selected as a wind, to which an expansion velocity of $25\!-\!50\kms$ (typical for the \ha\ velocity dispersion in the ISM of these systems) is assigned.
Such an approach has the advantage of relying on a visual identification of the wind component, intended as ionised gas beyond some scale radius. Shortcomings are that the definition of such radius is arbitrary and that the gas expansion speed must be assumed.
Velocities of $25\!-\!50\kms$ are typically insufficient to gravitationally unbind the wind material (see our $v_{\rm wind}$ estimates based on $v_{\rm esc}$ in Table \ref{tab:final}), which will eventually fall back onto the galaxy in a galactic fountain cycle or join the CGM.
Hence, $\beta$ values estimated with this approach likely refer to gas that participates to the disc-halo cycle rather than to baryons that get permanently expelled from galaxy halos.
Models of the galactic fountain \citep[e.g.][]{FB06,Marasco+19b} require $\beta$ greater than unity in order to reproduce the properties of extra-planar gas in nearby galaxies.

We stress that some of the galaxies in the \textsc{Dwalin-19} sample have been studied individually in separate works.
\citet{ThuanIzotov97} found P-Cygni profiles in the stellar UV absorption lines of SBS\,0335-052 and Tol\,1214-277, suggesting the presence of a stellar wind from massive stars in these two systems.
The terminal velocities measured for the stellar winds in SBS\,0335-052 and Tol\,1214-277 were $\sim500$ and $\sim2000\kms$, respectively.
Interestingly, while in our study we do not infer terminal velocities, our estimates for mass outflow rates and loading factors in Tol\,1214-277 are about one order of magnitude larger than for SBS\,0335-052.
\citet{Cresci+17} carried out a detailed investigation of the ionised gas properties in He\,2-10 using the same MUSE data analysed here.
They estimated a mass outflow rate of $0.3\msunyr$ that is consistent with our measurement, although our uncertainties are particularly large on this system.

\citet{Cohen+18} studied the kinematics of the Brackett $\alpha$ emission line towards a supernebula in NGC\,5253, in order to quantify the effects of feedback from its embedded super star cluster. Based on the absence of a massive outflow, they concluded that feedback is ineffective at dispersing gas around the cluster, in line with our findings. 
Using MUSE data and a dynamical approach that is conceptually similar to that adopted here, \citet{Menacho+19} inferred the ionised gas fraction that could escape the gravitational potential in Haro\,11, finding values between $0.1$ and $0.3$. 
In our study, instead, the flux fraction of the wind component in Haro\,11 is only $0.01$.
This discrepancy is largely driven by the different dark matter halo mass assumed for this galaxy: we use $4\times10^{11}\msun$ from the SHMR of \citet{Moster+13}, while \citet{Menacho+19} adopt $7$--$9\times10^{10}\msun$, from the estimate of \citet{Ostlin+15} based on the rotational speed of the ionised gas in the galaxy outskirts.
As discussed in Section \ref{s:kinematic}, the \textsc{Dwalin-19} galaxies are characterised by irregular velocity fields. 
This complicates any possible estimate for their circular velocity - hence, for their dynamical mass - from the ionised gas kinematics, which made us opt for a different approach.

Galaxy winds in J1044+0353 and J1418+2102, two galaxies in the \textsc{Dwalin} sample (but not in the \textsc{Dwalin-19} sub-sample), have been recently detected using deep optical slit-spectroscopy by \citet{Xu+22b}, who inferred mass loading factors of $0.44$ and $0.36$ for the two systems, respectively. 
These values are in excellent agreement with the $\beta$--SFR and $\beta$--$M_\star$ relations found in our study (Fig.\,\ref{fig:outflow_properties}), given the $M_\star$ and SFRs of these two galaxies listed in Table \ref{tab:dwalin}.

\subsection{Comparison with theoretical expectations}\label{ss:comparison_theory}

\begin{figure}
\begin{center}
\includegraphics[width=0.5\textwidth]{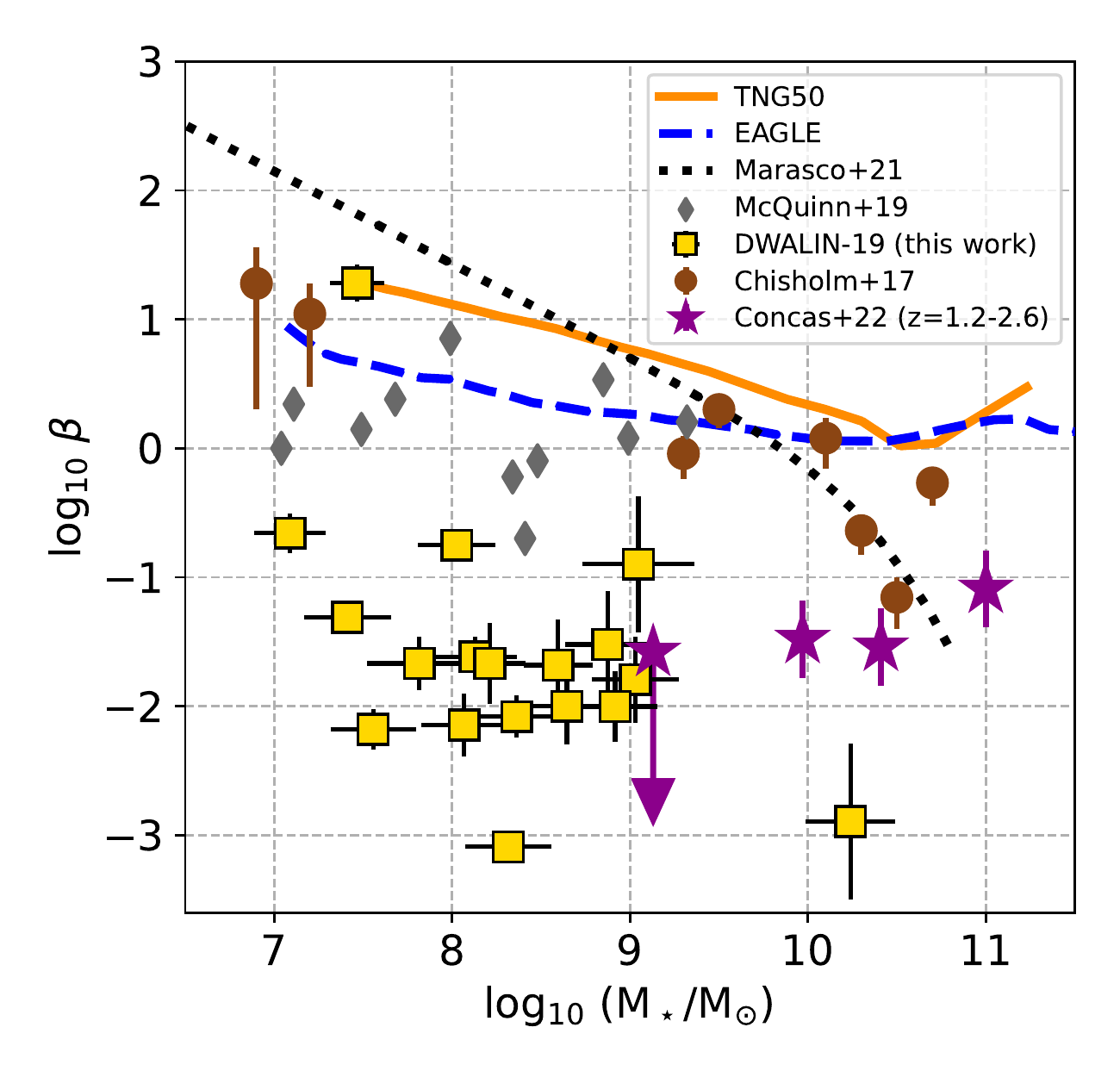}
\caption{Wind mass-loading factor $\beta$ as a function of galaxy $M_\star$ from different studies. Markers show observational results from \citet[][brown circles]{Chisholm+17}, \citet[][grey diamonds]{McQuinn+19}, \citet[][purple stars]{Concas+22}, and the present study (yellow squares). Lines show theoretical predictions from the EAGLE \citep[][blue-dashed]{Mitchell+20} and the Illustris TNG\,50 \citep[][orange-solid]{Nelson+19} cosmological simulations, and from the evolutionary model of \citet[][black-dotted]{Marasco+21}.}
\label{fig:outflow_comparison}
\end{center}
\end{figure}

A scenario where the outflow mass-loading anti-correlates with the galaxy stellar (or dynamical) mass, as we find here (Fig.\,\ref{fig:outflow_properties}), is supported by arguments based on energy- and momentum- driven winds and, in general, by theoretical models of galaxy evolution in the $\Lambda$CDM framework \citep[e.g.][]{Somerville+08,SomervilleDave15,Bower+17,Zhang18}.
This scenario is based on the simple expectation that more massive galaxies prevent gas from escaping due to the depth of their gravitational potential well. 
Unfortunately, the comparison between theoretical predictions and observational measurements of outflow rates is not trivial: observations are limited by projection effects and can provide only an instantaneous measurement of the outflow rate for a given gas phase (which is, in this study, the warm-ionised phase traced by optical emission lines), whereas theoretical predictions are robust only for time-averaged wind properties and seldom distinguish between the different gas phases.
It is very likely, therefore, that theory will provide higher outflow rates than observational determinations.

Fig.\,\ref{fig:outflow_comparison} shows the predictions for $\beta$ as a function of $M_\star$ from the EAGLE \citep{Schaye+15} and Illustris TNG50 \citep{Pillepich+18} suites of cosmological hydrodynamical simulations (dashed-blue and solid-orange lines, respectively), and that of the analytical model of galaxy evolution from \citet{Marasco+21}.
These predictions make use of very different prescriptions for determining $\dot{M}_{\rm wind}$. 
\citet{{Mitchell+20}} derive $\dot{M}_{\rm wind}$ in EAGLE using all gas particles whose time-averaged radial speeds exceed a given fraction of the halo maximum circular velocity.
\citet{Nelson+19} compute outflow rates in TNG50 at a fixed galactocentric radius ($r\!=\!10\kpc$), by considering all gas particles with radial velocity above $5\times$ the the halo virial velocity (see their Appendix A\footnote{Amongst the various prescriptions adopted by \citet{Nelson+19} this is the one that comes closer to our approach, as in our mass models $v_{\rm esc}\simeq4-5\times$ the halo virial velocity.}).
In the analytical evolutionary model of \citet{Marasco+21}, $\beta$ is parameterised as $(M_{\rm h}/M_{\rm crit})^{-\alpha}$, where $M_{\rm h}$ is the galaxy halo virial mass and $M_{\rm crit}$ and $\alpha$ are free parameters that control the efficiency of stellar feedback in driving winds.
These parameters, along with others controlling feedback from supermassive black holes, are adjusted to reproduce the relation between black hole masses, $M_\star$ and $M_{\rm h}$ observed in nearby galaxies ($M_{\rm crit}\!=\!2.5\times10^{11}$ and $\alpha\!=\!1.7$ in their fiducial model).
Yet, in spite of these diversities, all predictions must be re-scaled by at least two orders of magnitude in order to match the \textsc{Dwalin-19} data points.
A similar result seems to hold even at a higher redshift, as found by \citet{Concas+22} via the modelling of stacked optical emission lines from lensed KMOS data at $1.2\!<\!z\!<\!2.6$.
Taken at face value, this result suggests that either theory drastically over-predicts outflow rates in star-forming galaxies by more than two orders of magnitudes, or warm ionised gas accounts for less than $1\%$ of the wind mass, as it seems to be the case in local ultra-luminous infrared galaxies \citep{Fluetsch+21}.
We remark, though, that the tension between theory and observations is alleviated when the higher values of $\beta$ determined by \citet{Chisholm+17} and \citet{McQuinn+19} are considered.
On the other hand, recent high-resolution simulations of isolated galaxy formation indicate that a feedback efficiency lower than often employed in cosmological models is required in order to correctly reproduce the chemical and morphological properties of stellar discs in Milky Way-like systems \citep[e.g.][]{Clarke+19,BeraldoeSilva+20}.


\section{Conclusions}\label{s:conclusions}
Feedback from star formation and/or AGN (`baryonic' feedback) is expected to strongly affect the evolution of low-mass galaxies by launching multi-phase, galaxy-scale winds characterised by large ($1\!-\!50$) mass-loading factors \citep{Somerville+08,Muratov+15,Nelson+19,Mitchell+20}.
Feedback models predict that most of the wind mass is expected to be found in the warm ($T\!\sim\!10^4\K$) phase \citep{Kim+17, KimOstriker18}, thus spatially resolved optical spectroscopy of local starburst dwarfs has the potential to put key constraints on the wind properties - and therefore on the role of baryonic feedback - in galaxies at the low-mass end of the $M_\star$ function.

In this paper, we have studied the properties of the warm ionised winds in a sample of 19 nearby starburst galaxies, the \textsc{Dwalin-19} sample, using archival MUSE@VLT data.
Our results can be summarised as follows.
\begin{enumerate}
\item 
We have determined $M_\star$ and SFRs for all the galaxies in the \textsc{Dwalin} sample (Fig.\,\ref{fig:dwalin_all}) in a homogeneous way, using the method outlined by \citet{Leroy+19}.
This makes use of photometric measurements in various bands, ranging from the FUV to the MIR, which we have obtained by processing \emph{GALEX}, WISE and \emph{Spitzer} images with an ad-hoc pipeline based on the extraction of cumulative light profiles (Fig.\,\ref{fig:photo_example}).
We find that, as expected, the vast majority of our galaxies lie above the star-forming main-sequence, so they can be considered low-mass starbursts.
\item 
Detailed modelling of the \ha\ velocity profiles from the MUSE data shows that starburst galaxies feature complex velocity fields characterised by irregular velocity gradients (Fig.\,\ref{fig:moment1}), indicating the presence of non-circular motions with speeds of a few tens $\kms$ which are well below the galaxy escape speed.
The typical velocity dispersion for the ionised gas is $40\!-\!60\kms$ (Fig.\,\ref{fig:moment2}), slightly larger than that of typical star forming galaxies ($30\pm10\kms$), in line with the idea of feedback injecting turbulence into the ISM.
\item
A wind component for the ionised gas is determined spaxel-by-spaxel from the \ha\ velocity profiles, by comparing the gas distribution in the phase-space with simple models for the escape speed radial profile.
To better assess the uncertainties in our measurements we adopt two approaches to extract the wind component based on two different feedback scenarios (Fig.\,\ref{fig:phase_space_method}).
We find ionised gas outflow rates in the range of $10^{-4}\!-\!10^{-1}\msunyr$, corresponding to mass-loading factors of $10^{-3}\!-\!10^{1}$, with a typical value of $0.02$.
\item
Outflow rates (loading factors) are tightly correlated (anti-correlated) with $M_\star$, SFRs, SFR densities and specific SFRs (Fig.\,\ref{fig:outflow_properties}).
While these trends are in qualitative agreement with expectations from hydrodynamical and analytical models of galaxy evolution, model predictions exceed the observed values by at least two orders of magnitude.
\end{enumerate}

Our findings suggest that baryonic feedback in starburst dwarfs stimulates a gentle gas cycle rather than producing a large-scale blow out, in line with previous results based on interferometric \hi\ observations \citep{Lelli+14a} and deep \ha\ imaging \citep{McQuinn+19}.
This leaves open the question of whether most of the wind mass in these systems is confined to the colder, denser molecular phase.
Deep observations with radio or sub-mm interferometers like ALMA are available for some of the \textsc{Dwalin} galaxies. 
However, studies that used such data \citep[e.g.][]{Hunt+14,Hunt+15,Amorin+16,Cormier+17,Gao+22} have mostly focused on determining molecular gas fractions, gas depletion time-scales and dust properties, while little attention has been dedicated to quantify the properties of molecular outflows.
A homogeneous study of the molecular gas kinematics in \textsc{Dwalin}, analogous to that presented in this work, will be mandatory to infer whether the coldest gas phase plays a dominant role in starburst-driven galactic winds.

\begin{acknowledgements}
The authors thank an anonymous referee for a prompt and constructive report.
AM and GC acknowledge the support by INAF/Frontiera through the "Progetti Premiali" funding scheme of the Italian Ministry of Education, University, and Research.
\end{acknowledgements}


\bibliographystyle{aa} 
\bibliography{dwalin} 

\appendix
\section{Photometric analysis}\label{app:photometry}
\begin{figure*}
\begin{center}
\includegraphics[width=\textwidth]{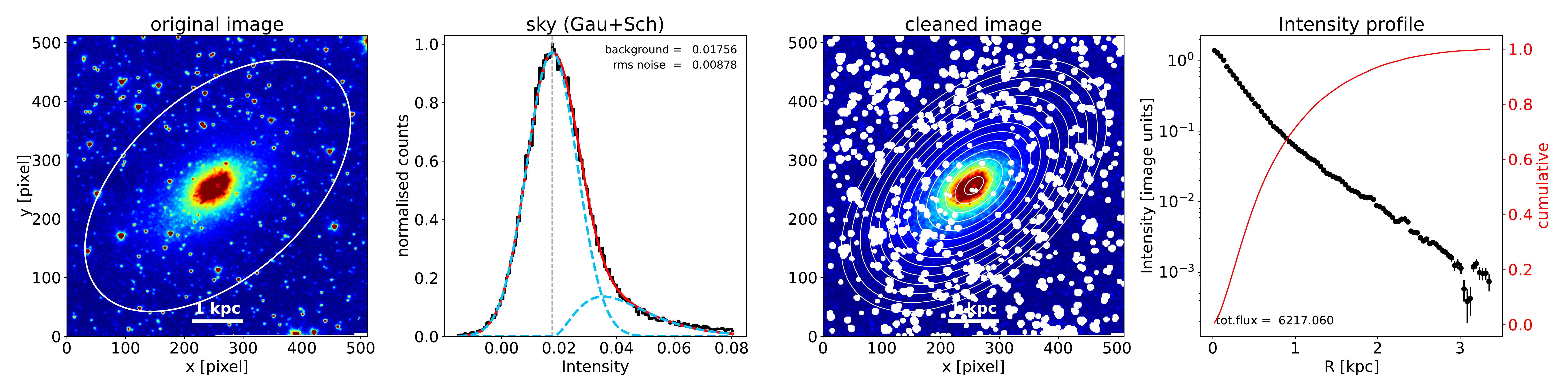}
\caption{Example of our photometric analysis on the IRAC $3.6$\um\ data of NGC\,2915. \emph{First panel}: IRAC image, with the ellipse marking the division between galaxy and sky regions. \emph{Second panel}: pixel intensity distribution within the sky region (black histogram). The red curve shows the best-fit model made by the sum of a Gaussian and a Schechter component (individually shown by light-blue dashed curves). The vertical dotted line shows the mean of the Gaussian component, corresponding to the sky background value. \emph{Third panel}: IRAC image filtered with our point-source masking technique (see text). The division in concentric annuli is also shown. \emph{Fourth panel:} final radial intensity profile in image units (black circles with error-bars), and normalised growth curve (red solid curve).}
\label{fig:photo_example}
\end{center}
\end{figure*}

The method employed for our photometric analysis is an upgraded version of that used by \citet{Marasco+19a}, and is based on the extraction of the cumulative light profile from sky-subtracted images after the removal of contamination from point-like sources such as foreground stars and background galaxies.
This approach is more refined than measurements based on traditional aperture photometry which, as we show below, may lead to significantly different results,

We describe our procedure below, and show an illustrative application to the IRAC $3.6$\um\ data of NGC\,2915 in Fig.\,\ref{fig:photo_example}.
We anticipate that several of the parameters that regulate our method are set by eye, image-by-image, on the basis of the credibility of the resulting mask and of the final light profile.
However, as we discuss below, variations in our choices are accounted for in the estimates of the uncertainties on our $M_\star$ and SFR measurements.

We first define a region of the image where the contribution of the galaxy is sufficiently small that it can be assumed to be largely dominated by the sky. 
This region is defined via an ellipse, centred at the coordinates given in Table \ref{tab:dwalin}, and with an axial ratio and orientation that are defined by eye from either the IRAC\,3.6\um\ image or, when this is not available, the W1 images (first panel in Fig.\,\ref{fig:photo_example}).
Ideally, the ellipse should define the overall galaxy radial extent, inclination and position angle.
Once set, the ellipse parameters are maintained also for the images of the same system in the other bands, while the radial extent of the ellipse is manually adjusted depending on the spatial extent and quality of the data.

Pixels external to the ellipse are used to characterise the `sky' region of the image.
This is composed by a combination of a smooth background, point-like sources (i.e. unresolved background stars and galaxies) and, in some rare cases, resolved nearby systems.
Point-like sources are very rare in UV images but strongly contaminate W1 and IRAC\,3.6\um\ images.
As we are mainly interested in determining the sky background $b$ and rms-noise $\sigma$, we employ an automatic approach that allows us to filter out the contamination from the other components.
We model the pixel intensity ($I$) distribution in the sky region, $n_{\rm sky}(I)$, with a two component model made by the sum of a Gaussian and a Schechter function:
\begin{equation}
  n_{\rm sky}(x)=\left\{
    \begin{array}{ll}
      n_{\rm G}\exp\left(-\frac{x^2}{2\sigma^2}\right), & \mbox{if $x\le0$}.\\
       n_{\rm G}\exp\left(-\frac{x^2}{2\sigma^2}\right) + n_{\rm S} \left(\frac{x}{I_{\rm S}}\right)^\alpha \exp\left(-\frac{x}{I_{\rm S}}\right), & \mbox{if $x>0$}.
    \end{array}
  \right.
\end{equation}
where $x\equiv(I-b)$, and $n_{\rm G}$, $n_{\rm S}$, $I_{\rm S}$ and $\alpha$ are free parameters of the fit along with $b$ and $\sigma$.
This approach allows us to account for the positive tail in the intensity distribution introduced by point-like sources: a Schechter function is the optimal choice for pure stellar contamination and, in general, is flexible enough to describe complex tails.
The second panel of Fig.\,\ref{fig:photo_example} shows that the observed distribution (black histogram) is well fitted by our two-component model (red curve). The determined background intensity $b$ is subtracted from the image before the next analysis step.

We stress that the fit of our model to the sky intensity distribution does not always converge.
In these occurrences we determine $b$ and $\sigma$ as the mean and standard deviation of the $n_{\rm sky}(I)$ after filtering the original distribution with a sigma-rejection method.
Finally, in some cases we are forced to adjust the background value manually until the cumulative intensity profile converges (see below).
This mainly occurs in the W4 band, where sky fluctuations across the image can be severe and difficult to deal with our automatic approach.

We now move to the analysis of the region within the initial ellipse (`galaxy' region).
Here, the galactic emission (ideally smooth and axi-symmetric) is also contaminated by point-like sources that, if not filtered out correctly, can significantly affect the radial profile especially in the outermost regions where the galaxy surface brightness is low. 
To tackle this, we first divide the galaxy region into a series of concentric ellipses, all with the same centre, orientation and axial ratios of the initial ellipse, and with inter-ellipse separation given by the image resolution.
The intensity distribution in each ring is filtered with a sigma-rejection technique (with a clip imposed at $3$ or $4$ rms, depending on the image), which allows us to mask pixels that are too bright or too dim with respect to the typical intensity value of that annulus.
To further clean the image, the mask obtained is then broadened by a few pixels.
The resulting `cleaned' image obtained for NGC\,2915 is shown in the third panel of Fig.\,\ref{fig:photo_example}.
Using this map, we extract the radial profile by computing the mean intensity of all the non-masked pixels in each ring.
The outermost ring considered is defined by the initial ellipse, but we stop tracking the profile when the signal-to-noise ratio\footnote{computed as $\max({\sigma,I_\sigma}/\sqrt{n})$, where $I_\sigma$ is the standard deviation of the pixel intensity within that ring and $n$ is the number of unmasked pixels considered} falls below unity.

Using the derived intensity profile we build the cumulative radial profile (or `growth curve'), for which the masked pixels in a given ring are replaced with the mean intensity computed in that ring.
The progressive flattening of the growth curve (shown as a solid red curve in the rightmost panel of Fig.\,\ref{fig:photo_example}) is a key check for the goodness of our photometry, as it indicates both that the sky background has been correctly determined and that no additional flux from the galaxy can be measured in regions beyond the outermost annulus considered.
This is only an a-posteriori check that is done by visual inspection, and we stress that we have not tuned our approach to output a flat cumulative profile a-priori in order not to bias our analysis.
The outermost value of the growth curve gives the galaxy flux in image units, which is then converted to physical units using conversion factors that depend on the instrument, and finally to a luminosity using distances reported in Table \ref{tab:dwalin}.

\begin{figure}
\begin{center}
\includegraphics[width=0.40\textwidth]{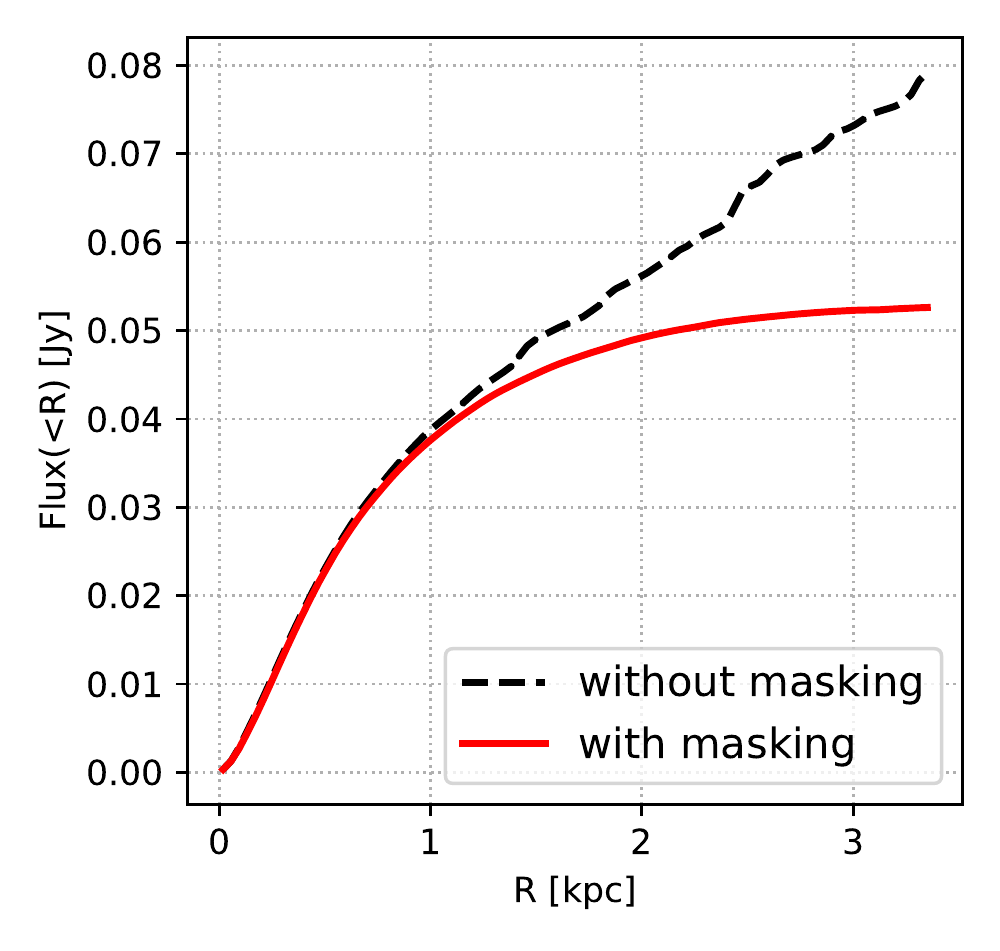}
\caption{Flux growth curves for NGC\,2915 in the IRAC $3.6$\um\ band. The solid red (dashed black) curve shows the growth determined with (without) masking of the point-like sources. The approach without masking outputs a growth curve that does not flatten and an overestimation of the total flux.}
\label{fig:cumul}
\end{center}
\end{figure}

The lack of flattening in the growth curve can be caused by different factors, the most crucial of which is the poor masking of point-like sources.
To appreciate this effect, in Fig\,.\ref{fig:cumul} we compare the growth curves obtained with and without masking.
The two curves are similar only for $R\!<\!1\kpc$, beyond which the one derived without masking, rather than flattening, grows linearly with $R$ due to the contribution of the many point-like sources that simulate the effect of an additional background.
The final flux determined with this approach is $\approx50\%$ larger than that computed with the mask, and can reach up to a factor of $2$ in some galaxies.
This exemplifies very well the importance of a correct treatment for the contamination of point sources.

Further complications are induced by small errors in the sky background estimate, which produce a flattening (if the background is underestimated) or a steepening (if the background is overestimated) of the outer data points in the radial profile.
While we cannot exclude a change of slope in the outer region of the galaxy surface brightness profile, a perfectly flat or an abruptly truncated profile indicate a mistake in the background calculation. 
In these rare occurrences, we manually adjust the background value (typically by a few percent only) so that the profile slope does not show strong discontinuities at large radii.

A risk associated with the masking of point-like sources within the galaxy region is the removal of bright star clusters that belongs to the galaxy itself. 
To minimise this risk, we have selected the parameter of our masking method so that the fractions of masked pixels in the sky region and in galaxy region are similar to each other. 
This ensures that, in a statistical sense, we are not filtering out genuine galaxy features. 
Clearly, less features are masked close to the galaxy centre, where the galaxy surface brightness is larger than that of possible contaminants.

\subsection{Estimate of the uncertainties}
We compute the uncertainty associated with our flux measurements ($\epsilon_{\rm flux}$) as the quadratic sum of two errors, the first due to the image noise ($\epsilon_\sigma$) and the second due to the method ($\epsilon_{\rm met}$).

To determine ($\epsilon_\sigma$), we produce a series of $N$ stochastic realisation of the cleaned, background-subtracted image by replacing each pixel intensity $I$ in the galaxy region with a value randomly extracted from a normal distribution with a mean equal to $I$ and variance given by $\sigma^2(I+b)/b$, where $b$ and $\sigma$ are the sky background and rms-noise determined as discussed above.
The formulation adopted for the variance ensures that the image noise scales as the square root of the signal, being equal to $\sigma$ at the background level, as expected.
For each of these $N$ images we determine a value for the galaxy flux as described above, and set $\epsilon_\sigma$ equal to the standard deviation of the resulting flux distribution.

\begin{table}
\caption{Parameters that are varied in the computation of $\epsilon_{\rm met}$, type of distribution adopted and range considered.}
\label{tab:eps_met} 
\centering
\begin{tabular}{ccc}
\hline\hline
Parameter & adopted  & standard deviation \\
          &  distribution &  or width \\
    \hline
inclination & normal & $5\de$ \\
position angle & normal & $5\de$ \\
ellipse size & normal & $10\%$ of the initial size\\
sigma-clipping threshold & uniform & $\pm1$ rms\\
mask broadening & uniform & $\pm2$ pixels \\
\hline
\end{tabular}
\end{table}

To determine $\epsilon_{\rm met}$, we repeat $M$ times the whole photometric analysis procedure using each time different values for the main parameters that regulate our method.
The parameters are randomly extracted from either normal or uniform distributions, centred around the values adopted in the initial photometric calculation. 
Table \ref{tab:eps_met} lists the parameters subject of this procedure and their variation range.
Also in this case we get $M$ flux measurements and set $\epsilon_{\rm met}$ equal to the standard deviation of these estimates.

We use $N\!=\!250$ and $M\!=\!100$, which we found to be a good compromise between sampling accuracy and computation speed. 
As expected, $\epsilon_{\rm met}$ is the dominant source of uncertainty in most images, while $\epsilon_\sigma$ is relevant only in images with very low signal-to-noise ratio, typical of FUV and W4 images of the faintest galaxies.

\section{Phase-space distribution and wind maps}\label{app:phasespace_wind_plots}
In Fig.\,\ref{fig:phase_space_appendix} we present the phase-space analysis used to extract the wind component from the \ha\ velocity cubes in the 19 galaxies of the \textsc{Dwalin-19} sample. 
For each system we show the same panels (a), (d1), (d2), (e1) and (e2) shown in Fig.\,\ref{fig:phase_space_method} for He\,2-10. 
We redirect the reader to the description of Fig.\,\ref{fig:phase_space_method} for detailed information on each panel.

\begin{figure*}
\begin{center}
\frame{\includegraphics[width=0.43\textwidth]{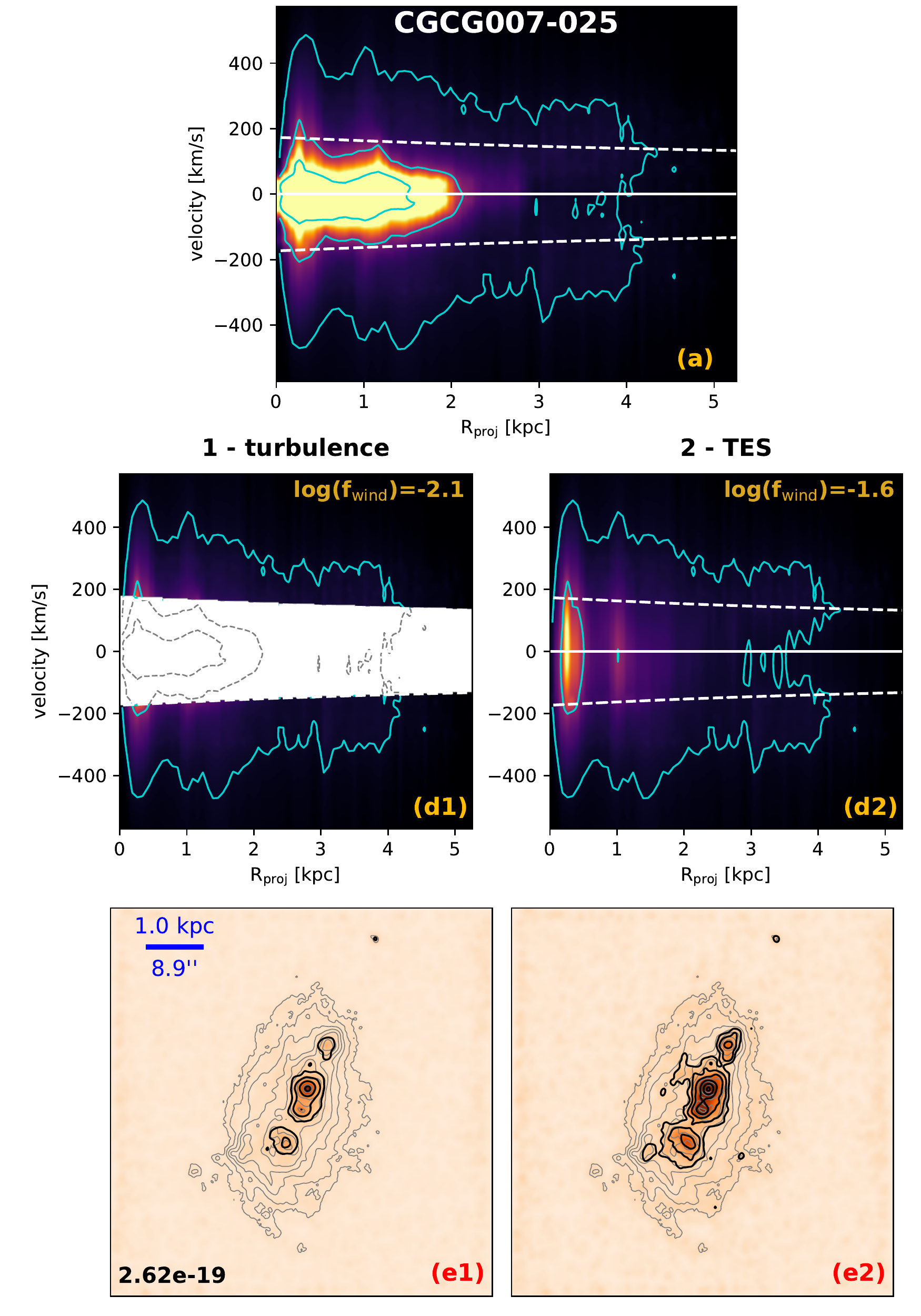}}
\frame{\includegraphics[width=0.43\textwidth]{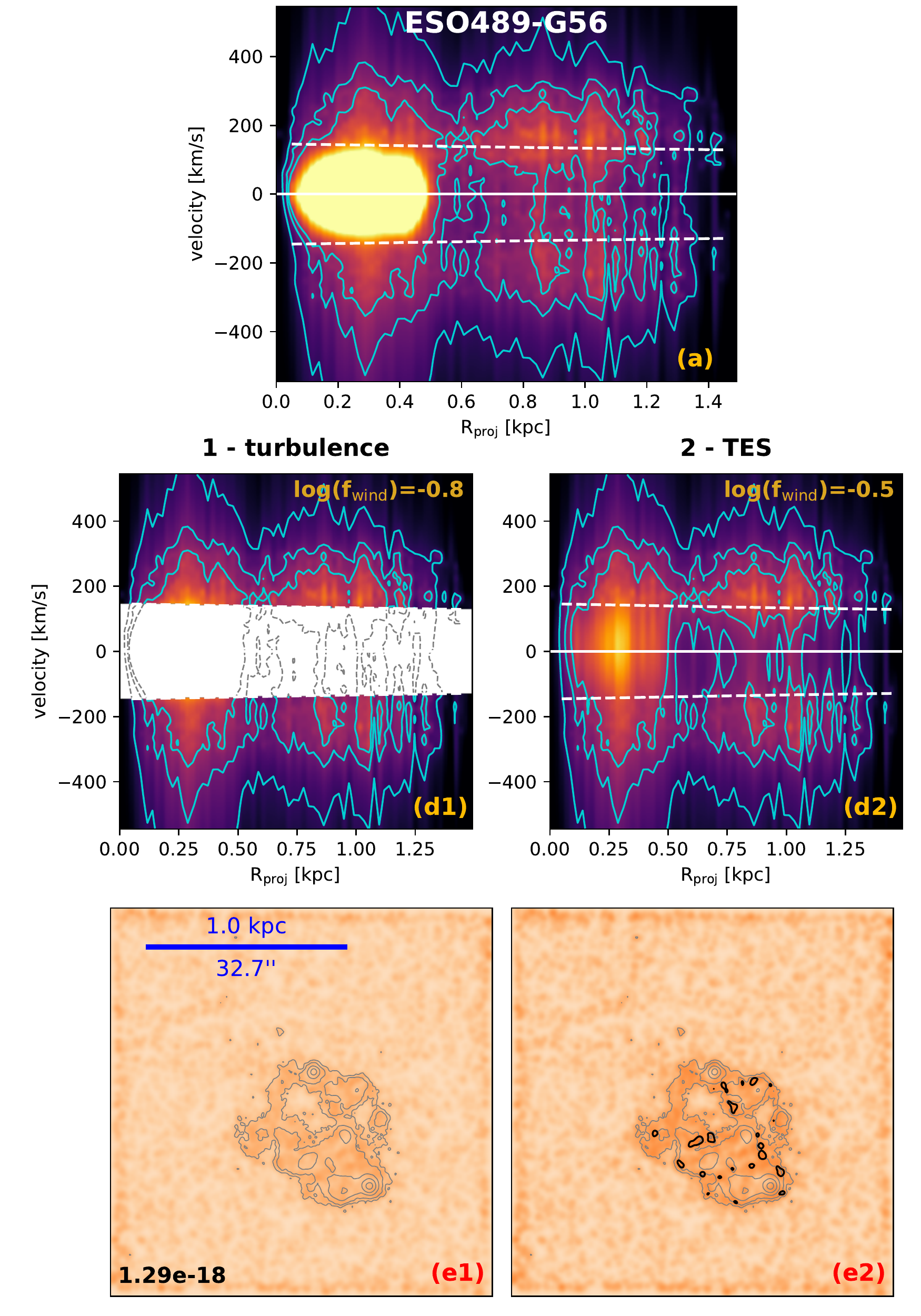}}
\frame{\includegraphics[width=0.43\textwidth]{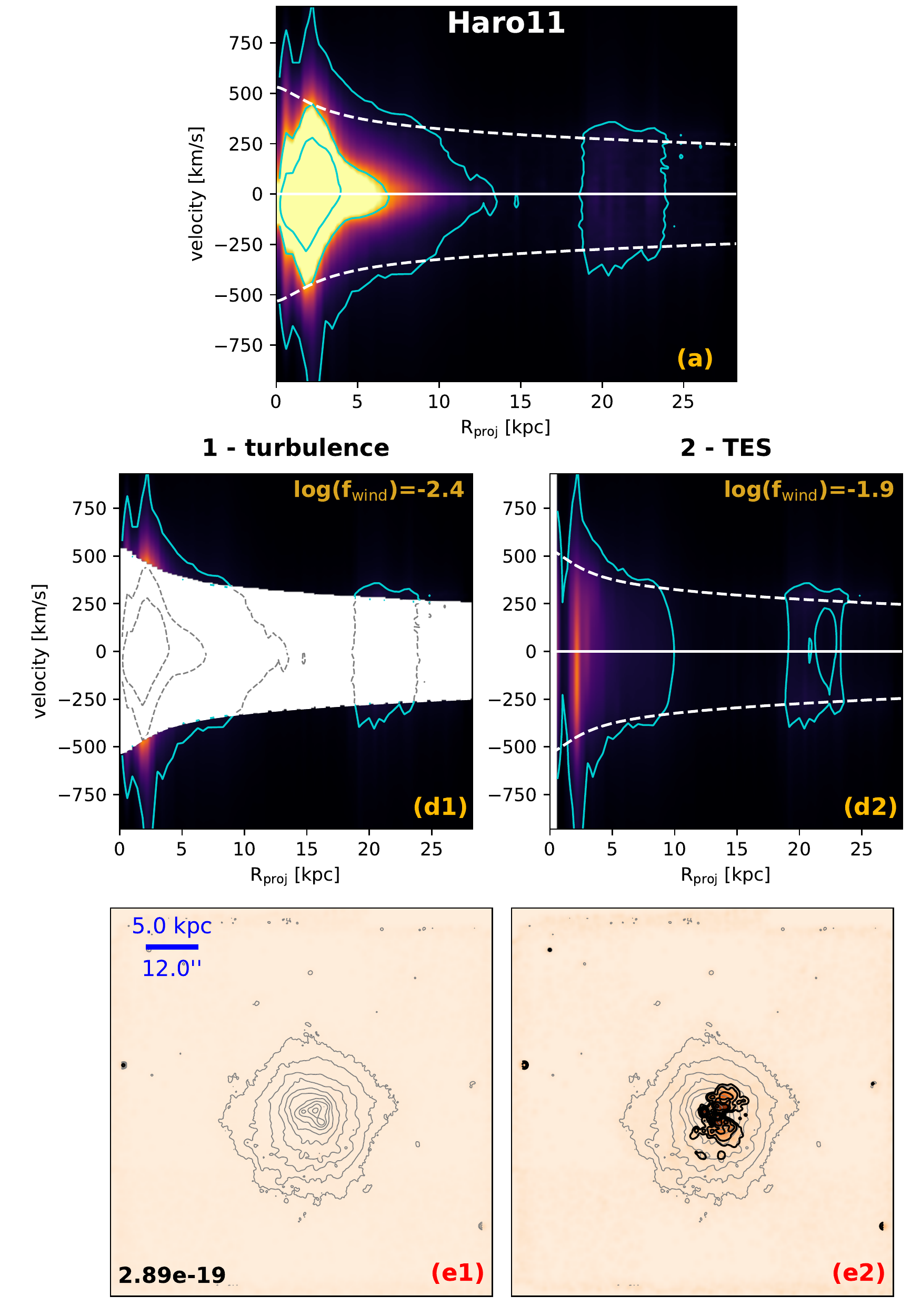}}
\frame{\includegraphics[width=0.43\textwidth]{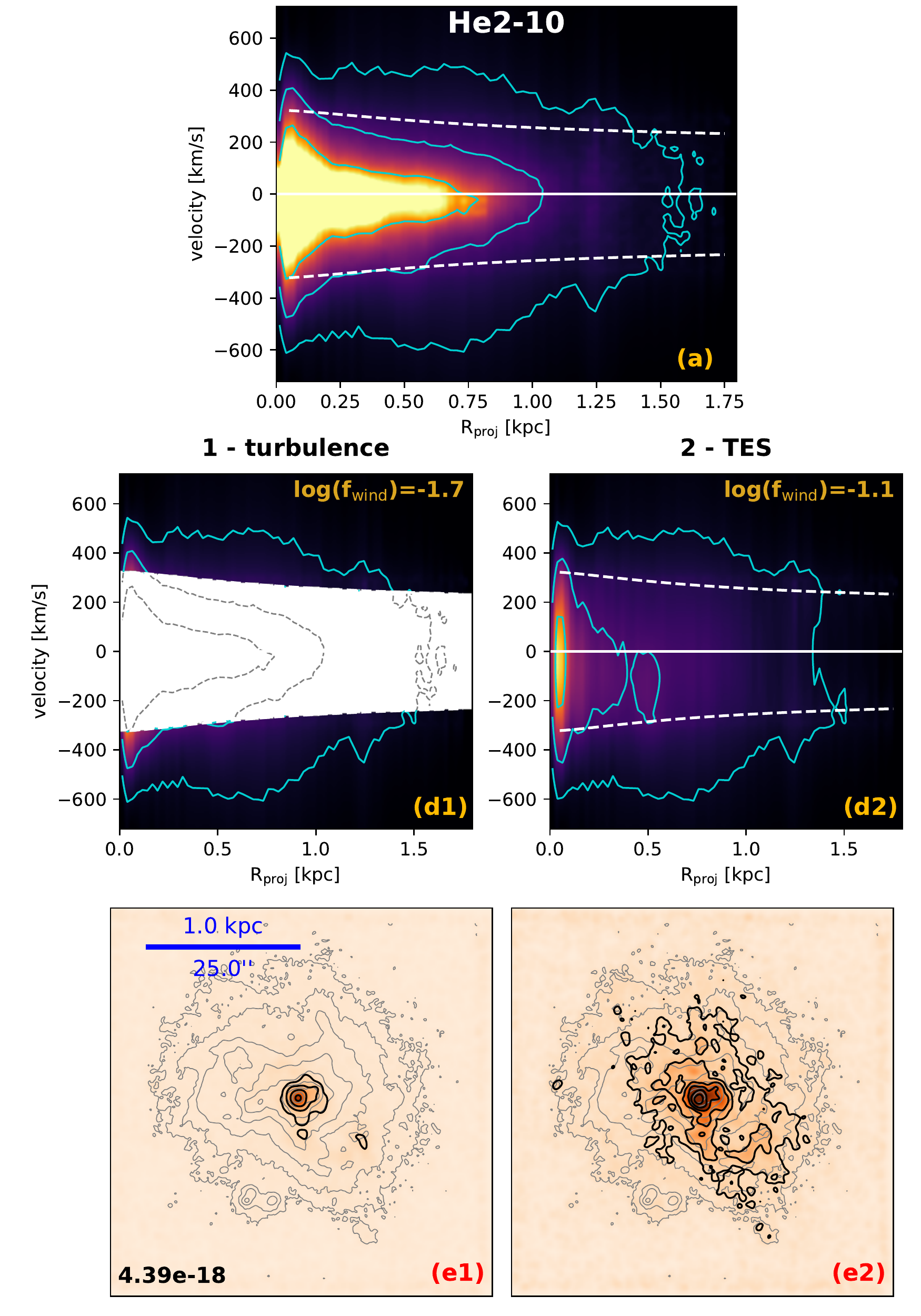}}
\caption{Extraction of the wind component from the \ha\ line in four galaxies of the \textsc{Dwalin-19} sample. The individual panels are analogous to those shown in Fig.\,\ref{fig:phase_space_method}, with the galaxy name indicated on top of panel (a).
To improve their visualisation, the intensity maps have been smoothed to a resolution (FWHM) of $\sim1.5"$. 
Iso-intensity contours in panels (e1) and (e2) are spaced by a factor of three, with the outermost being at an intensity level indicated in the bottom-left corner of panel (e1) (in units of erg\,s$^{-1}$\,cm$^{-2}$, corresponding to $4\sigma_{\rm noise}$), both for the whole emission (grey contours) and for the wind component alone (black contours). No contours are shown when fluxes are below such value.}
\label{fig:phase_space_appendix}
\end{center}
\end{figure*}
\addtocounter{figure}{-1}

\begin{figure*}
\begin{center}
\frame{\includegraphics[width=0.43\textwidth]{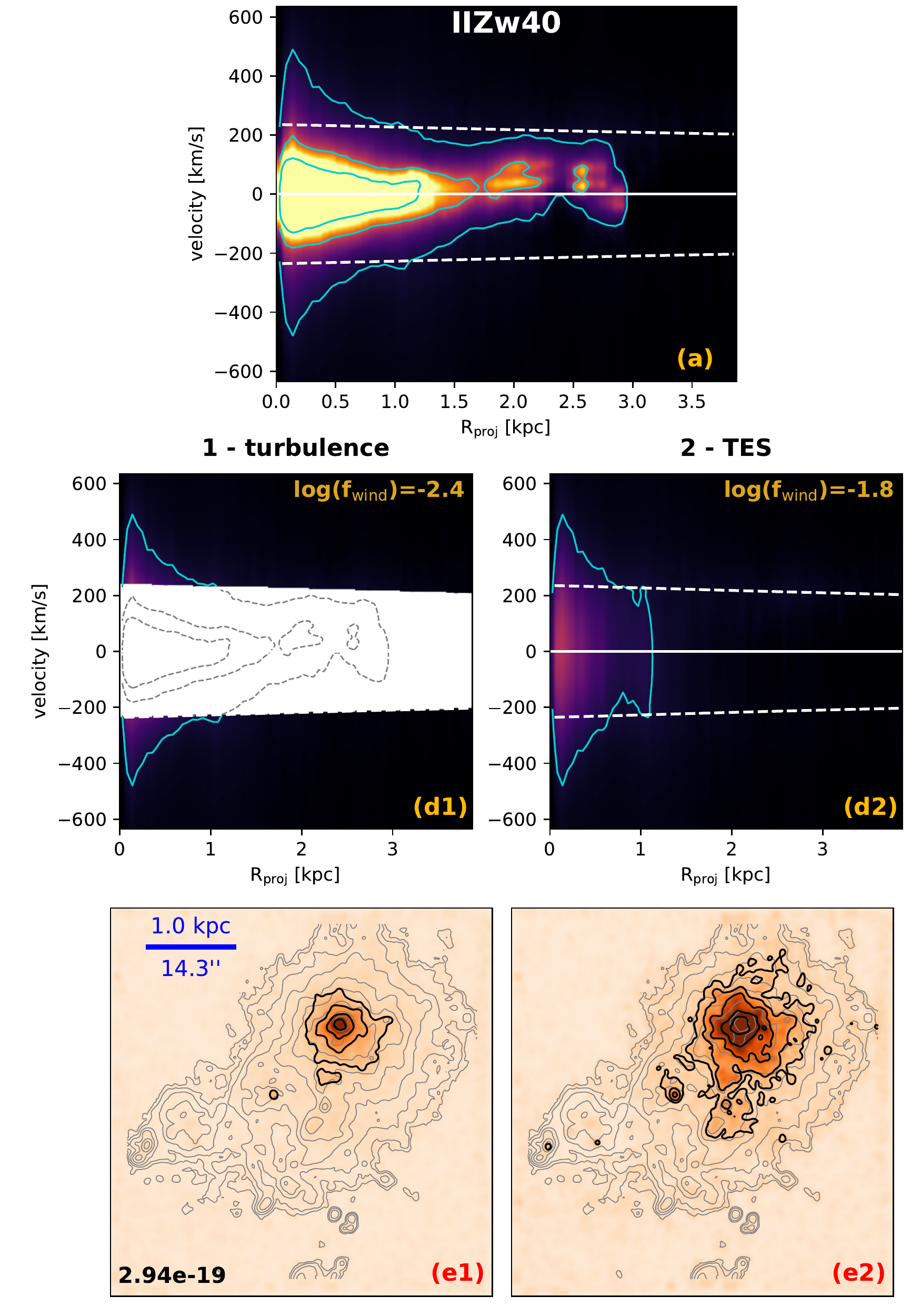}}
\frame{\includegraphics[width=0.43\textwidth]{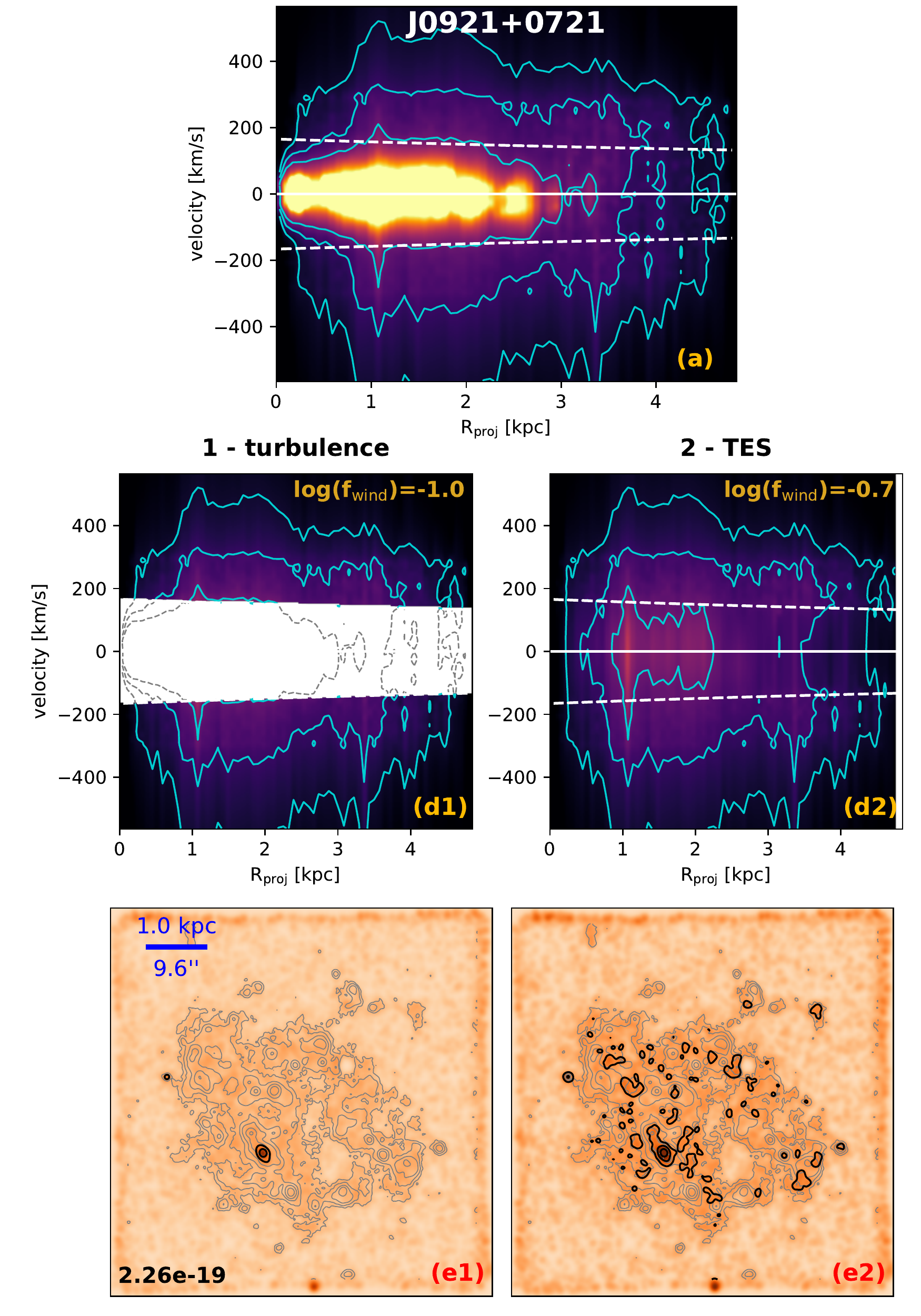}}
\frame{\includegraphics[width=0.43\textwidth]{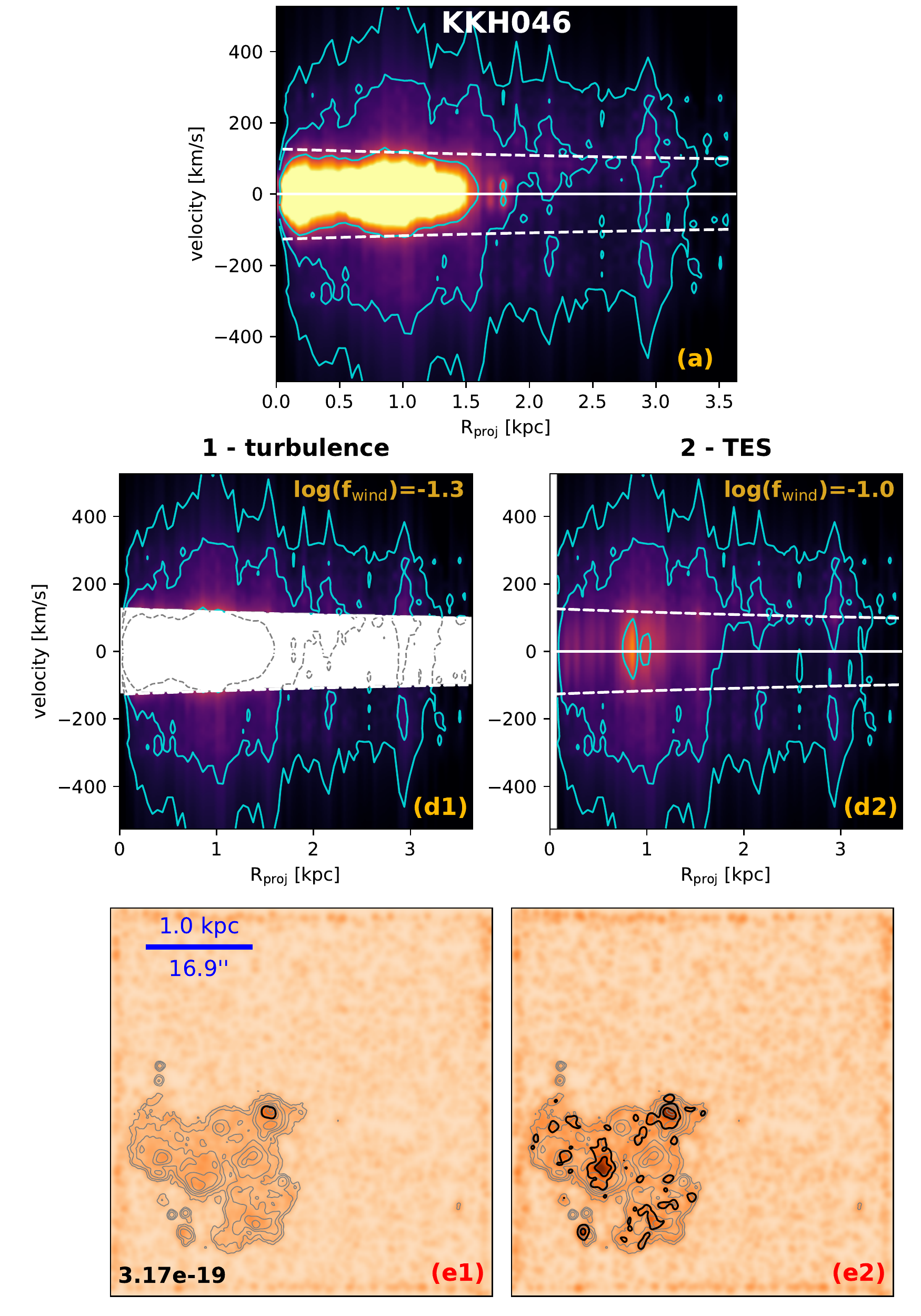}}
\frame{\includegraphics[width=0.43\textwidth]{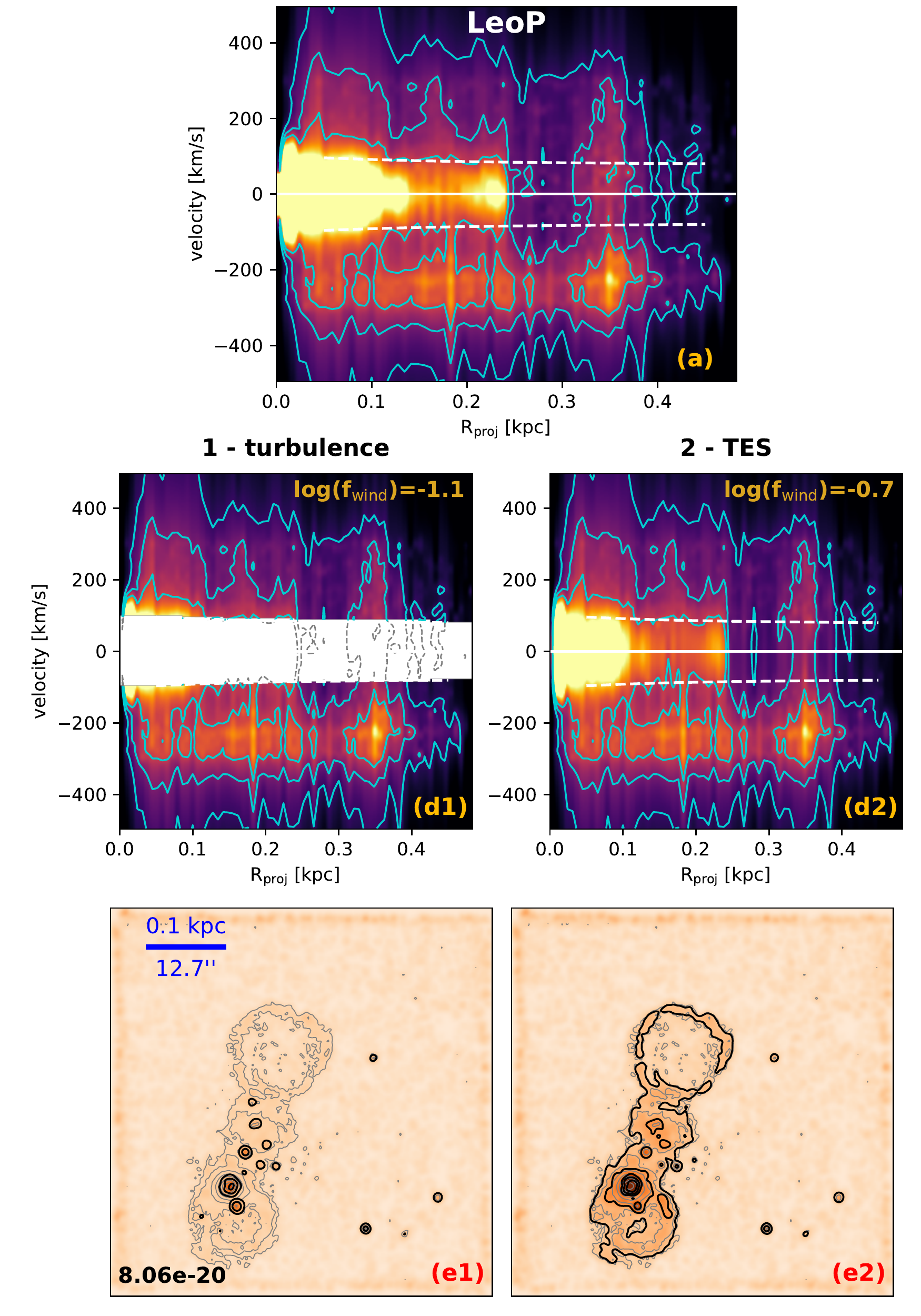}}
\caption{continued.}
\label{fig:phase_space_appendix}
\end{center}
\end{figure*}
\addtocounter{figure}{-1}

\begin{figure*}
\begin{center}
\frame{\includegraphics[width=0.43\textwidth]{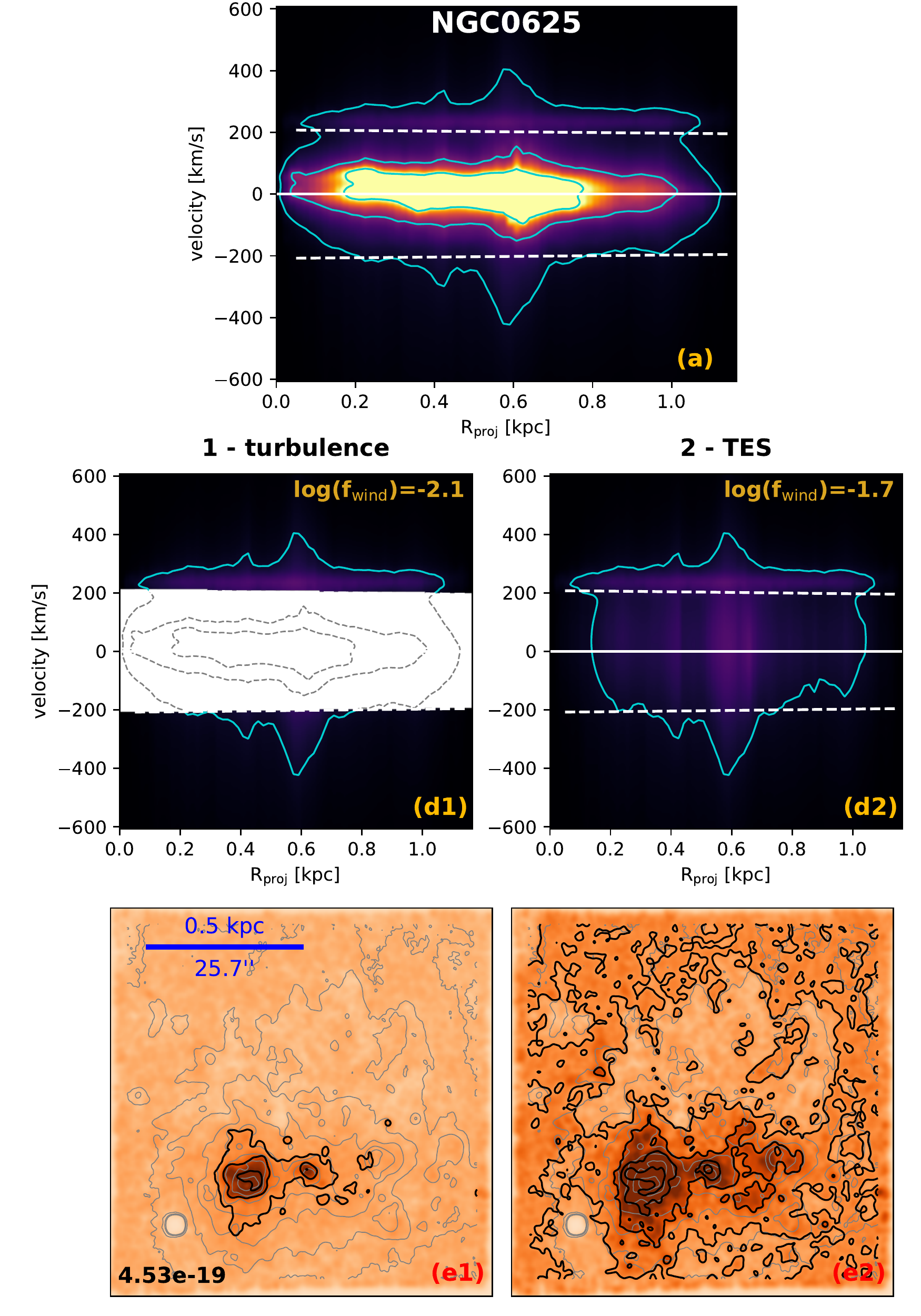}}
\frame{\includegraphics[width=0.43\textwidth]{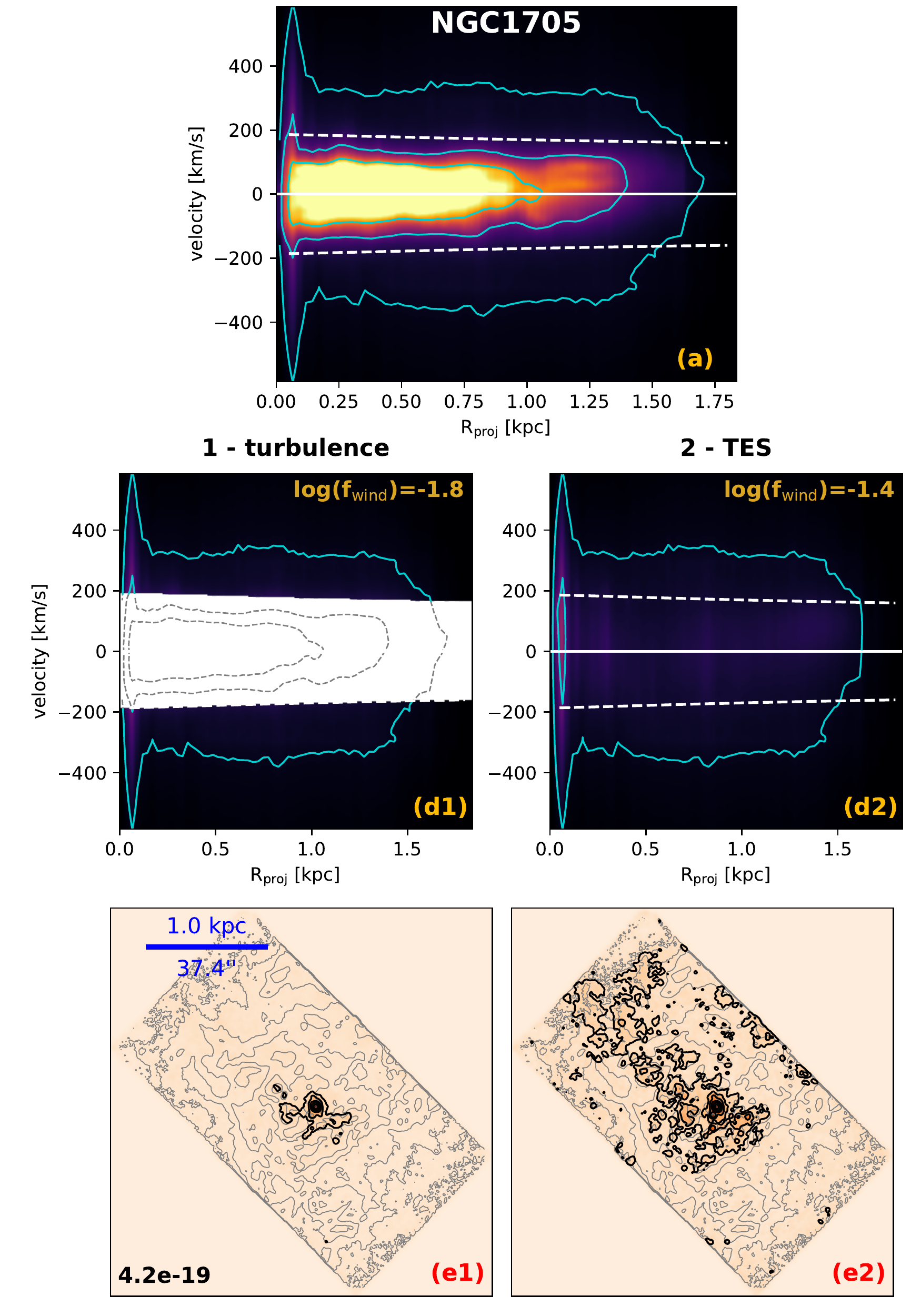}}
\frame{\includegraphics[width=0.43\textwidth]{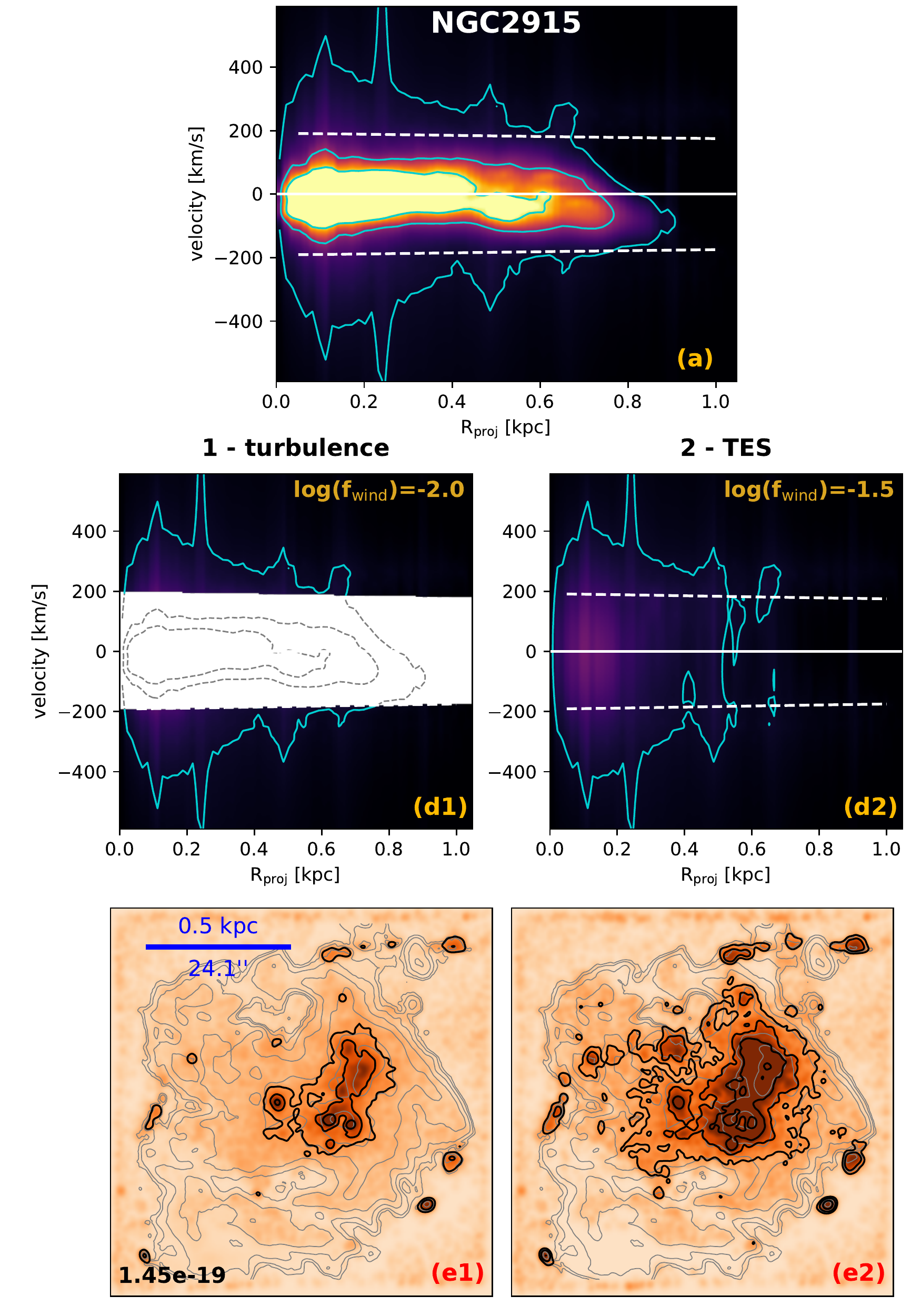}}
\frame{\includegraphics[width=0.43\textwidth]{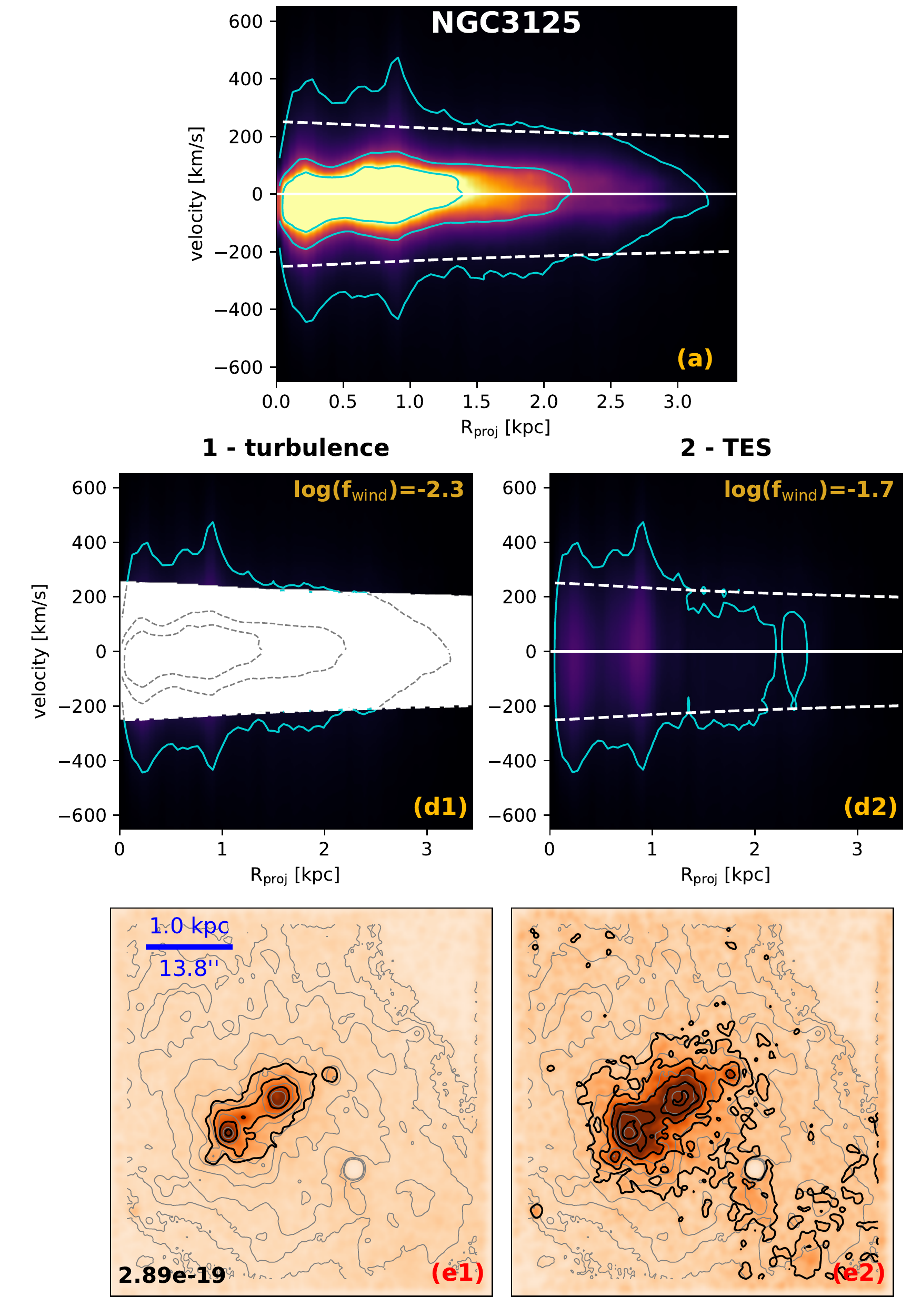}}
\caption{continued.}
\label{fig:phase_space_appendix}
\end{center}
\end{figure*}
\addtocounter{figure}{-1}

\begin{figure*}
\begin{center}
\frame{\includegraphics[width=0.43\textwidth]{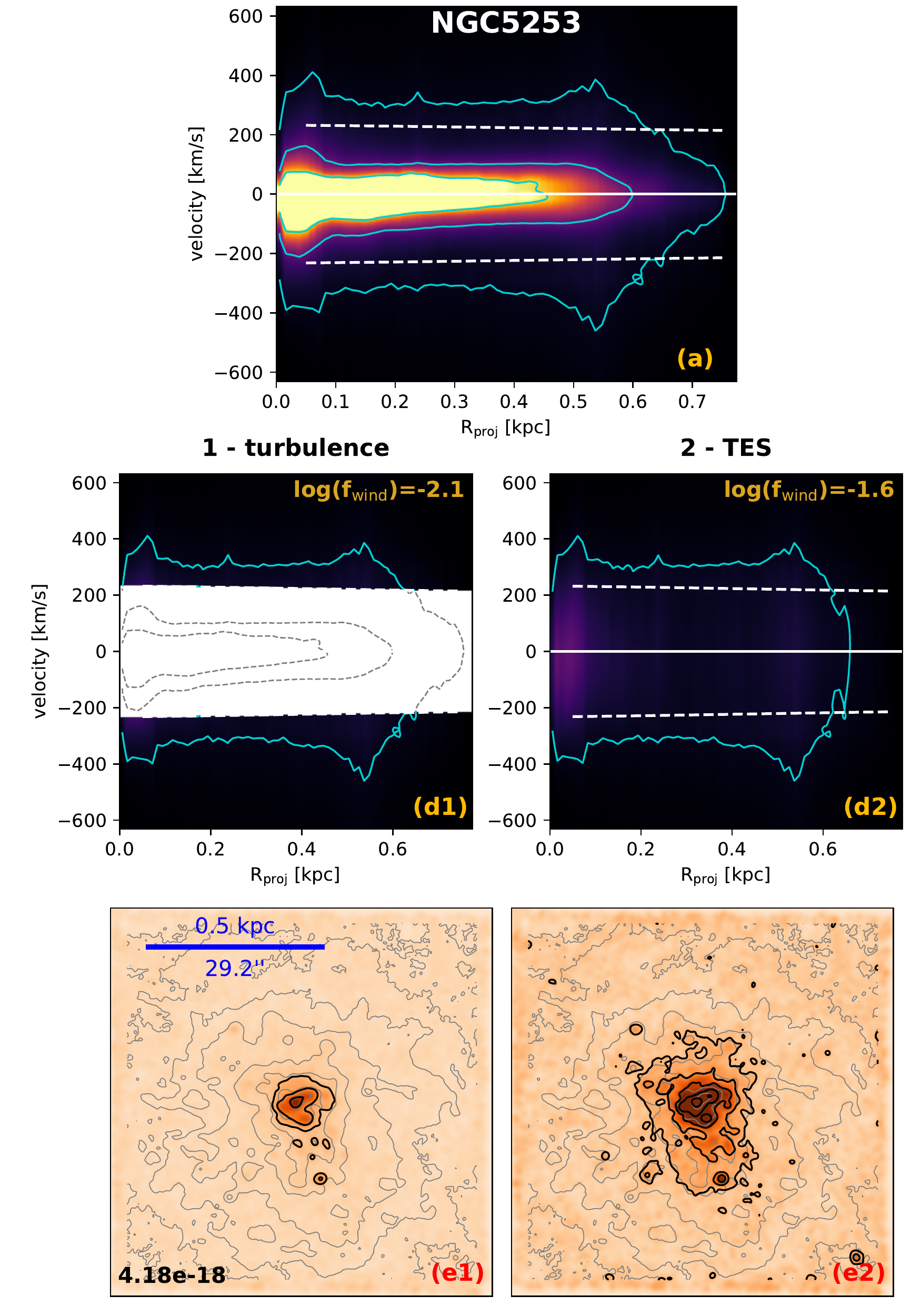}}
\frame{\includegraphics[width=0.43\textwidth]{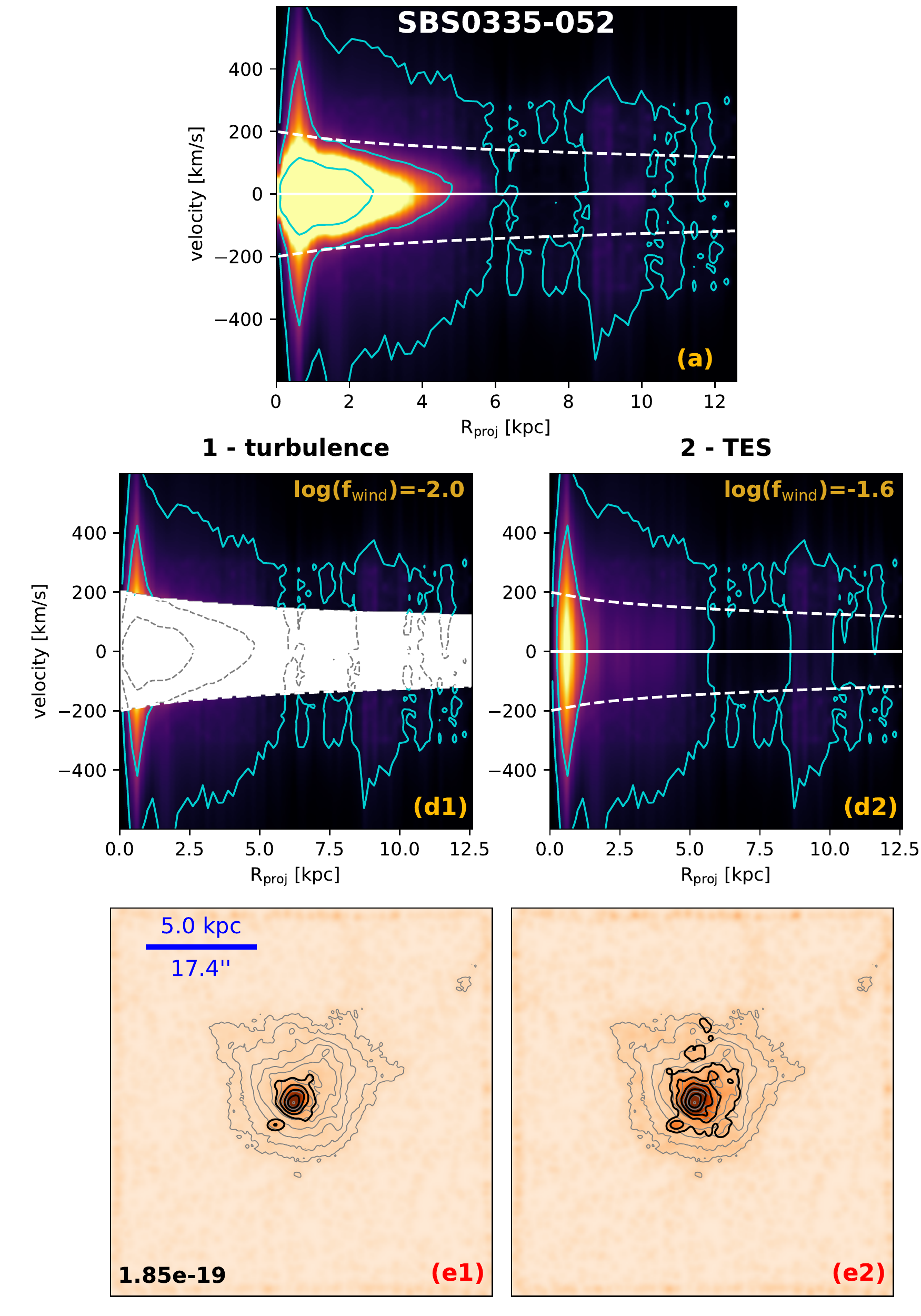}}
\frame{\includegraphics[width=0.43\textwidth]{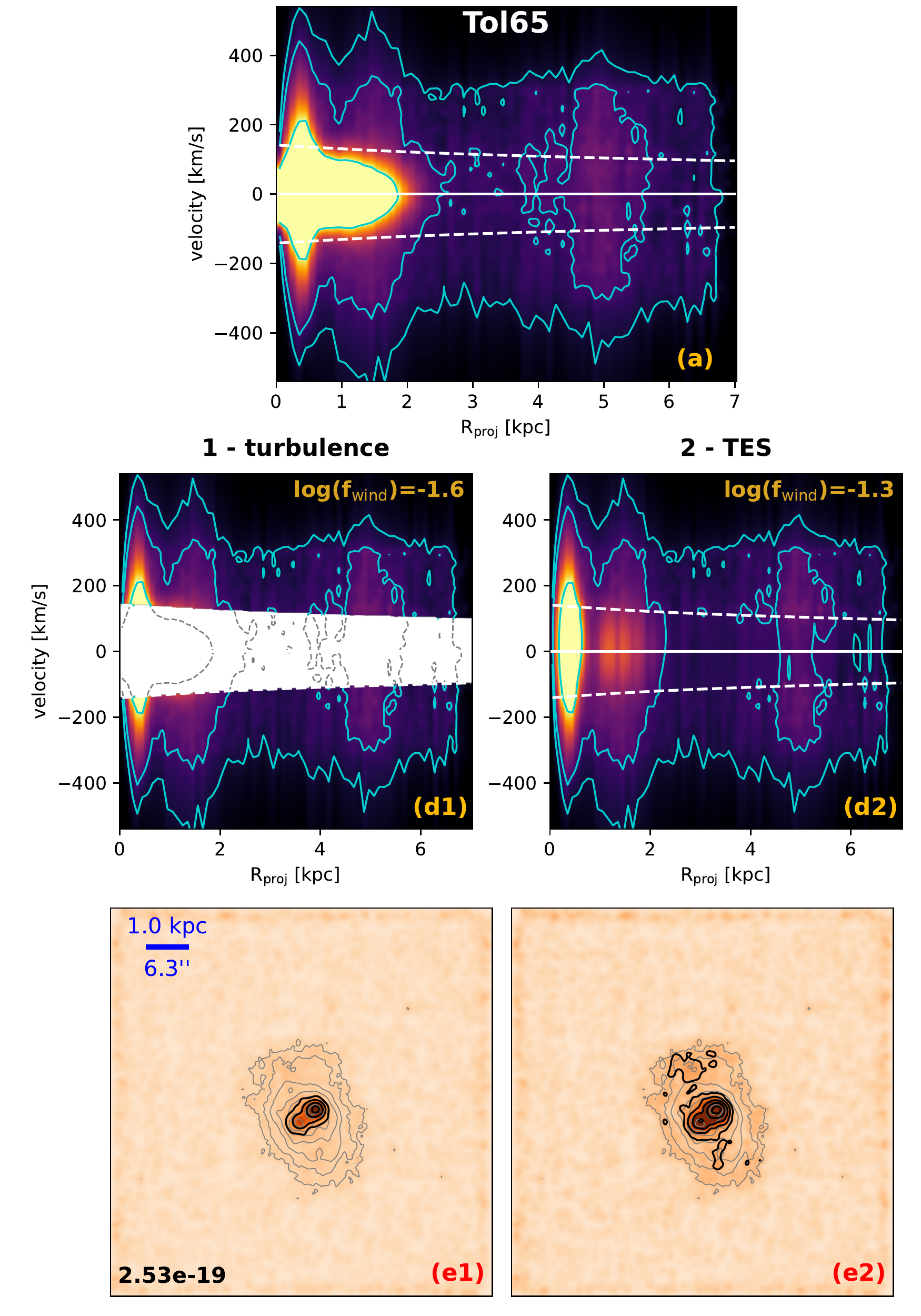}}
\frame{\includegraphics[width=0.43\textwidth]{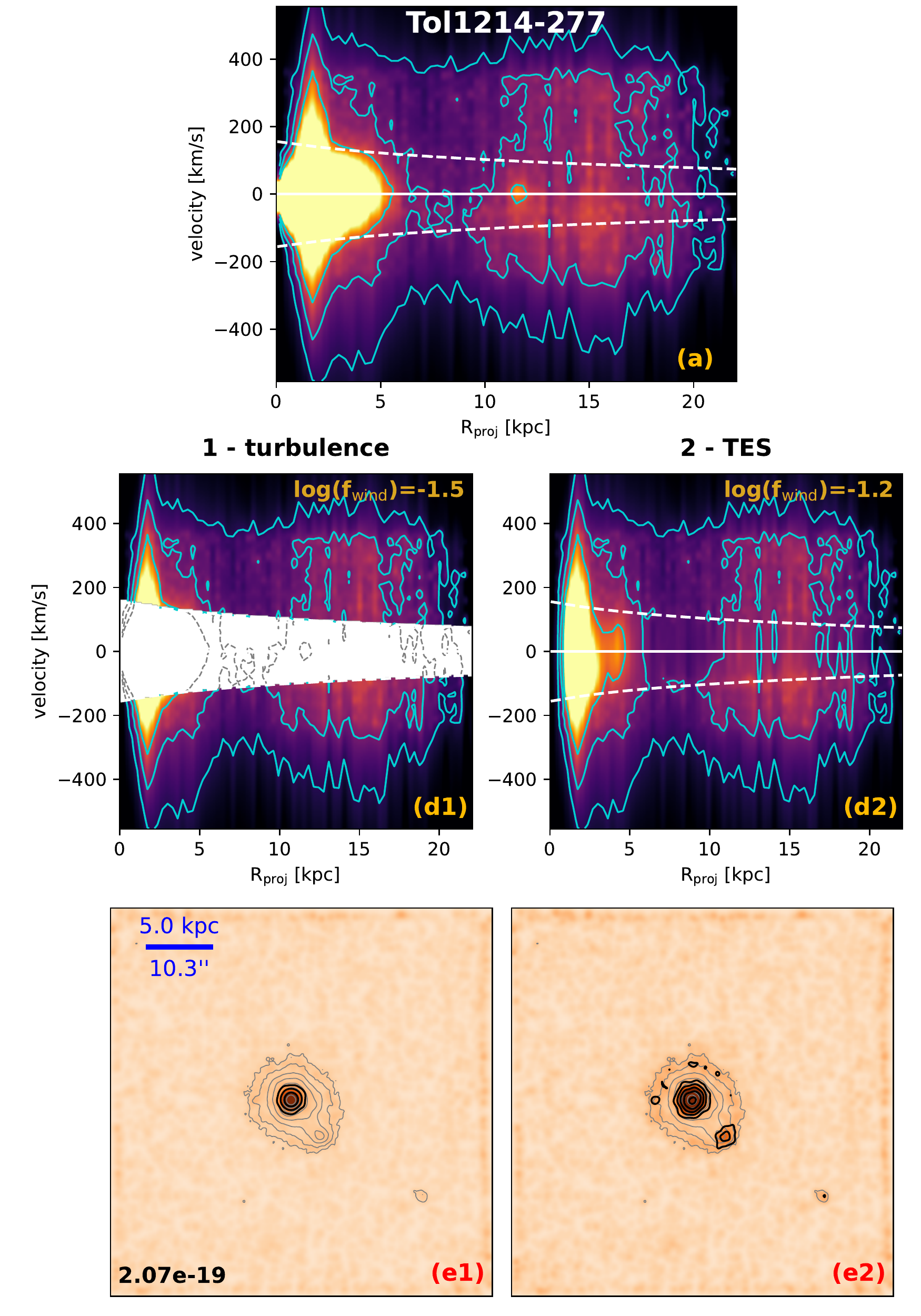}}
\caption{continued.}
\label{fig:phase_space_appendix}
\end{center}
\end{figure*}
\addtocounter{figure}{-1}

\begin{figure*}
\begin{center}
\frame{\includegraphics[width=0.43\textwidth]{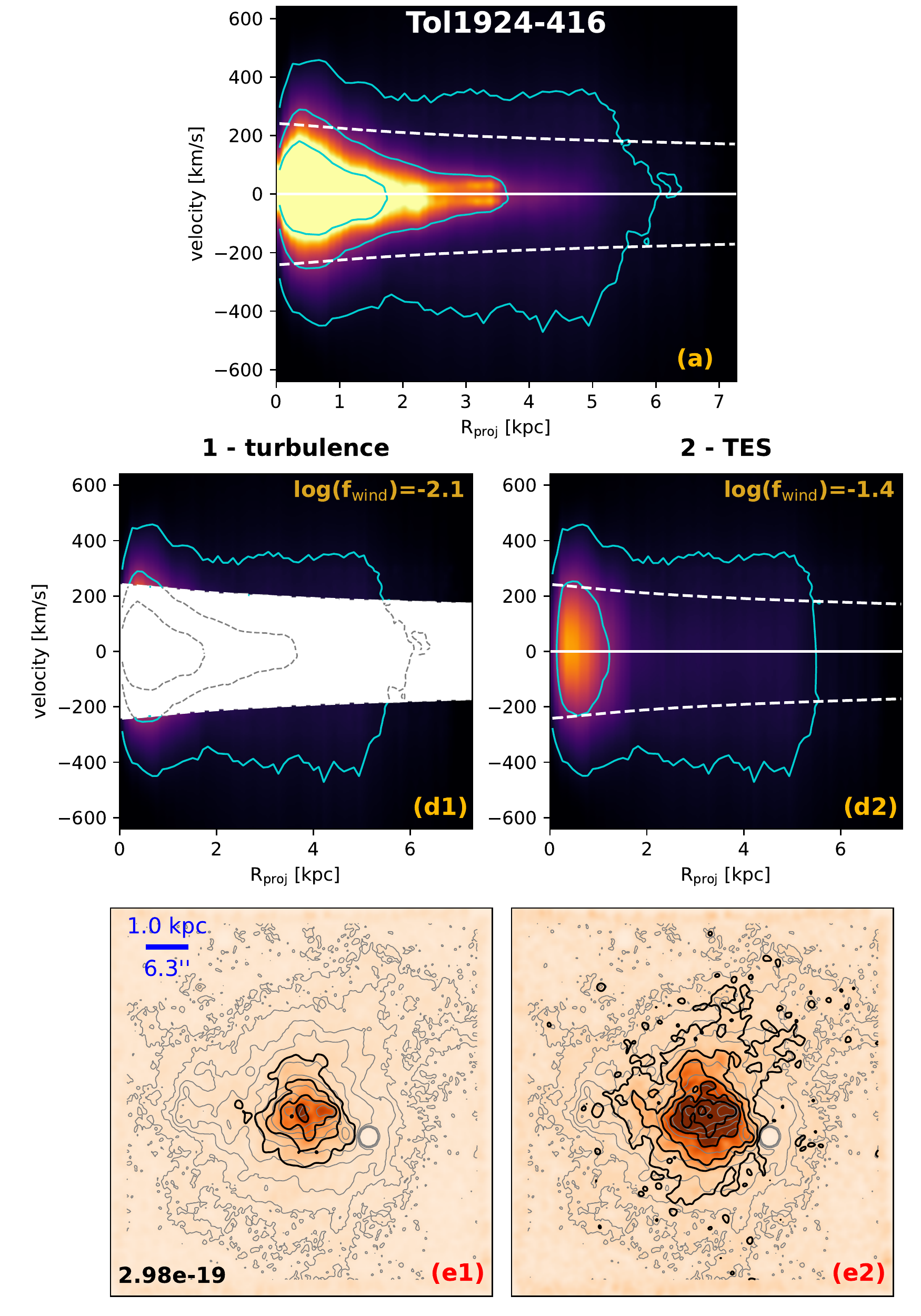}}
\frame{\includegraphics[width=0.43\textwidth]{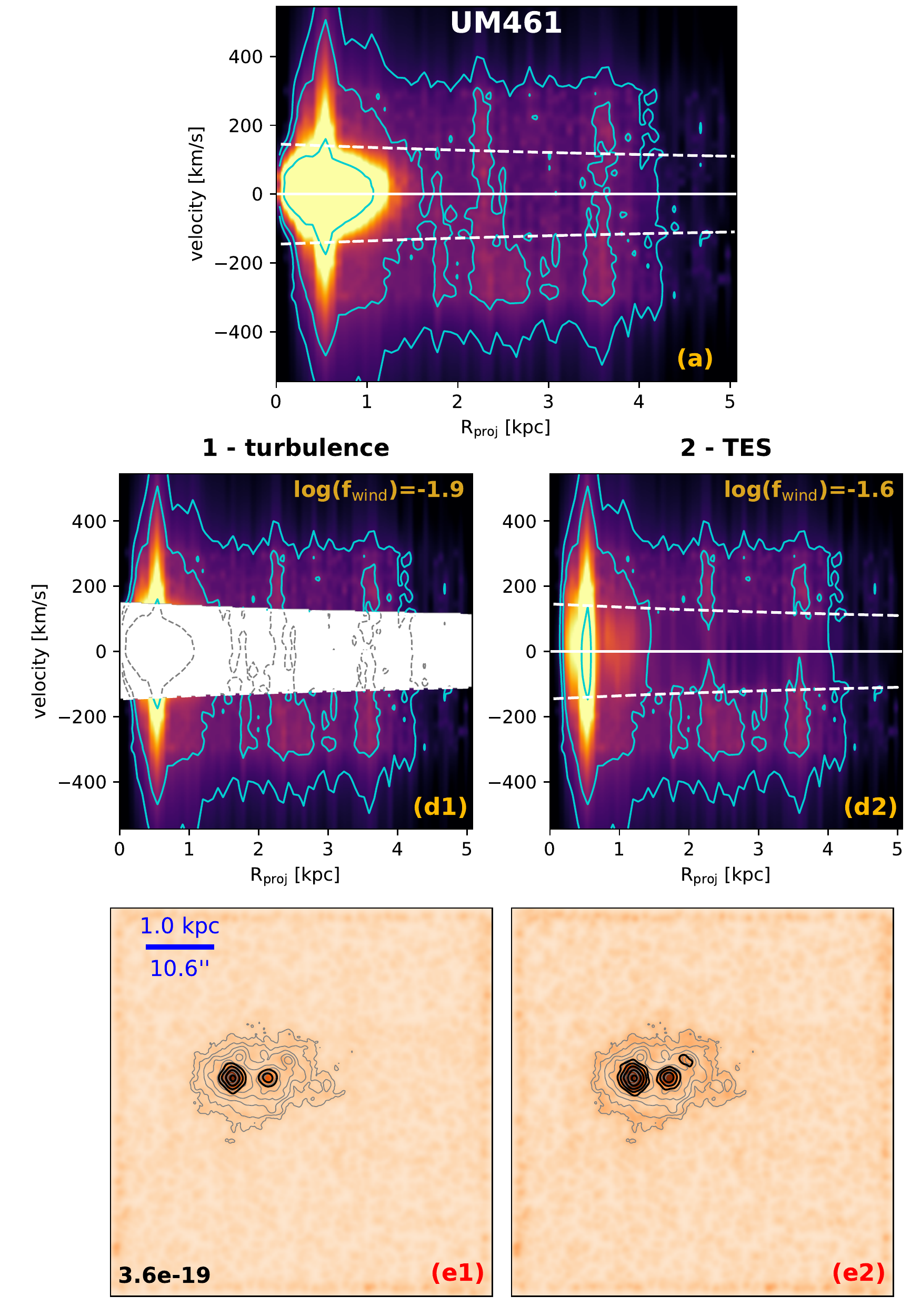}}
\frame{\includegraphics[width=0.43\textwidth]{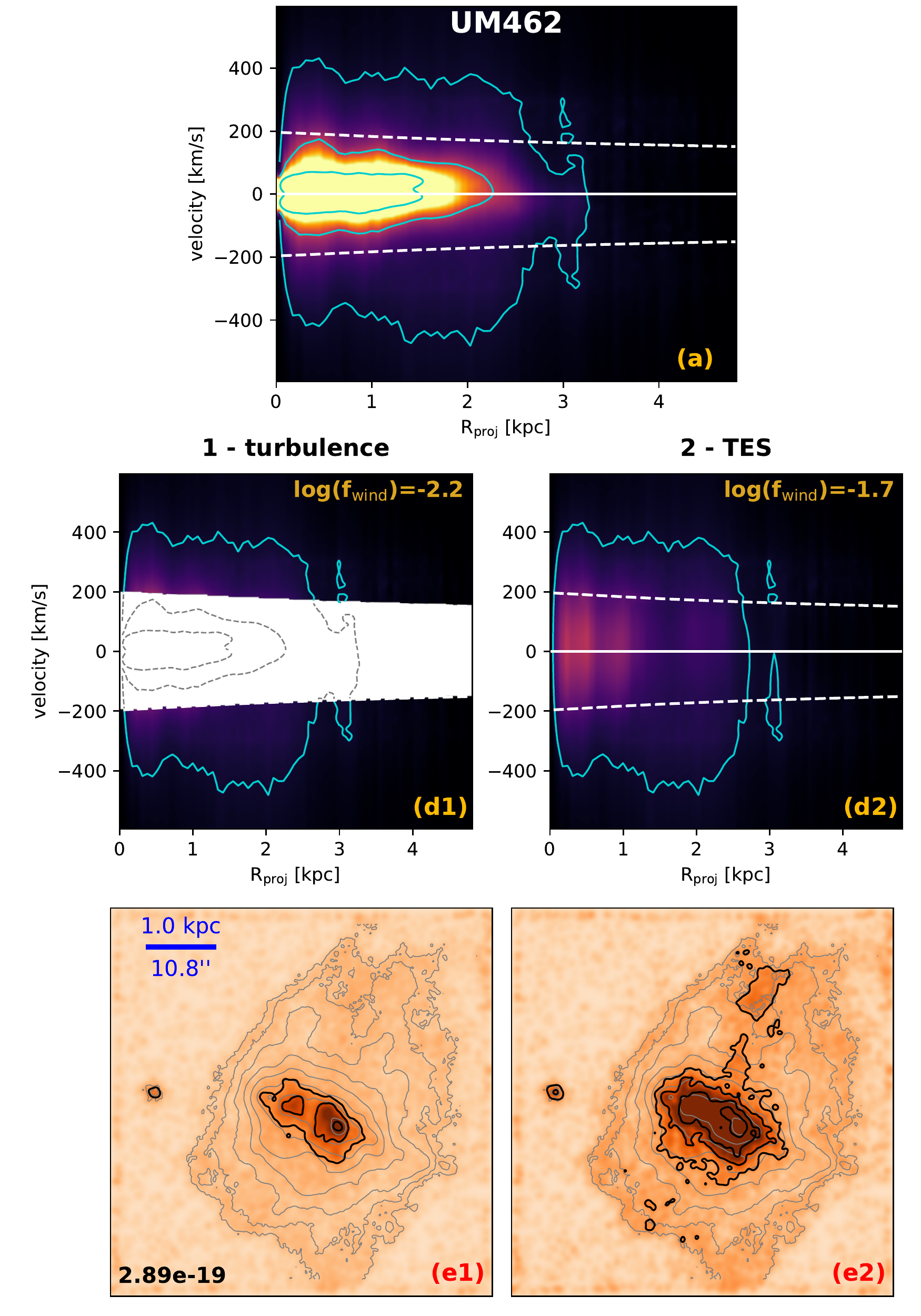}}
\caption{continued.}
\label{fig:phase_space_appendix}
\end{center}
\end{figure*}

\end{document}